\documentclass[aps, prb, singlecolumn, eqsecnum, showpacs,a4paper]{revtex4}

\usepackage{epsfig}
\usepackage{amsmath}
\usepackage{amssymb}
\usepackage{mathrsfs}
\usepackage{mathtools}
\usepackage{comment}
\usepackage{color}
\usepackage{braket}
\usepackage{nicefrac}
\usepackage{yfonts}
\usepackage{bm}
\usepackage{enumerate}
\newcommand{\be}{\begin{equation}}
\newcommand{\ee}{\end{equation}}
\newcommand{\ba}{\begin{aligned}}
\newcommand{\ea}{\end{aligned}}

\newcommand{\bea}{\begin{eqnarray}}
\newcommand{\eea}{\end{eqnarray}}
\def\nn{\nonumber\\}
\def\fr#1{(\ref{#1})}

\newcommand{\gothb}{\text{\textgoth{b}}}

\begin{document}
\title{Relaxation after quantum quenches in the spin-1/2 Heisenberg
  XXZ chain}  
\author{Maurizio Fagotti$^1$, Mario Collura$^2$, Fabian H.L. Essler$^1$
  and Pasquale Calabrese$^2$}
\affiliation{$^1$The Rudolf Peierls Centre for Theoretical Physics,
    Oxford University, Oxford, OX1 3NP, United Kingdom}
\affiliation{$^2$Dipartimento di Fisica dell'Universit\`a di Pisa and INFN,
  56127 Pisa, Italy}
\begin{abstract}
We consider the time evolution after quantum quenches in the spin-1/2
Heisenberg XXZ quantum spin chain with Ising-like anisotropy. The time
evolution of short-distance spin-spin correlation functions is studied
by numerical tensor network techniques for a variety of initial
states, including N\'eel and Majumdar-Ghosh states and the ground
state of the XXZ chain at large values of the anisotropy. The
various correlators appear to approach stationary values, which are
found to be in good agreement with the results of exact calculations
of stationary expectation values in appropriate generalized Gibbs
ensembles. In particular, our analysis shows how symmetries of the
post-quench Hamiltonian that are broken by particular initial states
are restored at late times.

\end{abstract}
\maketitle
\section{Introduction}

Recent years witnessed great advances in our understanding of
\emph{isolated} non-equilibrium many-particle quantum systems, mainly
triggered by ground-breaking experiments with ultra-cold, trapped atoms
\cite{kww-06,tetal-11,cetal-12,getal-11,shr-12}. One of the most
celebrated results is that expectation values of {\it local} observables  
generically approach stationary values at late times in the thermodynamic
limit, in spite of the time evolution being unitary and the entire system
concomitantly always being in a pure state (assuming that it started
out in a pure state). There is compelling evidence that these
stationary values can be predicted by statistical ensembles without
having to solve the complicated non-equilibrium dynamics.

For non-integrable models the appropriate statistical ensemble
is expected to be the standard Gibbs distribution with an effective
temperature fixed by the value of the energy in the initial state
\cite{nonint}. For integrable models, the existence of {\it local}
conservation laws strongly constrains the dynamics, and it has been
proposed\cite{GGE} that stationary values are described by a
generalised Gibbs ensemble (GGE). By now rather convincing evidence
supporting this proposal has
accumulated\cite{BS_PRL08,Cramer_10,CEF,FE_13a,CC_JStatMech07,IC_PRA09,BKL_PRL10,FM_NJPhys10,EEF:12,Pozsgay_JStatMech11,CCR_PRL11,CIC_PRE11,CK_PRL12,CE_PRL13,MC_NJPhys12,
Pozsgay:13a,FE_13b,m-13,CSC:13b,KCC:13a}, but a general
proof is still outstanding.
Whereas in equilibrium integrability may be viewed chiefly as powerful tool for
obtaining exact solutions of paradigmatic models, out of equilibrium
it is an essential physical feature. For models that can be mapped to
free fermions or bosons the stationary behaviour as well as
essentially the full dynamics has been obtained analytically 
\cite{BS_PRL08,CC_JStatMech07,IC_PRA09,CEF,FE_13a,EEF:12,CSC:13b,KCC:13a}.
Unfortunately the methods applicable to these cases do not generalise
to \emph{interacting} integrable models, i.e. models with
momentum-dependent (dressed) scattering matrices.

A crucial next step is therefore to calculate expectation values of local
observables in the stationary state after a quantum quench in interacting
integrable models, assuming that local properties can be described by
an appropriate GGE. This task has recently been undertaken by several
groups using different integrability-based
techniques\cite{Pozsgay:13a,FE_13b,mc-12b,ce-13,m-13,nwbc-13,KSCCI}. 
In some of these works, specific predictions for stationary values of
local observables have been made. Given the underlying assumption of
relaxation to a GGE these predictions need to be checked by
independent methods such as numerical simulations.

Here we focus on the non-equilibrium dynamics of the XXZ spin chain
described by the Hamiltonian  
\be
H^{(1)}(\Delta)=\frac{1}{4}\sum_{\ell=1}^L
\sigma_\ell^x\sigma_{\ell+1}^x+\sigma_{\ell}^y\sigma_{\ell+1}^y+\Delta(\sigma_\ell^z\sigma_{\ell+1}^z-1)\ ,\quad \Delta=\cosh(\eta)>1.
\label{HXXZ}
\ee 
We will consider a variety of initial states and consider the question
whether the dynamics of local observables exhibits relaxation to a
stationary state compatible with a GGE. The latter is of the form
\be
\rho_{\rm GGE}=\frac{1}{Z_{\rm GGE}}\exp\left(h S^z-\sum_{l=1}\lambda_lH^{(l)}\right).
\label{rhoGGE}
\ee
Here $S^z=\frac{1}{2}\sum_\ell \sigma_\ell^z$ is the z-component of
total spin, $H^{(1)}$ is the Hamiltonian and $H^{(l)}$ are
local\cite{CEF} integrals of motion that fulfil $[H^{(m)},H^{(n)}]=0$
and are  obtained by taking logarithmic 
derivatives of the transfer matrix of the six-vertex model
\cite{Korepinbook}. The Lagrange multipliers $h$ and
$\lambda_l$ are fixed by the requirement that the expectation values
of the conservation laws are the same at time $t=0$ and in the
stationary state
\be
\lim_{L\to\infty}\frac{\langle\Psi_0|S^z|\Psi_0\rangle}{L}
=\lim_{L\to\infty}\frac{{\rm Tr}\left[\rho_{\rm GGE}
S^z\right]}{L}\, ,\qquad \lim_{L\to\infty}\frac{\langle\Psi_0|H^{(l)}|\Psi_0\rangle}{L}
=\lim_{L\to\infty}\frac{{\rm Tr}\left[\rho_{\rm GGE}
H^{(l)}\right]}{L}.
\label{constraints}
\ee
We stress that locality of the integrals of motion is the key feature
which sets integrable models apart from generic ones. In fact,
\emph{any} quantum mechanical Hamiltonian $H$ has as many integrals of
motions as there are basis states in the Hilbert space, as the
one-dimensional projectors $P_n=|\psi_n\rangle\langle\psi_n|$ on energy
eigenstates are in involution and commute with $H$. However, they are
not local. The GGE built with all the $P_n$'s is by definition
equivalent to the so-called diagonal ensemble, which describes the
infinite time average of arbitrary observables (including non-stationary
ones) in a finite volume (assuming that spectral degeneracies do not
play a role).

The last few months have witnessed considerable progress in developing
analytic approaches to the quench problem in the XXZ spin chain
\cite{FE_13b,Pozsgay:13a,fp-13,p-13a,la-13}.  
In this manuscript we follow the route developed by two of the present
authors \cite{FE_13b}, which allows to calculate short distance
spin-spin correlation functions in the appropriate GGEs. As compared
to other methods, the approach of Ref.~\onlinecite{FE_13b} works
directly in the thermodynamic limit {\it at finite energy density}
compared to the ground state of the post-quench Hamiltonian. Here we
extend the calculations of Ref.~\onlinecite{FE_13b} to a variety of initial
states not previously considered. We also provide the details of how
to treat initial states of matrix-product form, which may be known only
numerically from a ground state tensor network computation.
\subsection{Symmetry restoration after quantum quenches}
\label{ssec:symm}
A key issue we will be investigating is that of \emph{symmetry restoration}.
The post-quench Hamiltonian exhibits a number of symmetries such as
\begin{enumerate}
\item{} $U(1)$ rotations around the z-axis in spin space:
\be
\sigma^\pm_\ell\longrightarrow e^{\pm
  i\varphi}\sigma^\pm_{\ell}\ ,\quad
\sigma^\pm_\ell=\frac{\sigma^x_\ell\pm i\sigma^y_\ell}{2}.
\ee
\item{} Translational invariance:
\be
\sigma^\alpha_\ell\longrightarrow\sigma^\alpha_{\ell+1}.
\ee
\item{} Bond inversion symmetry $P_B$:
\be
\sigma^\alpha_\ell\longrightarrow\sigma^\alpha_{L+1-\ell}.
\ee
\item{} Site inversion symmetry $P_S$:
\be
\sigma^\alpha_\ell\longrightarrow\sigma^\alpha_{L+2-\ell}.
\ee
\end{enumerate}
The local conservation laws of the XXZ chain, and hence the GGE
\fr{rhoGGE}, share the first two of these symmetries. However, the
higher conservation laws can be odd under the inversion
symmetries. For example, 
\be
H^{(2)}(\Delta)\propto\sum_\ell
\epsilon_{\alpha\beta\gamma}
{\sigma}_\ell^\beta{\sigma}_{\ell+2}^\gamma
{\sigma}_{\ell+1}^\alpha\left[\Delta+(1-\Delta)\delta_{\alpha,z}\right],
\ee
is parity-odd. The same is true for all $H^{(2n)}$. By construction of
the GGE, stationary values of local observables will exhibit the first
two symmetries, provided that the GGE indeed describes the late time
behaviour after the quench. 
This implies that if we start out in an initial state that breaks the
U(1) or translational symmetry, \emph{they must be restored in the
  course of the unitary time evolution}. The situation is different
for $P_{B,S}$: these are not necessarily
restored. However, for the initial states that are invariant under
at least one of the reflection symmetries, they will be. This can be
seen by noting that as $|\Psi_0\rangle$ is even and $H^{(2n)}$ is odd
under the symmetry, we have
\be
\langle\Psi_0|H^{(2n)}|\Psi_0\rangle=0.
\ee
The constraints \fr{constraints} are then fulfilled by setting all
Lagrange multipliers $\lambda_{2n}$ to zero, as can be checked by
taking the traces in a basis of simultaneous eigenstates of the
Hamiltonian and parity. In other words, the GGE only contains
the parity-even conservation laws $H^{(2n+1)}$ in this case, which
implies the restoration of the reflection symmetries in the stationary
state. All initial states studied below by means of numerical
tensor-product methods are of this kind.
\section{Generalized Gibbs Expectation Values by the Quantum Transfer
  Matrix Method} \label{s:nlies}
In a recent paper Fagotti and Essler \cite{FE_13b} developed an approach for
calculating expectation values of local operators in generalised
Gibbs ensembles describing the stationary states for quenches from
matrix product states. Here we summarise some key results.
The density matrix \fr{rhoGGE} can be viewed as describing a thermal
(Gibbs) ensemble for an integrable Hamiltonian with long-range
interactions. For a given set of Lagrange multipliers $\{\lambda_j\}$
it is then straightforward \cite{FE_13b} to generalise the quantum
transfer matrix approach \cite{QTM} for calculating thermal
expectation values to \fr{rhoGGE} (some remarks regarding the
structure of the largest eigenvalue of the quantum transfer matrix in
our cases are given in Appendix~\ref{app:lambdamax}).
In particular, we may use 
the explicit expressions for short-distance correlators given in
Ref.~[\onlinecite{thermal}], which involve three functions
$\varphi(\mu)$, $\omega(\mu_1,\mu_2)$ and $\omega^\prime(\mu_1,\mu_2)$
that encode the necessary information on the density matrix.
Examples are
\be
\langle\sigma^z_1\sigma^z_2\rangle={\rm
  cth}(\eta)\omega+\frac{\omega'_x}{\eta}\ ,\quad
\langle\sigma^x_1\sigma^x_2\rangle=-\frac{\omega}{2\sinh(\eta)}-
\frac{\cosh(\eta)\omega'_x}{2\eta}\ ,
\label{zzxx}
\ee
where $\omega=\omega(0,0)$, $\omega'_x=\partial_x\omega'(x,y)|_{x,y=0}$.
Determining the Lagrange multipliers by solving the system
\fr{constraints} is a difficult problem \cite{Pozsgay:13a}. In
Ref.~\onlinecite{FE_13b} a method was introduced that avoids having to
calculate them explicitly by working with the generating function
\be
\Omega_{\Psi_0}(\lambda)=-i\sum_{k=1}
\left(\frac{\eta}{\sinh\eta}\right)^k\frac{\lambda^{k-1}}{(k-1)!}
\frac{\langle\Psi_0|H^{(k)}|\Psi_0\rangle}{L}\, .
\label{eq:Om0}
\ee
Given $\Omega_{\Psi_0}(\lambda)$, the largest eigenvalue of the
quantum transfer matrix, and concomitantly the generalised Gibbs
ensemble, is obtained by solving the system of nonlinear integral
equations 
\bea
\label{eq:systemh}
\log\gothb(x)-\log\bar\gothb(x)+h&=&
[(k_++k)\ast\log(1+\gothb)](x)-[(k_-+k)\ast\log(1+\bar\gothb)](x)\ ,\nn
g_\mu^+(x)&=&-d(x-\mu)+\Bigl[k\ast\frac{g_\mu^+}{1+\gothb^{-1}}\Bigr](x)-\Bigl[k_-\ast\frac{g_\mu^-}{1+\bar{\gothb}^{-1}}\Bigr](x)\ ,\nn
g_\mu^-(x)&=&-d(x-\mu)+\Bigl[k\ast\frac{g_\mu^-}{1+\bar\gothb^{-1}}\Bigr](x)-\Bigl[k_+\ast\frac{g_\mu^+}{1+\gothb^{-1}}\Bigr](x)\ ,\nn
4k(\mu)+\frac{4 i}{\eta}\Omega_{\Psi_0}(-2\mu/\eta)&=&
-\int_{-\frac{\pi}{2}}^{\frac{\pi}{2}}\frac{\mathrm d
  x}{\pi}d(x)\Bigl(\frac{g^+_{\mu}(x)}{1+\gothb^{-1}(x)}+\frac{g^-_{\mu}(x)}{1+\bar\gothb^{-1}(x)}\Bigr)\ ,\nn
4m^z&=&\int_{-\frac{\pi}{2}}^{\frac{\pi}{2}}\frac{\mathrm d
  x}{\pi}\Bigl(\frac{g_{0}^{+}(x)}{1+\gothb^{-1}(x)}-\frac{g_{0}^{-}(x)}{1+\bar\gothb^{-1}(x)}\Bigr)\ ,
\eea
where $m^z$ is the magnetisation per site in the initial state, and
\bea
[f_1\ast  f_2](x)&=&\int_{-\frac{\pi}{2}}^{\frac{\pi}{2}}\frac{\mathrm dy}{\pi}
f_1(x-y)f_2(y)\ ,\nn
d(x)&=&\sum_{n=-\infty}^{\infty}\frac{e^{2 i n x}}{\cosh(\eta n)}\ ,
\quad k(x)=\sum_{n=-\infty}^{\infty}\frac{e^{2 i n x}}{e^{2\eta
    |n|}+1}\ ,
\quad k_\pm(x)=k(x \pm i[\eta-\epsilon])\, .
\label{dk}
\eea
Here $\epsilon$ is a positive infinitesimal. The first three equations
of \eqref{eq:systemh} have the same form as for the equilibrium
problem at finite temperature\cite{thermal}, while the last two
equations are different and encode the quench setup and the associated
constraints on the expectation values of the higher integrals of
motion \eqref{constraints}. 
In general the system \fr{eq:systemh} has to be solved numerically by
iteration, and some details on how to do this are presented in section
\ref{ssec:NumInt}. The structure of \fr{eq:systemh} is such that
the second and third equations can be straightforwardly inverted (as
they are linear) in order to express $g_\mu^\pm$ as functions of
$\gothb$ and $\bar\gothb$. The first equation of \fr{eq:systemh} is
nonlinear, but nonetheless can be used to
express $\bar\gothb$ in terms of $\gothb$. The last equation, which is more
conveniently analyzed in Fourier space, can finally be inverted to
obtain the remaining unknown $\gothb$.
The three functions 
$\varphi(\mu)$, $\omega(\mu_1,\mu_2)$ and $\omega^\prime(\mu_1,\mu_2)$
that enter the expressions for the spin correlation functions are
\bea
\varphi(\mu)&=&\int_{-\frac{\pi}{2}}^{\frac{\pi}{2}}\frac{\mathrm d
  x}{2\pi}\Bigl(\frac{g_{-i
    \mu}^{-}(x)}{1+\bar\gothb^{-1}(x)}-\frac{g_{-i
    \mu}^{+}(x)}{1+\gothb^{-1}(x)}\Bigr)\ ,\nn
\omega(\mu_1,\mu_2)&=&-4k(i\mu_1-i\mu_2)+K_{2\eta}(i\mu_1-i\mu_2)
-\Bigl[d\ast\Bigl(\frac{g_{-i\mu_1}^+(x)}{1+\gothb^{-1}(x)}
+\frac{g_{-i\mu_1}^-(x)}{1+\bar \gothb^{-1}(x)}\Bigr)\Bigr](-i\mu_2)\ ,\nn
\omega^\prime(\mu_1,\mu_2)&=&-4\eta\ell(i\mu_1-i\mu_2)+\eta K_{2(\mu_1-\mu_2)}(i\eta)
-\Bigl[d\ast\Bigl(\frac{g_{-i\mu_1}^{\prime +}}{1+\gothb^{-1}}
+\frac{g_{-i\mu_1}^{\prime -}}{1+\bar\gothb^{-1}}\Bigr)\Bigr](-i\mu_2)\nn
&&-\eta
\Bigl[c_-\ast\frac{g_{-i\mu_1}^+}{1+\gothb^{-1}}\Bigr](-i\mu_2)-\eta\Bigl[c_+\ast\frac{g^-_{-i\mu_1}}{1+\bar\gothb^{-1}}\Bigr](-i\mu_2)\ , 
\label{eq:Smirnov}
\eea
where
\bea
K_\eta(x)&=&\frac{\sinh\eta}{\cosh\eta-\cos(2x)}\ ,\quad
c_{\pm}(x)=\pm\sum_{n=-\infty}^{\infty}\frac{e^{\pm \eta n+2i n x}}{2\cosh^2(\eta n)},\nn
\ell(x)&=&\sum_{n=-\infty}^{\infty}\frac{\mathrm{sgn}(n)e^{2i n
    x}}{4\cosh^2(\eta n)}\, ,\quad \ell_\pm(x)=\ell(x\pm i[\eta-\epsilon])\ ,
\eea
and the auxiliary functions $g^{\prime\pm}_\mu(x)$ are solutions to
the integral equations
\bea
g_\mu^{\prime +}(x)&=&-\eta c_+(x-\mu)+\eta\Bigl[\ell\ast
  \frac{g^+_\mu}{1+\gothb^{-1}}\Bigr](x)-\eta\Bigl[\ell_-\ast
  \frac{g^-_\mu}{1+\bar\gothb^{-1}}\Bigr](x)+
\Bigl[\kappa\ast\frac{g^{\prime+}_\mu}
{1+\gothb^{-1}}\Bigr](x)-\Bigl[\kappa_-\ast\frac{g^{\prime-}_\mu}
{1+\bar\gothb^{-1}}\Bigr](x)\ ,\nn 
g_\mu^{\prime -}(x)&=&-\eta c_-(x-\mu)+\eta\Bigl[\ell\ast
  \frac{g^-_\mu}
{1+\bar\gothb^{-1}}\Bigr](x)-\eta\Bigl[\ell_+\ast 
\frac{g^+_\mu}{1+\gothb^{-1}}\Bigr](x)+
\Bigl[\kappa\ast\frac{g^{\prime-}_\mu}
{1+\bar \gothb^{-1}}\Bigr](x)-\Bigl[\kappa_+\ast\frac{g^{\prime+}_\mu}
{1+\gothb^{-1}}\Bigr](x).
\label{gprime}
\eea
The method proposed in Ref.~[\onlinecite{FE_13b}] for computing
spin-spin correlations functions then consists of three main steps:
\begin{enumerate}
\item Calculate the generating function $\Omega_{\Psi_0}$. While this
is difficult in general, it was pointed out in
Ref.~[\onlinecite{FE_13b}] that it can be done efficiently for initial
states that are of matrix-product form.
\item Solve the system \fr{eq:systemh} of nonlinear integral equations
for the auxiliary functions $\gothb(x)$ and $\bar\gothb(x)$.
\item Use the auxiliary functions to determine the functions
  $\varphi(\mu)$, $\omega(\mu_1,\mu_2)$ and $\omega'(\mu_1,\mu_2)$
  and in turn the spin-spin correlation functions.
\end{enumerate}
In the next subsection we provide the details of how to calculate
$\Omega_{\Psi_0}$ for translationally invariant initial states of
matrix-product form. The generalisation to certain initial states that
break translational invariance, e.g. states with N\'eel order, is
considered in subsection \ref{ssec:breakT}. In subsection
\ref{ssec:NumInt} we present an efficient numerical algorithm for
solving our system of nonlinear integral equations.

Readers not interested in details pertaining to the computation of the
generating function \eqref{eq:Om0} and the numerical solution of
the system of nonlinear integral equations \eqref{eq:systemh}
may proceed directly to Sec.~\ref{s:num}.

\subsection{Generating function for translationally invariant 
matrix-product initial states}
\label{ssec:GenFun}
\begin{figure}[!htbp]
\begin{center}
\begin{tabular}{cc}
(a)\includegraphics[width=0.45\textwidth]{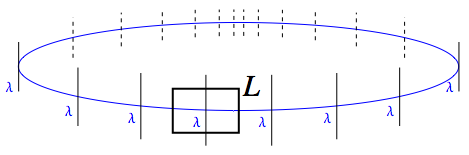}&(c)\includegraphics[width=0.45\textwidth]{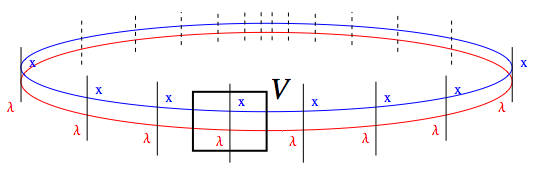}\\
(b)\includegraphics[width=0.45\textwidth]{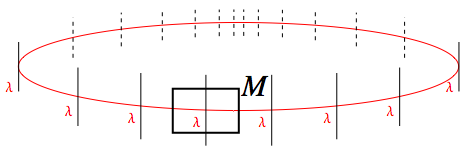}\end{tabular}
\caption{Pictorial representation of the transfer matrices underlying
the construction of the generating function $\Omega_{\Psi_0}(\lambda)$.
The ``two-layer'' transfer matrix shown in panel (c) is formed by
multiplying the transfer matrices made from the L-operators
$L(\lambda;\sigma)$ and $M(\lambda;\sigma)$ shown in panels (a) and
(b). }\label{fig:V}
\end{center}
\end{figure}
Our starting point is the following representation of the generating
function derived in Ref.~[\onlinecite{FE_13b}]
\bea\label{eq:Om}
\Omega_{\Psi_0}(\lambda)&=&\lim_{L\rightarrow\infty}\frac{1}{L}\frac{\partial}{\partial x}\Big|_{x=\lambda}{\rm Tr}
\langle\Psi_0|V_L(x,\lambda)\ldots
V_1(x,\lambda)|\Psi_0\rangle\ ,
\eea
where $V_j(x,\lambda)$ are $4\times 4$ matrices with entries 
$\left(V_j(x,\lambda)\right)^{ab}_{cd}$ that are
operators acting on the two-dimensional quantum space on site $j$
\be
\left(V_j(x,\lambda)\right)^{ab}_{cd}=(L(x;\sigma_j))^{ab}\left(M(\lambda;\sigma_j)\right)^{cd}\, .
\label{V}
\ee
Here $L$ is the L-operator of the XXZ model \cite{Korepinbook} and $M$
is the corresponding matrix associated with the inverse transfer
matrix, i.e.
\be\label{eq:LM}
\begin{aligned}
L(\lambda;\sigma)&=\frac{1+\tau^z\sigma^z}{2}-i\frac{\sin(\frac{\eta\lambda}{2})}{\sinh(\eta-i\frac{\eta\lambda}{2})}\frac{1-\tau^z\sigma^z}{2}+\frac{\sinh(\eta)}{\sinh(\eta-i\frac{\eta\lambda}{2})}(\tau^+\sigma^-+\tau^-\sigma^+)\\
M(\lambda;\sigma)&=\frac{1+\mu^z\sigma^z}{2}+i\frac{\sin(\frac{\eta \lambda}{2})}{\sinh(\eta+i\frac{\eta\lambda}{2})}\frac{1-\mu^z\sigma^z}{2}+\frac{\sinh(\eta)}{\sinh(\eta+i \frac{\eta \lambda}{2})}(\mu^+\sigma^++\mu^-\sigma^-)\, ,
\end{aligned}
\ee
where the Pauli matrices $\tau^\alpha$ and $\mu^\alpha$ act on
distinct auxiliary spaces, which we denote by $\mathcal T$ and $\mathcal M$
respectively, while $\sigma^\alpha$ act on the ``quantum space'' at a
given site of the lattice. The trace in \fr{eq:Om} is over the tensor
product ${\mathcal T}\otimes{\mathcal M}\sim\mathbb{C}^4$ of auxiliary
spaces, i.e. ${\rm Tr}(M)=\sum_{a,b}M^{aa}_{bb}$. A pictorial
representation of transfer matrices underlying the construction of the
generating function is shown in Fig.~\ref{fig:V}.

Denoting the two possible spin states at
site $j$ (corresponding to spin up and spin down in the z-direction
respectively) by $|\pm\rangle_j$, we can construct a basis of the
Hilbert space ${\cal H}$ of our $L$-site lattice by $\{|s_1s_2\ldots
s_L\rangle=\otimes_{j=1}^L|s_j\rangle\ ,\ s_\ell=\pm\}$. The most
general translationally invariant matrix-product state then can be
expressed as 
\be\label{eq:staterep}
\ket{\Psi_0}=\sum_{s_1=\pm}\ldots\sum_{s_L=\pm}{\rm Tr}_{\mathcal A}
\left[A^{(s_1)}A^{(s_2)}\ldots A^{(s_L)}\right]
\ket{s_1\cdots s_L}\equiv \mathrm{Tr}_{\mathcal A}[\ket{\hat \Psi_{\mathcal A}}]\, ,
\ee
where $A^{(\pm)}$ are matrices acting on some auxiliary space
$\mathcal A$, which we take to be isomorphic to $\mathbb{C}^{m}$ for some
integer $m$. The normalisation condition $\langle\Psi_0|\Psi_0\rangle=1$
implies that
\be\label{eq:norm}
\parallel [A^{(+)}]^\ast\otimes A^{(+)}+[A^{(-)}]^\ast\otimes A^{(-)}
\parallel_{\rm op}=1,
\ee
where $||\cdot||_{\rm op}$ is the operator norm, \emph{i.e.} the
absolute value of the maximal eigenvalue, which we assume to be
non-degenerate. It is customary to restrict the form of the
matrix-product state by replacing the condition \eqref{eq:norm} with
the stronger requirement
\be\label{eq:norm1}
A^{(+)}[A^{(+)}]^\dag+A^{(-)}[A^{(-)}]^\dag=\mathrm I,
\ee
and we adopt this convention in the following. In order to calculate
$\Omega_{\Psi_0}(\lambda)$ we require
\be
{\rm Tr}
\langle\Psi_0|V_L(x,\lambda)\ldots
V_1(x,\lambda)|\Psi_0\rangle=
{\rm Tr}_{\mathcal{T}\otimes\mathcal{M}\otimes\bar{\mathcal A}\otimes{\mathcal A}}
\braket{\hat \Psi_{\bar{\mathcal A}}|V_L(x,\lambda)\cdots
  V_1(x,\lambda)|\hat \Psi_{\mathcal A}}.
\ee
As a consequence of translational invariance of \fr{eq:staterep} we have
\be
\label{eq:expM}
\braket{\hat \Psi_{\bar{\mathcal A}}|V_L(x,\lambda)\cdots V_1(x,\lambda)|\hat \Psi_{\mathcal A}}= [\mathrm V(x,\lambda)]^L\, ,
\ee
where 
\be\label{eq:V}
{\rm V}(x,\lambda)=\sum_{s=\pm}\langle s|\left[\sum_{\alpha,\beta=\pm}
e_{\alpha\beta}A^{(\alpha)}\otimes (A^{(\beta)})^\ast\otimes L(x;\sigma)\otimes M(\lambda;\sigma)\right]|s\rangle\, .
\ee
Here $A^{(\pm)}$ act on $\mathcal A$, $(A^{(\pm)})^\ast$ on $\bar{\mathcal A}$, and we defined
\be
e_{++}=\frac{\mathrm I+\sigma^z}{2}\quad e_{+-}=\sigma^+\quad e_{-+}=\sigma^-\quad e_{--}=\frac{\mathrm I-\sigma^z}{2}\, .
\ee

By construction, the matrix $\mathrm V(x,x)$ has an eigenvalue equal to $1$. 
The corresponding right eigenvector $\ket{1;R}$ is independent of $x$,
$\eta$, and $A^{(\pm)}$ and can be determined explicitly. If the matrices
$A^{(\pm)}$ satisfy \eqref{eq:norm1} it has the form
\be
\label{eq:Rvect}
\ket{1;R}= \frac{\ket{\uparrow}_{\mathcal
    T}\otimes\ket{\uparrow}_{\mathcal M}
+\ket{\downarrow}_{\mathcal T}\otimes\ket{\downarrow}_{\mathcal
  M}}{\sqrt{2}}\otimes
\Bigl(\sum_{j=1}^m \ket{a_j}_{\mathcal A}\otimes\ket{\bar{a}_j}_{\bar{\mathcal A}} \Bigr)\, ,
\ee 
where the vectors $\{\ket{a_1},\dots,\ket{a_m}\}$ form a orthonormal
basis of $\mathcal A$ and $\ket{\bar{a}_j}=\ket{a_j}^\ast$. In
contrast, the left eigenvector of $\mathrm V(x,x)$ corresponding to
the eigenvalue $1$ depends in a nontrivial way on $x$, $\eta$, and the
initial state. In order to proceed, we assume that the eigenvalue $1$
of $\mathrm V(x,x)$ is non-degenerate, and that there exists an eigenvalue
$v(x,\lambda)$ of $\mathrm V(x,\lambda)$ that smoothly approaches $1$
in the limit $\lambda\to x$. The generating function is then simply
equal to
\be\label{eq:OmfromD}
\Omega(\lambda)=\frac{\partial}{\partial x}\Bigr|_{x=\lambda}v(x,\lambda)\, .
\ee
In practice, the eigenvalue $v(x,\lambda)$ can be calculated in closed
form only for very simple initial states. To deal with more general
cases the following representation of the generating function turns
out to be very useful
\be\label{eq:OmegaCP}
\Omega(\lambda)=\frac{\mathrm{Tr}\Bigl[\mathrm{adj}\big[\mathrm
      V(\lambda,\lambda)-\mathrm I\big]\ \mathrm
\partial_x\Big|_{x=\lambda}V(x,\lambda)
\Bigr]}{\mathrm{Tr}\Bigl[\mathrm{adj}[\mathrm
      V(\lambda,\lambda)-\mathrm I]\Bigr]}\, .
\ee
Here $\mathrm{adj}[M]$ denotes the adjugate, i.e. the transpose of the
matrix of cofactors of the matrix $M$. The representation
\fr{eq:OmegaCP} can be established as follows. Since $v(x,\lambda)$ is
eigenvalue of $V(x,\lambda)$ we have 
\be
\det\bigl|v(x,\lambda)\mathrm I-V(x,\lambda)\bigr|=0\qquad \forall x,\lambda\, .
\label{det1}
\ee
The derivative with respect to $x$ is related to the adjugate by
Jacobi's formula 
\be
\partial_x \det\bigl|v(x,\lambda)\mathrm I-V(x,\lambda)\bigr|=\mathrm{Tr}\Bigl[\mathrm{adj}[v(x,\lambda)\mathrm I-V(x,\lambda)]\partial_x\bigl(v(x,\lambda)\mathrm I-V(x,\lambda)\bigr)\Bigr].
\ee
By virtue of \fr{det1} the derivative of the determinant vanishes, so that
in the limit $\lambda\rightarrow x$ we arrive at
\be
0=\partial_x\mathrm{Tr}\Bigl[\mathrm{adj}[\mathrm I-V(\lambda,\lambda)]\bigl(v(x,\lambda)\mathrm I-V(x,\lambda)\bigr)\Bigr]_{\lambda=x}\, .
\label{det2}
\ee
Combining \fr{det2} with \fr{eq:OmfromD} we obtain \fr{eq:OmegaCP}.

\subsubsection{Explicit expressions for $\Omega(\lambda)$}
\label{ssec:explicit}
In order to determine our generating function from \fr{eq:OmegaCP} we
require the two quantities $\mathrm V(\lambda,\lambda)$ and
$\partial_x\Big|_{x=\lambda}\mathrm V(x,\lambda)$. It is convenient to
employ \fr{eq:LM} in order to rewrite $\mathrm V(x,\lambda)$ in the
form 
\be
\mathrm V(x,\lambda)=\frac{1}{e^{\frac{i\eta
      (\lambda-x)}{2}}\sinh(\eta-\frac{i\eta
    x}{2})\sinh(\eta+\frac{i\eta \lambda}{2})}
\begin{pmatrix}
\mathbb{I},& e^{-i\eta x/2}\mathbb{I},&e^{-i\eta
  x}\mathbb{I}\end{pmatrix}
\bar{V}\begin{pmatrix}\mathbb{ I}\\ e^{i\eta \lambda/2}\mathbb{ I}\\e^{i\eta
    \lambda}\mathbb{ I}\end{pmatrix}\, , 
\label{vbar}
\ee
where $\bar V$ is a $(12 m^2)\times (12 m^2)$ matrix \emph{independent
of $x$ and $\lambda$}, and the identities $\mathbb{ I}$ are $4m^2\times
4m^2$ matrices. The two matrices we need in order to calculate the
generating function can be expressed as
\bea
\mathrm V(\lambda,\lambda)&=&\frac{2}{\cosh(2\eta)-\cos(\eta \lambda)}
\begin{pmatrix}\mathbb{I},& e^{-i\eta \lambda/2}\mathbb{I},&e^{-i\eta
    \lambda}\mathbb{I}\end{pmatrix}\bar
V\begin{pmatrix}\mathbb{I}\\ e^{i\eta \lambda/2}\mathbb{I}\\e^{i\eta
  \lambda}\mathbb{I}\end{pmatrix}\ ,\nn
\partial_x\Big|_{x=\lambda} \mathrm V(x,\lambda)&=&\frac{i\eta}{1-e^{-2\eta}e^{i\eta \lambda}}V(\lambda,\lambda)-\frac{i\eta}{\cosh(2\eta)-\cos(\eta \lambda)}\begin{pmatrix}\mathrm 0,& e^{-i\eta \lambda/2}\mathbb{I},&2 e^{-i\eta \lambda}\mathbb{I}\end{pmatrix}\bar V\begin{pmatrix}\mathbb{I}\\ e^{i\eta \lambda/2}\mathbb{I}\\e^{i\eta \lambda}\mathbb{I}\end{pmatrix}\, .
\eea
Instead of working with $\mathrm V(\lambda,\lambda)$ and
$\partial_x\Big|_{x=\lambda} \mathrm V(x,\lambda)$ it is convenient to
consider the matrix-valued functions
\bea
P(\lambda)&=&e^{i\eta \lambda}(\cosh(2\eta)-\cos(\eta \lambda))(\mathrm
V(\lambda,\lambda)-\mathbb{I})\ ,\nn
Q(\lambda)&=&(e^{i\eta \lambda}-e^{-2\eta}e^{2i\eta \lambda})
(\cosh(2\eta)-\cos(\eta \lambda))\
\partial_x\Big|_{x=\lambda} \mathrm V(x,\lambda)\ .
\eea
The generating function is related to these functions by
\be
\Omega_{\Psi_0}(\lambda)=\frac{1}
{(e^{i\eta \lambda}-e^{-2\eta}e^{2i\eta \lambda})
(\cosh(2\eta)-\cos(\eta \lambda))}
\frac{{\rm Tr}\left({\rm
    adj}\big[P(\lambda)\big]\ Q(\lambda)\right)}{
{\rm Tr}\left({\rm adj}\big[P(\lambda)\big]\right)}.
\label{eq:Omegarep}
\ee
It follows from \fr{vbar} that $P(\lambda)$ is a polynomial (with matrix-valued
coefficients) of degree four in the variable $e^{i\frac{\eta
    \lambda}{2}}$, and its adjugate is a polynomial of degree
$4(4m^2-1)$. Hence the latter is fully determined by evaluating it at
$16m^2-3$ different values of $\lambda$, which we choose as
\be
\lambda_j=\frac{4\pi j}{\eta(16m^2-3)}\ ,\quad j=0,1\ldots,16m^2-4.
\ee
The adjugate matrix then takes the form
\be
\label{eq:adjugate}
\mathrm{adj}[P(\lambda)]=\frac{1}{16 m^2-3}\sum_{n,\ell=0}^{16 m^2-4}
\mathrm{adj}\left[P(\lambda_\ell)\right]\ e^{i\frac{n \eta (\lambda-\lambda_\ell)}{2}},
\ee
which allows us to obtain $\left.\mathrm{adj}[\mathrm V(\lambda,\lambda)
-\mathbb{I}]\right/\mathrm{Tr}\Bigl[\mathrm{adj}[\mathrm
    V(\lambda,\lambda)-\mathbb{I}]\Bigr]$ with a numerical effort that
scales as $m^{8}$. Similarly, the function $Q(\lambda)$
is a fifth degree polynomial in $e^{i\frac{\eta x}{2}}$, and hence can
be expressed as
\be
Q(\lambda)=\frac{1}{6}\sum_{j,n=0}^5Q(\kappa_j)\ e^{i\frac{n\eta
(\lambda-\kappa_j)}{2}}\ ,\quad
\kappa_j=\frac{2\pi j}{3\eta}.
\label{Q}
\ee
Given an initial matrix-product state of the form \fr{eq:staterep}, we numerically compute the matrices
$\mathrm{adj}\left[P(\lambda_\ell)\right]$ and $Q(\kappa_j)$ either
exactly or to very high precision, and then use \fr{eq:adjugate} and
\fr{Q} to obtain the functions $\mathrm{adj}\left[P(\lambda)\right]$
and $Q(\lambda)$. 
In this way we can extract the value of the finite number of free
parameters of the representation \fr{eq:Omegarep} for the generating
functions. We stress that the functional form of
$\Omega_{\Psi_0}(\lambda)$ is fixed by the structure of the matrix
product state, and potential inaccuracies of the computation are therefore
practically independent of $\lambda$.

\subsubsection{Numerical computation of the generating function}
The method discussed in the previous subsection is most appropriate
for exact matrix-product states. In the following we will be
interested in situations where the initial state is only
\emph{approximately} of matrix-product form. An example would be the
ground state of the Heisenberg chain for large anisotropy $\Delta$. 
In such cases we resort to a faster, fully numerical computation of
the generating function by means of the representation
\eqref{eq:OmegaCP}. Employing a singular value decomposition we have
\be
\mathrm V(\lambda,\lambda)-\mathbb{I}=\mathcal U(\lambda)\mathcal
D(\lambda)\mathcal V^\dag(\lambda)\, , 
\ee
where $\mathcal U(\lambda)$ and $\mathcal V(\lambda)$ are unitary matrices and
$\mathcal D(\lambda)$ is a positive semidefinite diagonal matrix. As
$\mathrm V(\lambda,\lambda)$ has a non-degenerate eigenvalue equal to
$1$, the adjugate of $\mathrm V(\lambda,\lambda)-\mathbb{I}$ is of
rank one, so that the singular value decomposition becomes
\be
\mathrm{adj}[\mathrm V(\lambda,\lambda)-\mathbb{I}]
=\mathrm{det}_+\bigl|\mathcal
D(\lambda)\bigr|\vec{v}_R(\lambda)\vec{v}_L^\dag (\lambda)\, .
\ee
Here $\mathrm{det}_+$ denotes the pseudo-determinant, i.e. the
product of the nonzero eigenvalues, and $\vec v_{R/L}(x)$ are the
normalised right and left singular vectors corresponding to the unique
non-zero singular value. The generating function takes the form
\be\label{eq:Omrough}
\Omega_{\Psi_0}(\lambda)=\frac{\vec{v}_L^\dag(\lambda)\left[\partial_x\Big|_{x=\lambda}
  \mathrm V(x,\lambda)\right]\vec{v}_R(\lambda)}{\vec{v}^\dag_L(\lambda)\vec{v}_R(\lambda)}\, .
\ee
and, for a given $\lambda$, can be straightforwardly computed with a
computational effort that scales as $m^6$ (using the detailed
structure of the matrices it is in principle possible to significantly
reduce the numerical complexity\cite{TTN}). Compared to \fr{eq:Omegarep} the 
representation \fr{eq:Omrough} is numerically better behaved. On the
other hand, the representation \fr{eq:Omegarep} has the advantage of
providing the exact form of the generating function. In order to solve
the system \fr{eq:systemh} of nonlinear 
integral equations, we will require the values of $\Omega(\lambda)$ in
a complex domain. For the initial states we consider this domain is
the strip $|{\rm Im}(\lambda)|<1$. As $\Omega(\lambda)$ is a
$\frac{2\pi}{\eta}$-periodic function, it can be conveniently expanded
in a Fourier series
\be
\Omega_{\Psi_0}(\lambda)=i\eta\sum_n \omega_n e^{i \eta n \lambda}.
\ee
In practice we retain only a finite number of Fourier coefficients,
which we determine using \fr{eq:Omrough}.

\subsection{Initial states that break translational invariance}
\label{ssec:breakT}
In the previous subsection we showed how to determine the generating
function for translationally invariant matrix-product initial states.
Here we consider generalizations to certain simple classes of states
that break translational invariance.
\subsubsection{States with N\'eel order}
\label{sssec:Neel}
In the ground state of XXZ chain at $\Delta>1$ translational
invariance is broken spontaneously, and in order to describe
interaction quenches $H^{(1)}(\Delta_0)\rightarrow H^{(1)}(\Delta)$ in the
antiferromagnetic phase we therefore need to generalise the analysis
of section \ref{ssec:GenFun}. The spontaneously breaking of
translational symmetry to translations by two sites can be addressed
by employing a simple unitary transformation
\be
 H^{(1)}(\Delta)\longrightarrow
\left[\prod_{\ell}\sigma_{2\ell}^x\right]
 H^{(1)}(\Delta)
\left[\prod_{\ell}\sigma_{2\ell}^x\right]
=
\frac{1}{4}\sum_{\ell=1}^L
\sigma_\ell^x\sigma_{\ell+1}^x-\sigma_{\ell}^y\sigma_{\ell+1}^y-\Delta(\sigma_\ell^z\sigma_{\ell+1}^z+1).
\label{HXXF_{Z2}}
\ee
In the limit of large $\Delta$ the ground states of the transformed
Hamiltonian are ferromagnetic with all spins up or down
respectively. Spontaneous symmetry breaking selects one of them, but
crucially the resulting ground state is translationally invariant, and
can be approximated by a matrix product state of the
form~\eqref{eq:staterep}. Reversing the unitary transformation, we
are led to consider matrix-product states of the form
\be
\ket{{\rm GS};\Delta_0}=\left[\prod_{\ell}\sigma_{2\ell}^x\right]
\sum_{s_1,\hdots, s_L}{\rm Tr}_{\mathcal A}\left[A^{(s_1)}\ldots
  A^{(s_L)}\right]\ket{s_1\cdots s_L}\equiv
\left[\prod_{\ell}\sigma_{2\ell}^x\right]|\Psi_0\rangle\, . 
\ee
The corresponding generating function is then
\bea\label{eq:Om1}
\Omega_{|{\rm GS},\Delta_0\rangle}(\lambda)&=&
\lim_{L\rightarrow\infty}\frac{1}{L}\frac{\partial}{\partial x}
\Big|_{x=\lambda}{\rm Tr}
\langle{\rm GS};\Delta_0|V_L(x,\lambda)\ldots
V_1(x,\lambda)|{\rm GS};\Delta_0\rangle\ ,
\eea
where $V_n(x,\lambda)$ are given in \fr{V}. The evaluation of
\fr{eq:Om1} can be reduced to the same calculation as in the
translationally invariant case by noting that
\bea
{\rm Tr}
\langle{\rm GS};\Delta_0|V_L(x,\lambda)\ldots
V_1(x,\lambda)|{\rm GS};\Delta_0\rangle&=&
{\rm Tr}
\langle\Psi_0|\left[\prod_{\ell}\sigma_{2\ell}^x\right]V_L(x,\lambda)\ldots
V_1(x,\lambda)\left[\prod_{\ell}\sigma_{2\ell}^x\right]|\Psi_0\rangle\nn
&=&
{\rm Tr}
\langle\Psi_0|\widetilde{V}_L(x,\lambda)\ldots
\widetilde{V}_1(x,\lambda)|\Psi_0\rangle\ ,
\label{Vtilde0}
\eea
where
\be
\left(\widetilde{V}_j(x,\lambda)\right)^{ab}_{cd}=
(\tau^xL(x;\sigma_j))^{ab}\left(\mu^xM(\lambda;\sigma_j)\right)^{cd}\, .
\label{Vtilde}
\ee
In order to derive \fr{Vtilde0} we have used the property
\be
[L(x;\sigma) M(\lambda;\sigma),\sigma^x\tau^x\mu^x]=0\, ,
\ee
which follows from the definitions \fr{eq:LM} of $L$ and $M$.
We note that the simple reduction \fr{Vtilde0} does not generalise
straightforwardly to states with N\'eel order in a direction tilted
away from the z-axis.

\subsubsection{Matrix-product states obtained via DMRG}
Matrix-product states obtained by density-matrix renormalisation group
methods on open chains lack translation invariance. Such computations
typically result in states of the form 
\bea
\ket{\rm MPS}&=&\left[\prod_{\ell}\sigma_{2\ell}^x\right]
\sum_{s_1,\hdots, s_L}{\rm Tr}_{\mathcal A}\left[\tilde A_1^{(s_1)}
\tilde A_2^{(s_2)}\ldots \tilde A_L^{(s_L)}\right]\ket{s_1\cdots
  s_L}\, ,\label{MPS}\\
\tilde A_j^{(s_j)}&=&P_jA^{(s_j)}P^{-1}_{j+1}\qquad |2j- L|\ll L\, 
\label{Pj}
\eea
where, in the bulk of the system, the matrices $P_j$ are often
diagonal with elements $\pm 1$. In the following we will assume this
property to hold. The ``gauge transformation'' \fr{Pj} obscures
translational invariance, and in order to apply our method for
calculating the generating function we would like to make the state
manifestly invariant. By virtue of \eqref{Pj}, the state \eqref{MPS}
has the same bulk properties as 
\be\label{MPSpbc}
\left[\prod_{\ell}\sigma_{2\ell}^x\right]
\sum_{s_1,\hdots, s_L}{\rm Tr}_{\mathcal A}\left[ A^{(s_1)}
 A^{(s_2)}\ldots  A^{(s_L)}\right]\ket{s_1\cdots
  s_L}\, ,
\ee
which can be dealt with by the method outlined in section
\ref{sssec:Neel}. This leaves us with the problem of how to obtain the
matrices $A^{(\pm)}$ for a state of the form \fr{MPS}, \fr{Pj}. This
can be done by picking a site $\bar j$ that is sufficiently far 
away from the boundaries. The matrix 
\be\label{eq:Bs}
B^{(s)}=\tilde A_{\bar
  j}^{(s)}P_{\bar j}P_{\bar j+1}
  \ee
is related to $A^{(s)}$ by the similarity transformation
$B^{(s)}=P_{\bar{j}} A^{(s)}P_{\bar{j}}$ (note that
$P_j^2=\mathbb{I}$),  
and can be used to replace $A^{(s)}$ in \eqref{MPSpbc}. Since both
$P_{\bar j-1}P_{\bar j}$ and $P_{\bar j}P_{\bar j+1}$ are simple
diagonal matrices with elements $\pm 1$, i.e. at most $2m$ unknowns,
the (approximate) $m\times m$ matrix equation (\emph{cf.} \eqref{Pj})
\be
\tilde A_{\bar j-1}^{(s)}\approx P_{\bar j-1}P_{\bar j}\tilde A_{\bar
  j}^{(s)}P_{\bar j} P_{\bar j+1}
\ee
can generally be used to extract both $P_{\bar j}P_{\bar j+1}$ and 
$P_{\bar j-1}P_{\bar j}$. Knowing $P_{\bar j}P_{\bar j+1}$,
\eqref{eq:Bs} gives $B^{(s)}$ in turn, and therefore we obtained a
translation invariant representation of the state.

\subsection{Closed-form expressions for the generating function of some
simple initial states} 
We have calculated $\Omega_{\Psi_0}(\lambda)$ analytically for a
variety of initial states, which are invariant under translations by
$n$ sites. The case $n>1$ is dealt with by
a straightforward generalisation of Eq.~\eqref{eq:expM}, in which the
matrix $\rm V(x,\lambda)$ is associated with a block of $n$ adjacent
spins. Among the states we considered are
\begin{enumerate}
\item{}N\'eel state in the z-direction (in spin space) 
$\ket{\uparrow\downarrow\uparrow\downarrow\dots}$: 
\be
\Omega_{\ket{\uparrow\downarrow\uparrow\downarrow\dots}}(\lambda)=\frac{i
  \eta}{2}\frac{\sinh(2\eta)}{\cosh(2\eta)+1-2\cos(\eta\lambda)}.
\ee
\item{}N\'eel state in the x-direction 
$\ket{\rightarrow\leftarrow\rightarrow\leftarrow\dots}$:
\be
\Omega_{\ket{\rightarrow\leftarrow\rightarrow\leftarrow\dots}}=
\frac{i\eta}{2}\frac{\sinh(\eta)}{2\cosh(\eta)-1-\cos(\eta \lambda)}.
\ee
\item{} Ferromagnet along the x-direction $\ket{\rightarrow
\rightarrow\dots}$:
\be
\Omega_{\ket{\rightarrow\rightarrow\dots}}=\frac{i
  \eta}{2}\frac{\sinh(\eta)}{2\cosh(\eta)+1+\cos(\eta \lambda) }.
\ee
\item{} Majumdar Ghosh dimer product state
$\displaystyle|{\rm MG}\rangle=\prod_{j=1}^{L/2}
\frac{|\uparrow\rangle_{2j-1}\otimes|\downarrow\rangle_{2j}-
|\downarrow\rangle_{2j-1}\otimes|\uparrow\rangle_{2j}}{2}$:
\be
\Omega_{|\rm MG\rangle}=
\frac{i\eta\sinh(\eta)}{2}\frac{4\cos(\eta \lambda)
  (\sinh^2(\eta)-\cosh(\eta))+\cosh(\eta)+2\cosh(2\eta)+3\cosh(3\eta)-2}{4
  (\cosh(2\eta)-\cos(\eta \lambda))^2}.
\ee
We note that $|{\rm MG}\rangle$ is one of two ground states of the
Hamiltonian
\be
H_{\rm MG}=J_1\sum_{j}{\bf S}_j\cdot{\bf S}_{j+1}+
\frac{J_1}{2}\sum_{j}{\bf S}_j\cdot{\bf S}_{j+2}.
\ee
\item{}
Ferromagnetic domain state
$|{\rm FD}\rangle
=\ket{\dots
\underbrace{\downarrow\cdots\downarrow}_{s}
\underbrace{\uparrow\cdots\uparrow}_{s}
\underbrace{\downarrow\cdots\downarrow}_{s}
\underbrace{\uparrow\cdots\uparrow}_{s}\dots
}$: 
\be
\Omega_{|{\rm FD}\rangle}=
\frac{i \eta}{2s}\coth(\eta)\tanh\bigl(\frac{s}{2}\log\bigl(\frac{\cosh(2\eta)-\cos(\eta\lambda)}{1-\cos(\eta \lambda)}\bigr)\bigr).
\ee
\item{} Tilted ferromagnet $\ket{\theta;\nearrow\nearrow\dots}=
e^{i\theta\sum_j S^y_j}\ket{\uparrow\uparrow\dots}$
\be
\Omega_{\ket{\theta;\nearrow\nearrow\dots}}=
\frac{i \eta  \sinh\eta\sin^2 \theta }{(\cos(2\theta)+3)\cos(\eta
  \lambda)+4 \cosh \eta +2\sin^2\theta}.
\ee
This state has a magnetisation per site in z-direction of
$\frac{\cos\theta}{2}$.
\item{} Tilted N\'eel state 
$\ket{\theta;\nearrow\swarrow\dots}=
e^{i\theta\sum_j S^y_j}\ket{\uparrow\downarrow\dots}$
\bea
\Omega_{\ket{\theta;\nearrow\swarrow\dots}}&=&\frac{i \eta
  \sinh\eta\big[2 \sin^2 \theta 
  \cos(\eta \lambda)+\cosh \eta
  \big(\cos(2\theta)+3\big)\big]}{D(\theta)}\ ,\nn
D(\theta)&=&-2\sin^2 \theta   \cos(2\eta \lambda)+2\cos (\eta
  \lambda)\Bigl(-2\cosh^2(\frac{\eta }{2}) \cos(2\theta)\nn
&&+\cosh \eta  -3\Bigr)+2\cos(2 \theta) \cosh \eta-2 \cosh\eta+4\cosh(2 \eta)+\cos(2 \theta)+3.
\eea
\end{enumerate}
These initial states break some of the continuous or discrete
symmetries of the post-quench Hamiltonian discussed in subsection
\ref{ssec:symm}. Table \ref{tab:symm} summarises their symmetry
properties. 
\begin{table}[ht]
\begin{center}
\begin{tabular}{|l|l|l|l|l|l|l|}
\hline
& State  & $U(1)$ & $m^z$ & translations &site inversion & bond inversion\\ \hline\hline
1.&$|\uparrow\downarrow\dots\rangle$ & Yes & $0$ & by $2$ sites& Yes&No\\ \hline
2. &$|\rightarrow\leftarrow\dots\rangle$ & No & $0$ & by $2$ sites&Yes &No\\ \hline
3. &$|\rightarrow\rightarrow\dots\rangle$ & No & $0$ & Yes&Yes &Yes\\ \hline
4. &$|{\rm MG}\rangle$ & Yes & $0$ & by $2$ sites&No &Yes\\ \hline
5. &$|{\rm FD}\rangle$ & Yes & $0$ & No&No &No\\ \hline
6. &$|\theta;\nearrow\nearrow\dots\rangle$ & No & $\frac{1}{2}\cos\theta$ 
& Yes&Yes &Yes\\ \hline
7. &$|\theta;\nearrow\swarrow\dots\rangle$ & No & $0$ 
& by $2$ sites&Yes &No\\ \hline
\end{tabular}
\caption{Symmetries on the various initial states.}\label{tab:symm}
\end{center}
\end{table}

\subsection{Numerical solution of the system of nonlinear integral equations}
\label{ssec:NumInt}
The system \eqref{eq:systemh} of nonlinear integral equations
generally needs to be solved by iteration. A convenient limit in which
this can be done ``by hand'' is the case of a ``small'' quench for $m^z=0$
considered in Ref.~[\onlinecite{FE_13b}]. This corresponds to the regime
$|\gothb|,|\bar \gothb|\ll 1$, because in the relevant domain
$|\mathrm{Im}[x]|<1$ the generating function $\Omega(x)$ is close to
the analogous quantity evaluated in the ground state of the
post-quench Hamiltonian (and concomitantly the expectation values of
the integrals of motion deviate only slightly from their ground state
values). At the lowest order in the iteration one finds\cite{FE_13b}
\be\label{eq:rho1}
\gothb(x)\approx\bar\gothb(x)\approx  \rho^{(1)}(x)= \frac{\frac{i}{\eta}\Omega_{\Psi_0}(-\frac{2x}{\eta}+i)+\frac{i}{\eta}\Omega_{\Psi_0}(-\frac{2x}{\eta}-i)+\frac{\sinh(\eta)}{\cosh(\eta)-\cos(2x)}}{d(x)}\, ,
\ee
where $d(x)$ has been defined above in \eqref{dk}. For more general
quenches the system \eqref{eq:systemh} needs to be solved numerically,
and we now provide some details about how this can be done. We find it
convenient to work in the Fourier space, where \eqref{eq:systemh}
turns into nonlinear system of equations for the Fourier coefficients
of $\gothb$ and $\bar\gothb$. Since the latter are generally smooth
functions, a good approximation can be achieved by retaining only a
finite number $\sim n_{\rm max}$ of Fourier coefficients (the error being
exponentially small in $n_{\rm max}$). Although formally there
is no problem in writing the equations in different ways, in practice
the objects defined in each step of the process must have Fourier
coefficients that can be safely neglected at high frequency. This is
an important point to which we return later.

We start by introducing some useful notations. We denote by $[f]$ the
Toeplitz matrix with elements
\be
[f]_{n}^\ell=\int_{-\frac{\pi}{2}}^{\frac{\pi}{2}}\frac{\mathrm d
  x}{\pi}e^{-2i(n-\ell)x}f(x)\, .
\ee
By extension, $[f]^0$ is the vector of Fourier coefficients
\be
[f]_{n}^0=\int_{-\frac{\pi}{2}}^{\frac{\pi}{2}}\frac{\mathrm d
  x}{\pi}e^{-2i n x}f(x)\, .
\ee
In these notations we have e.g.
\be
[d]_n^\ell=\frac{1}{\cosh(\eta(n-\ell))}\, .
\ee
We further introduce the following matrices constructed from the
Fourier coefficients of $k(x)$, $k_\pm(x)$, and $d(x)$ (\emph{cf.}
\eqref{dk} respectively
\bea
\mathrm K_{n}^\ell&=&\frac{\delta_{n}^\ell}{e^{2\eta |n|}+1}\ ,\quad
(\mathrm K_\pm)_{n}^\ell=\frac{\delta_{n}^\ell e^{\mp 2\eta n}}
{e^{2\eta |n|}+1}\ ,\nn
\mathrm D_{n}^\ell&=&\frac{\delta_n^\ell}{\cosh(\eta n)}\ ,
\quad (\mathbb{I}_{\mathrm H})_{n \ell}=\delta_n^{-\ell}\, .
\eea
We note that in these expressions round brackets have no special meaning.
We further define two matrices $G_\pm$ with elements
\be
{}
(G_\pm)_n^\alpha=\chi^{\mp 1}\iint\limits_{[-\pi/2,\pi/2]^2}\frac{\mathrm d x\mathrm d \mu}{\pi^2} e^{-2i n x-2i\alpha \mu}g_\mu^{\pm }(x)
\ee
where $\chi$ is a real, positive parameter that equals $1$ if $m^z=0$. 
Finally, we parametrize the auxiliary functions $\gothb(x)$ and
$\bar{\gothb}(x)$ as follows\cite{FE_13b}
\be
\frac{1}{1+\gothb^{-1}(x)}=\chi \rho(x) e^{\zeta(x)/2}\ ,\
\quad\frac{1}{1+\bar \gothb^{-1}(x)}=\chi^{-1} \rho(x) e^{-\zeta(x)/2}\, .
\label{parametrization}
\ee
The system~\eqref{eq:systemh} of nonlinear integral equations can then
be recast in the compact form 
\bea
\label{eq:System}
\chi&=&\frac{1}{2m^z-1}\sum_\alpha (G_+)_0^\alpha\equiv-\frac{2m^z+1}{
\sum_\alpha (G_-)_0^\alpha}\ ,\label{first}\\
\lbrack\zeta\rbrack_n^0&=&2e^{-\eta|n|}\sinh(\eta n)
\big\lbrack\log(1-\chi^{\mathrm{sgn}(n)}
e^{\mathrm{sgn}(n)\zeta/2}\rho)\big
\rbrack_n^0\ ,\label{second}\\
\lbrack\rho\rbrack^0&=&2 [1/d]\ \mathbb{I}_{\mathrm
  H}\Lambda_-\left(\Lambda_-+\Lambda_+\right)^{-1}\Lambda_+
\mathbb{I}_{\mathrm H}\ [d]\ [\rho^{(1)}]^0\, \label{third},\\
G_\pm&=&-[e^{\mp\zeta/2}]\left(\Lambda_\pm^T\right)^{-1}
\mathbb{I}_{\mathrm H}\ \mathrm D,
\label{fourth}
\eea
where $\Lambda_\pm$ are the transposes of the matrices
\be
\Lambda^T_\pm=\bigl(\chi^{\pm 1}[
  e^{\pm\zeta/2}]-\mathrm{K}[\rho ]\bigr)
\left(\chi^{\pm 1}[ e^{\pm\zeta/2}]-(\mathrm K+\mathrm{K}_\mp)[\rho]\right)^{-1}
\bigl(\chi^{\mp 1}[e^{\mp\zeta/2}]-(\mathrm{K}+
\mathrm{K}_{\pm})[\rho ]\bigr)+\mathrm{K}_\pm[\rho]\, .
\ee
An approximate solution of \eqref{eq:System} can now be obtained 
as follows
\begin{enumerate}
\item 
\label{reg}
The system \fr{eq:System} is finitized by constraining the indices of the
infinite dimensional matrices and vectors to be contained in the
set $n\in[-n_{\rm max},n_{\rm max}]$; 
\item The resulting nonlinear equations are solved by iteration.  In
each step only a single function is updated, in particular the second (third)
equation of \eqref{eq:System} is used to iterate $\zeta$
($\rho$). Each equation is solved separately by iteration in order to
reach an intermediate accuracy goal, which is updated only after all
unknowns have met the same criterion.
\end{enumerate}
Some comments are in order. In all quenches we considered, the density
$\rho^{(1)}(x)$ given in~\eqref{eq:rho1} is a smooth function. Hence
the elements of the vector $[\rho^{(1)}]^0$ ``decay exponentially from the
center'', i.e. only the elements $[\rho^{(1)}]^0_n$ with $n\approx 0$
are significantly different from zero. The same holds true for the vector
$\mathbb{I}_{\mathrm H}[d][\rho^{(1)}]^0$.   
In order to be able to neglect the high frequency contribution and
approximate $[\rho^{(1)}]^0$ by a vector with a finite number of elements, the
matrix $\Lambda_-\left(\Lambda_-+\Lambda_+\right)^{-1}\Lambda_+$
(\emph{cf.} the third equation) should not have large elements that
connect high frequency components with low frequency ones.  
This is indeed what we observe in all cases considered. This justifies
our finitization procedure for all quantities appearing in 
\fr{third}. Similarly, we find that finitizing the matrix $G_\pm$
in \fr{fourth} induces only exponentially small (in $n_{\rm
  max}$) errors in \fr{first}. Equation \fr{second} is more
problematic. Although in all cases we considered we succeeded in
obtaining solutions such that the function $\zeta(x)$ is smooth, in
the course of the computation the function inside the logarithm can
develop zeroes, with catastrophic consequences. This can be controlled
using under-relaxation and/or pre-conditioning the equation (the
appropriate transformations are quench-dependent).   

In order to obtain results for short-distance correlation functions we
require expressions for the auxiliary functions $g^{\prime\pm}_\mu(x)$
defined in \fr{gprime}. In Fourier space equations \fr{gprime} take
the form
\bea
G^{\prime}_\pm
&=&[e^{\mp \zeta/2}]\left(\chi^{\mp 1}[e^{\mp \zeta/2}]-\mathrm
  K[\rho]-\mathrm K_{\mp}[\rho]
\left(\chi[e^{\pm \zeta/2}]-\mathrm K[\rho]\right)^{-1}
\mathrm K_\pm [\rho]\right)^{-1}\nn
&&\times\Bigl(M_{\pm}-\mathrm K_{\mp}
[\rho]\left(\chi^{\pm 1}[e^{\pm\zeta/2}]-\mathrm K[\rho]\right)^{-1}
M_{\mp}\Bigr)\, ,
\eea
where
\bea
(G^\prime_\pm)_n^\alpha&=&\chi ^{\mp 1}\int_{-\frac{\pi}{2}}^{\frac{\pi}{2}}
\frac{\mathrm d x}{\pi} 
\int_{-\frac{\pi}{2}}^{\frac{\pi}{2}}
\frac{\mathrm d \mu}{\pi} 
e^{-2i n x-2i\alpha \mu}g_\mu^{\prime\pm }(x)\ ,\nn
(M_\pm)_n^\alpha&=&\frac{\eta(\mathrm{sgn}(n) e^{\eta|n|}\mp e^{\pm \eta n})}{2\cosh^2(\eta n)}\delta_n^{-\alpha}+\eta \chi^{\mp 1}\frac{\mathrm{sgn}(n) e^{-\eta|n|}}{2\cosh(\eta n)}(G_\pm)_n^\alpha\, .
\eea
\subsubsection{Simplifications for parity symmetric states}
Parity symmetric initial states have the property
\be
\Omega(-x)=\Omega(x)\, ,
\ee
which leads to a number of simplifications. From \eqref{eq:rho1} it follows 
that 
\be
[\rho^{(1)}]_n^0=([\rho^{(1)}]_n^0)^\ast=[\rho^{(1)}]^0_{-n}\ ,
\ee
which in turn permits a solution of \eqref{eq:System} in terms of real
$[\rho]_n$, $[\zeta]_n$, and $\chi$. If the magnetisation $m^z$
vanishes, additional simplifications occur:
\begin{itemize}
\item The parameter $\chi$ is equal to $1$.
\item $[\rho]_n^0=[\rho]_{-n}^0=([\rho]_n^0)^\ast$, \emph{i.e.} $\mathbb{I}_{\mathrm H}[f(\rho)]\mathbb{I}_{\mathrm H}=[f(\rho)]$;
\item $[\zeta]_n^0=-[\zeta]_{-n}^0=([\zeta]_n^0)^\ast$, \emph{i.e.} $\mathbb{I}_{\mathrm H}[f(\zeta)]\mathbb{I}_{\mathrm H}=[f(-\zeta)]$;
\item $\Lambda_-=\mathbb{I}_{\mathrm H} \Lambda_+\mathbb{I}_{\mathrm H}$, $M_-=-\mathbb{I}_{\mathrm H} M_+\mathbb{I}_{\mathrm H}$, $G_-=\mathbb{I}_{\mathrm H} G_+\mathbb{I}_{\mathrm H}$ and $G^\prime_-=-\mathbb{I}_{\mathrm H} G^\prime_+\mathbb{I}_{\mathrm H}$.
\end{itemize} 
In particular, we no longer require the equation for $\chi$ 
and we can write the second equation of \eqref{eq:System} as follows 
\be
[\zeta]_n^0=\frac{2\tanh(\eta n)}{1+\kappa \tanh(\eta |n|)}\bigl[\log\bigl(e^{(\kappa-1)\frac{\zeta}{2}}-\rho e^{\frac{\kappa\zeta}{2}}\bigr)\bigr]^0_{|n|}\, ,
\ee
where $\kappa$ is a auxiliary parameter aimed at stabilising the iterative process and/or enhancing the convergence rate (notice that the solution does not depend on $\kappa$). 

\subsubsection{Results for some initial states}
In this section we present results of the numerical solution of our
system~\eqref{eq:systemh} of nonlinear integral equations for several
quenches with $m_z=0$. We focus on the auxiliary functions $\rho(x)$
and $\zeta(x)$ defined in \fr{parametrization}.
Let us define
\be\label{eq:dressed}
\mathcal E(x)=\frac{\sinh \eta}{2}\sum_{j=0}\lambda_{j+1}\Bigl(\frac{\sinh \eta}{2}\frac{\rm d}{\rm d x}\Bigr)^j d(x)\equiv [k\ast\log(1+\gothb)](x)-[k_-\ast\log(1+\bar\gothb)](x)-\log \gothb(x)\, ,
\ee
where the second equality is a by-product of the integral
equations~\cite{FE_13b}.  
We note that if the Lagrange multipliers fulfil certain conditions, cf
Refs~[\onlinecite{higherCL}], (which in particular should lead to $\mathcal
E(x)$ being positive), $\mathcal E(x)$ can be interpreted as a
\emph{dressed energy} associated with the ``Hamiltonian''
characterizing the GGE density matrix \eqref{rhoGGE}
\be\label{Hgen}
H_{\rm GGE}=\sum_{\ell=1} \lambda_{\ell} H^{(\ell)}\, .
\ee
Having possible interpretations of this kind in mind, we quote the 
analogous result for the Gibbs ensemble at temperature $T$ as a point
of reference\cite{QTM}
\be
\mathcal E_{\rm Gibbs}(x)=\frac{1}{T}\frac{\sinh\eta}{2}d(x)\, .
\label{Egibbs}
\ee
By construction this is proportional to the usual zero temperature
dressed energy of the Heisenberg XXZ chain\cite{Takahashi}. 
Below we compare $\rho(x)$ and $\zeta(x)$ to the corresponding
functions that solve the equilibrium \emph{finite temperature}
nonlinear integral equations\cite{thermal}. The latter
can be obtained by replacing the fourth equation of \eqref{eq:systemh}
by \eqref{eq:dressed}, with $\mathcal E_{\rm Gibbs}(x)$ of
Eq.~\fr{Egibbs} taking the place of $\mathcal E(x)$; the temperature
$T$ is fixed via the requirement that the average energy in the Gibbs
ensemble associated with the post-quench Hamiltonian is equal to the
energy in the initial state after the quench 
\be
\langle \Psi_0|H^{(1)}|\Psi_0\rangle=\frac{{\rm Tr}\left[
e^{-H^{(1)}/T}H^{(1)}\right]}{{\rm Tr}\left[e^{-H^{(1)}/T}\right]}.
\ee

\begin{figure}[ht]
\includegraphics[width=0.45\textwidth]{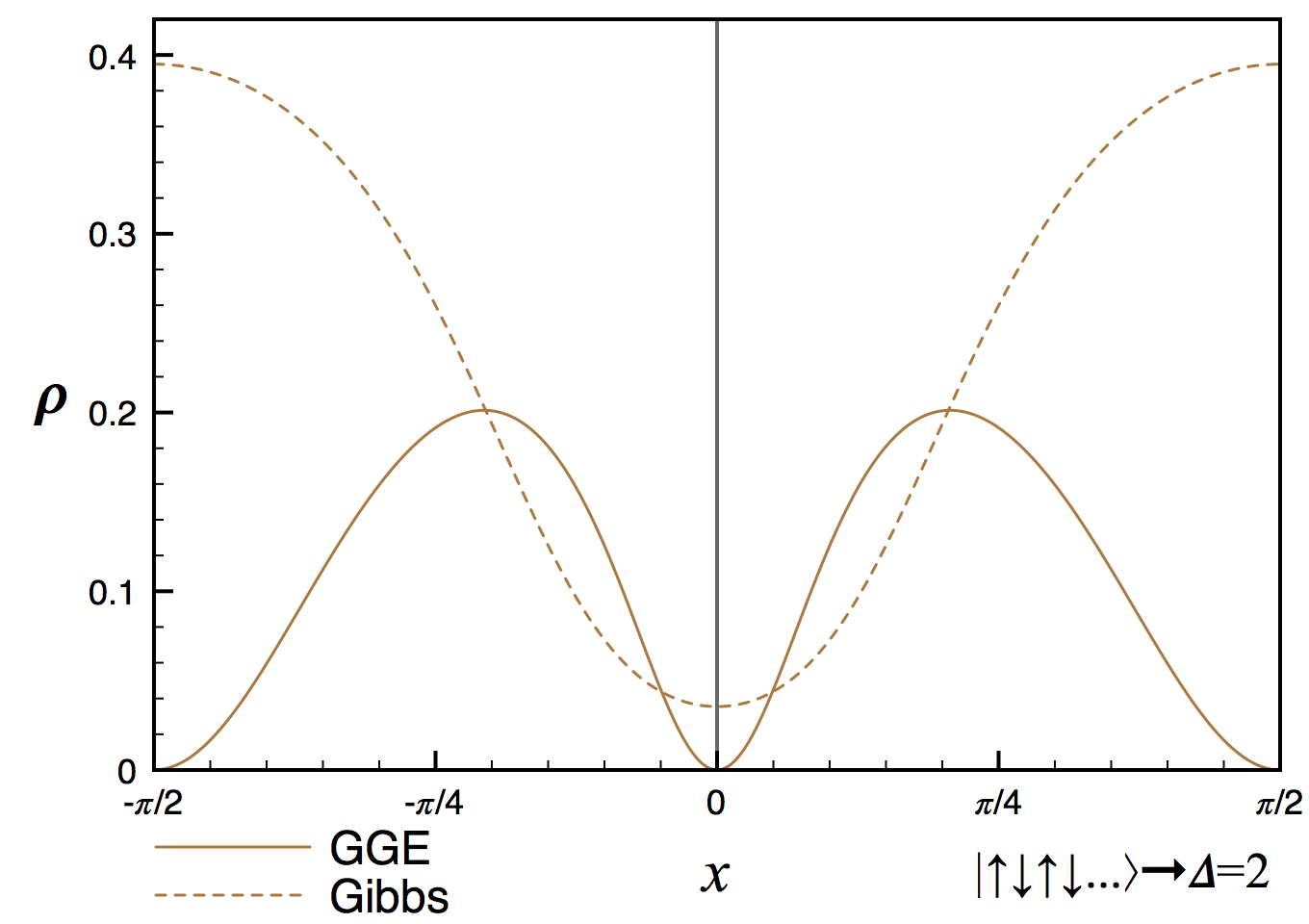}
\includegraphics[width=0.45\textwidth]{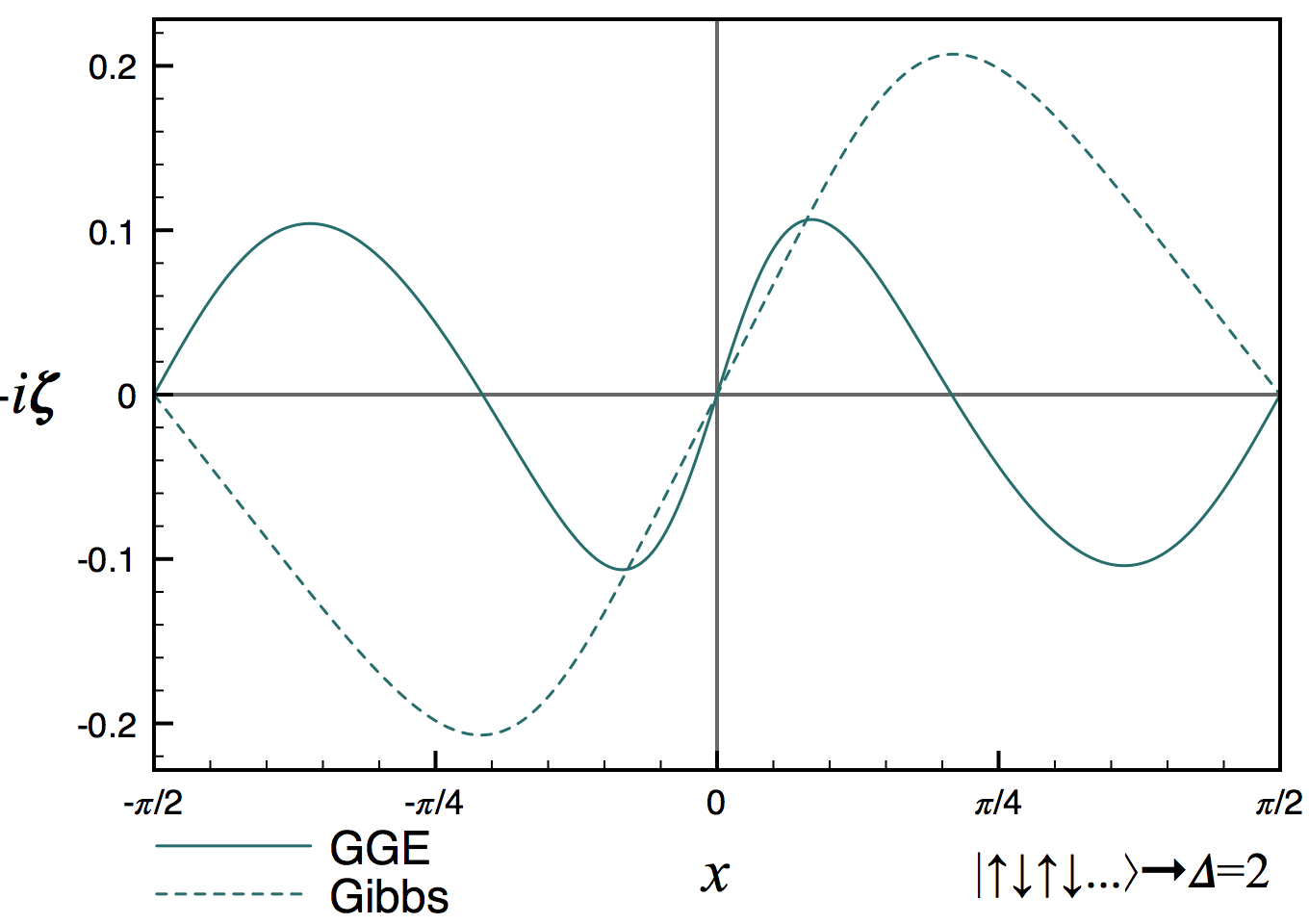}
\caption{Interaction quench from $\Delta_0=\infty$. The auxiliary
 function $\rho(x)$ vanishes at $x=0,\frac{\pi}{2}$. The dashed lines show $\rho$ (left) and $\zeta$ (right)  
at a temperature $T$ corresponding to the expectation value of energy in the initial state. 
The shape of the auxiliary functions does not change significantly if $\Delta_0$
is taken to be large but finite.
}\label{fig:NS} 
\end{figure}

\begin{figure}[ht]
\includegraphics[width=0.45\textwidth]{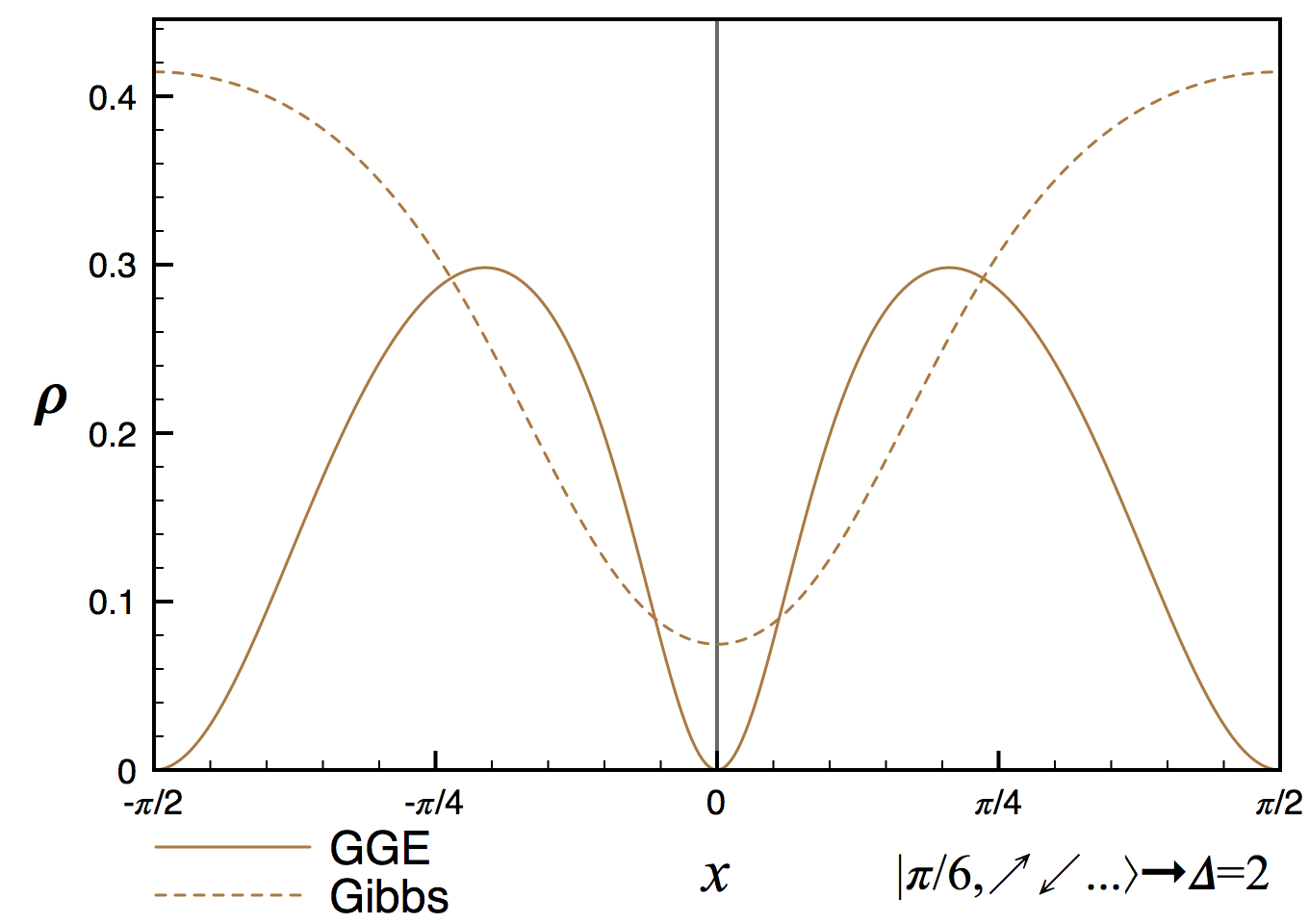}
\includegraphics[width=0.45\textwidth]{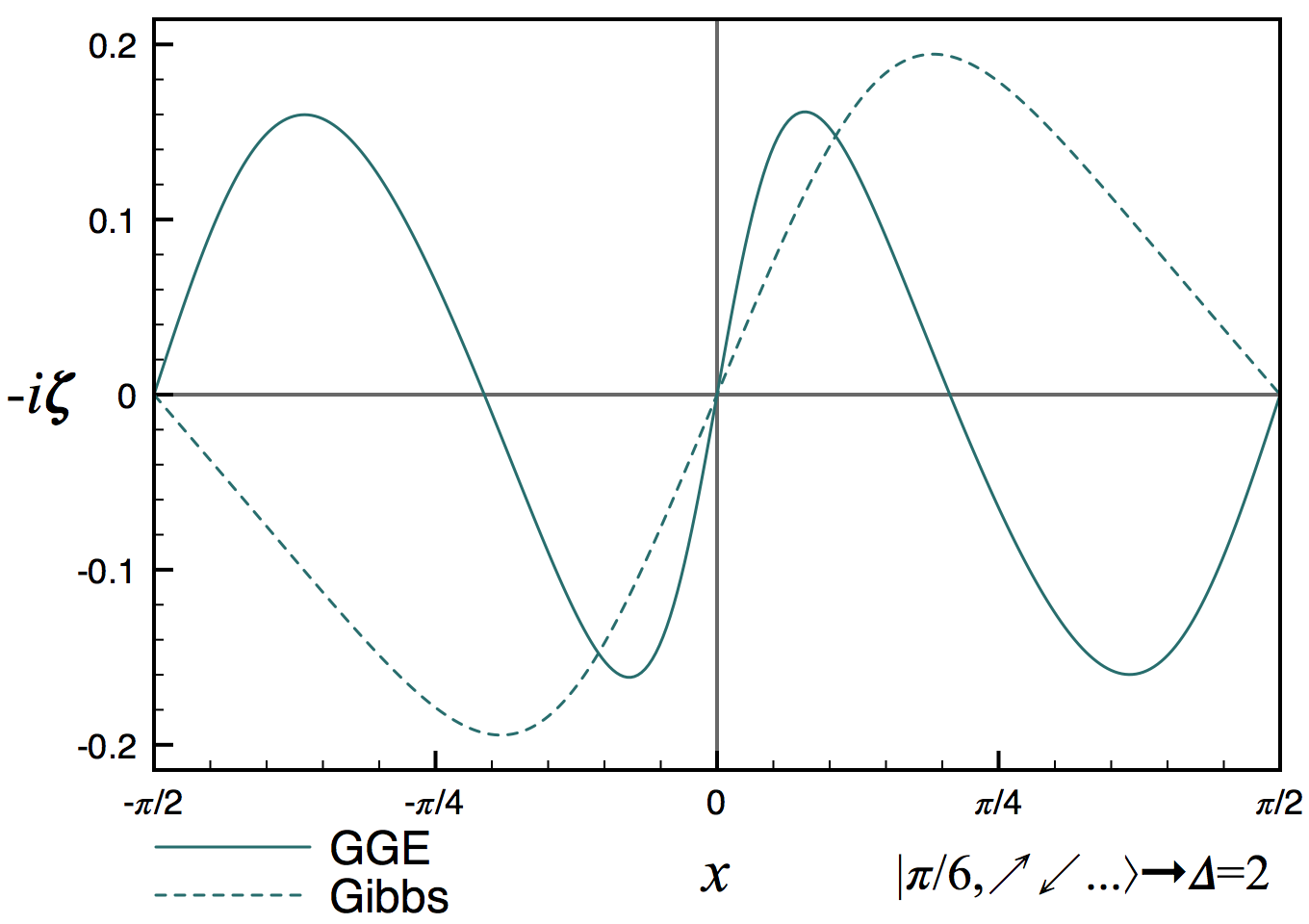}
\caption{Quench from the tilted N\'eel state $|\pi/6,\nearrow\swarrow\dots\rangle$. The
auxiliary function $\rho(x)$ vanishes at $x=0,\frac{\pi}{2}$. }\label{fig:XNS30}  
\end{figure}

\begin{figure}[ht]
\includegraphics[width=0.45\textwidth]{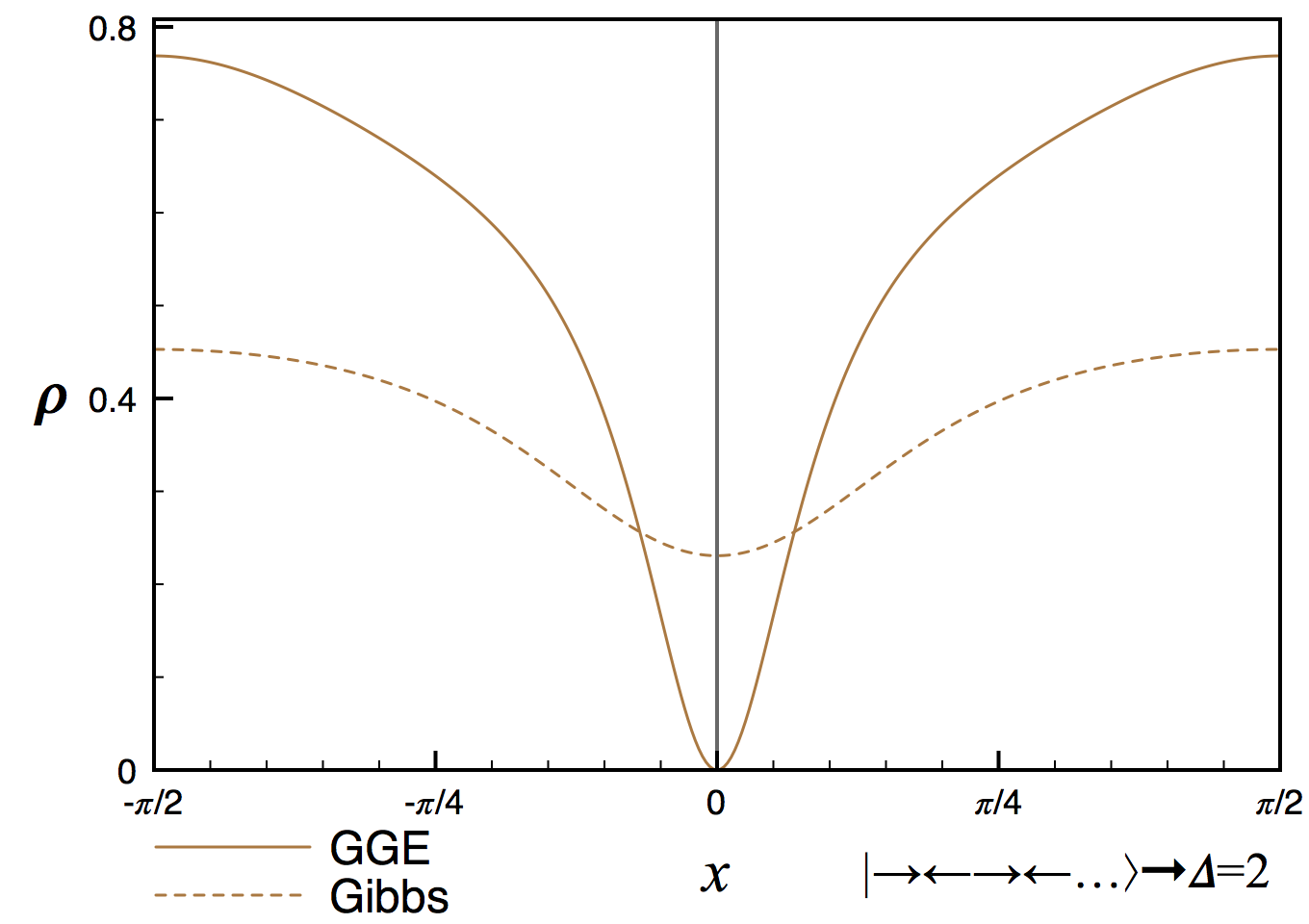}
\includegraphics[width=0.45\textwidth]{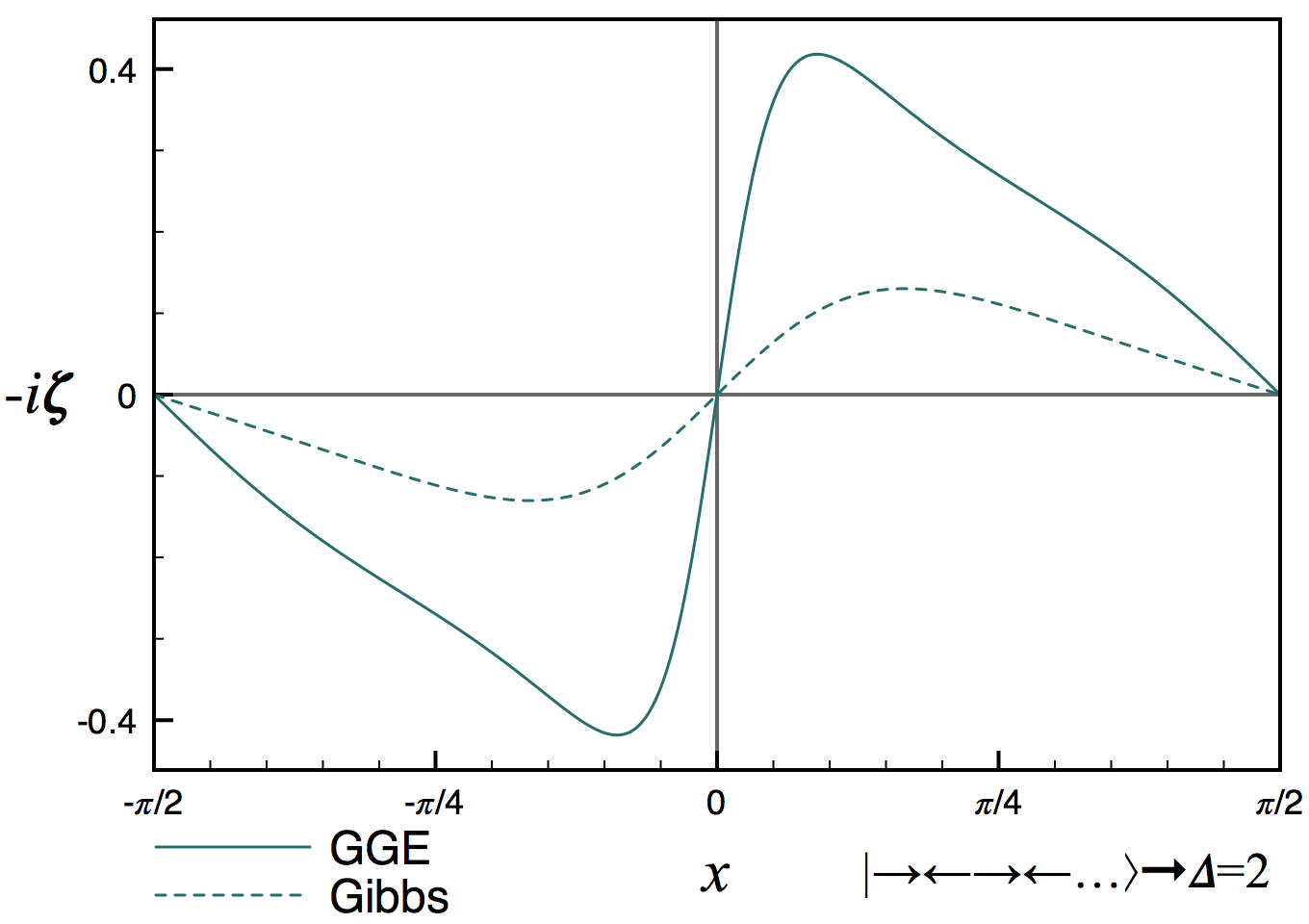}
\caption{Quench from the N\'eel state in $x$-direction
$|\rightarrow\leftarrow\rightarrow\leftarrow\dots\rangle$. The
auxiliary function $\rho(x)$ vanishes at $x=0$. }\label{fig:XNS}  
\end{figure}

\begin{figure}[ht]
\includegraphics[width=0.45\textwidth]{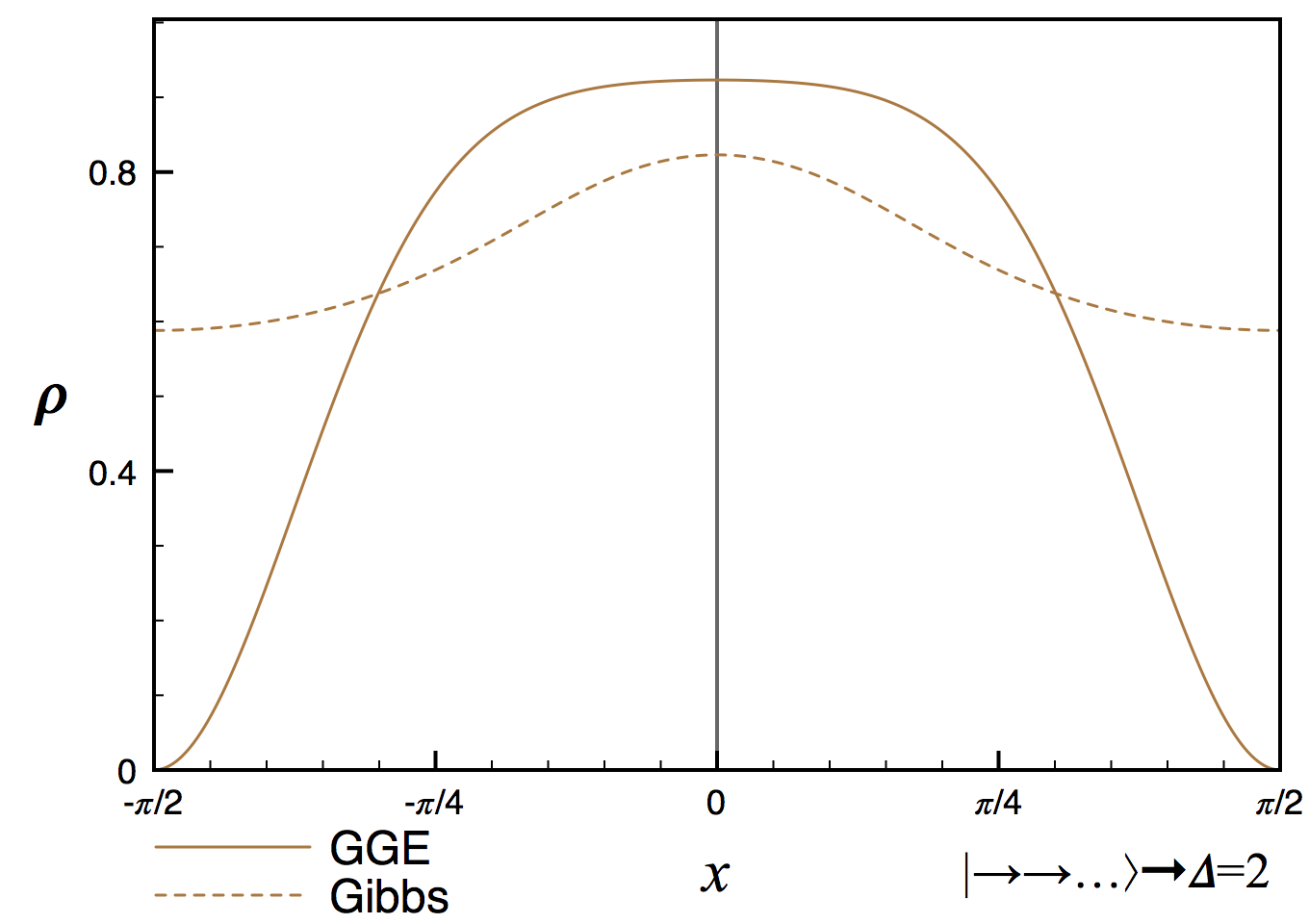}
\includegraphics[width=0.45\textwidth]{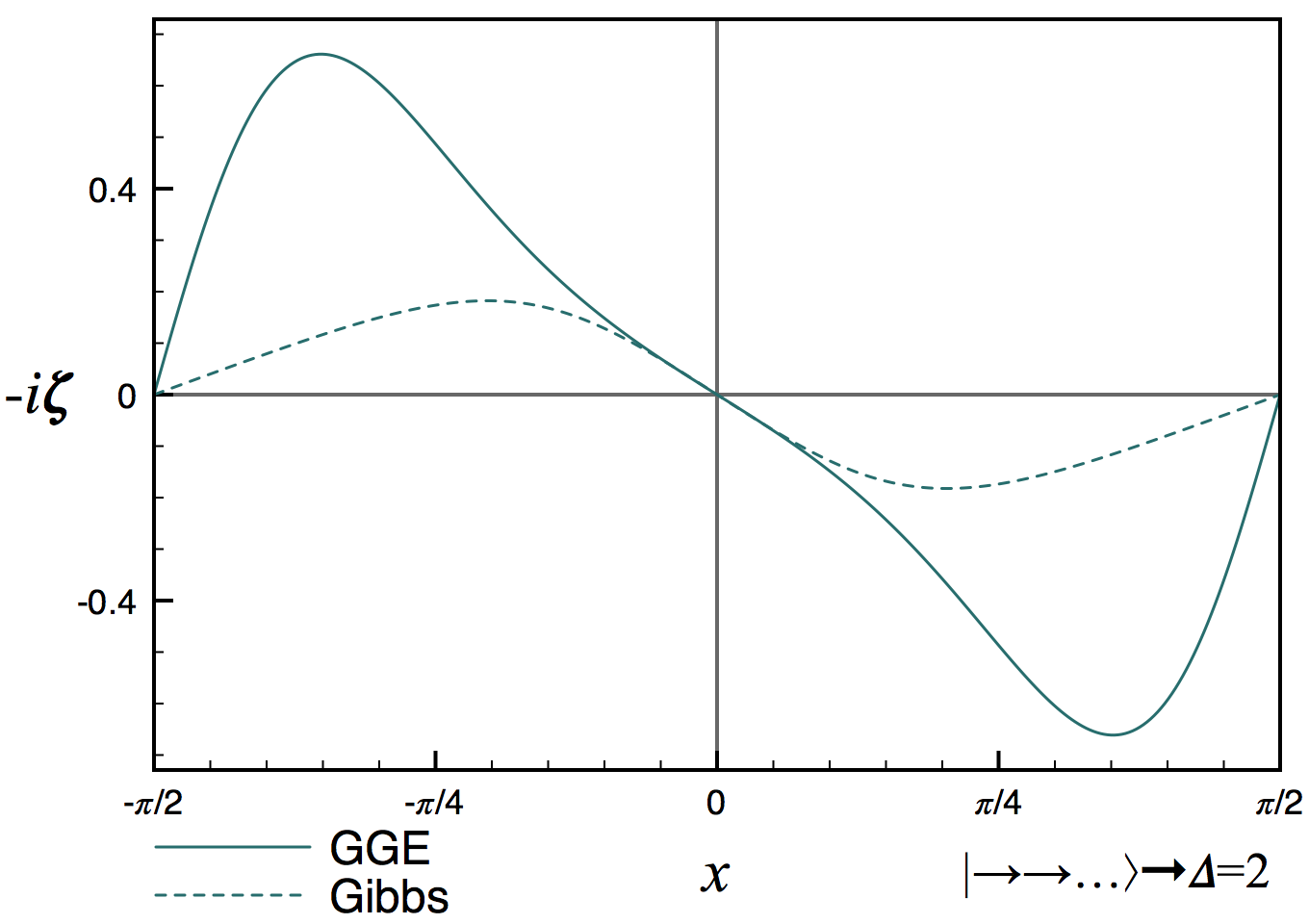}
\caption{Quench from the ferromagnetic state in x-direction
$|\rightarrow\rightarrow\dots\rangle$. The auxiliary function
  $\rho(x)$ vanishes at   $x=\frac{\pi}{2}$.}\label{fig:XUP}  
\end{figure}

\begin{figure}[ht]
\includegraphics[width=0.45\textwidth]{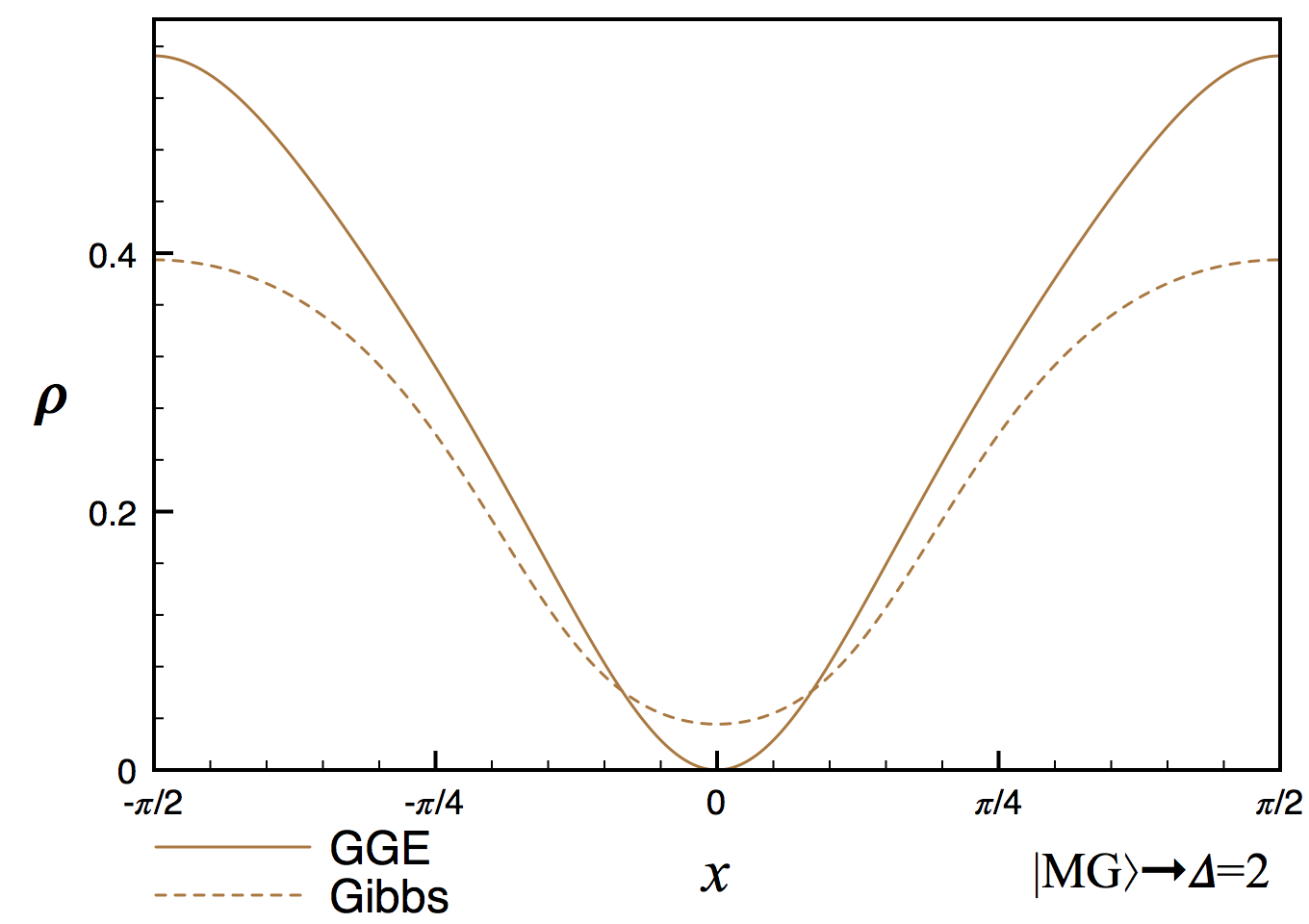}
\includegraphics[width=0.45\textwidth]{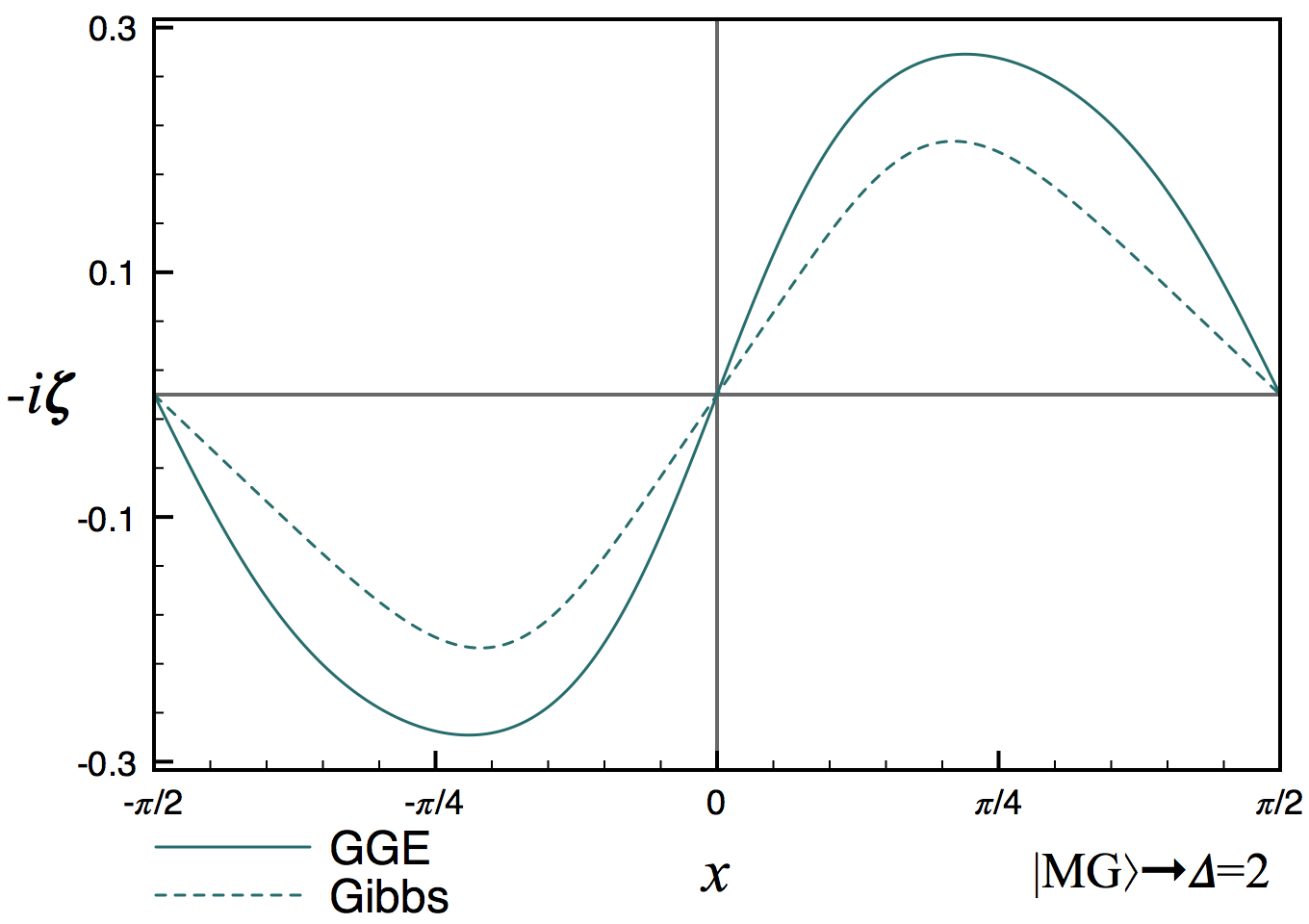}
\caption{Quench from the ground state of the Majumdar-Ghosh model. The
auxiliary function $\rho(x)$ vanishes at $x=0$.}\label{fig:MG} 
\end{figure}

\begin{figure}[ht]
\includegraphics[width=0.45\textwidth]{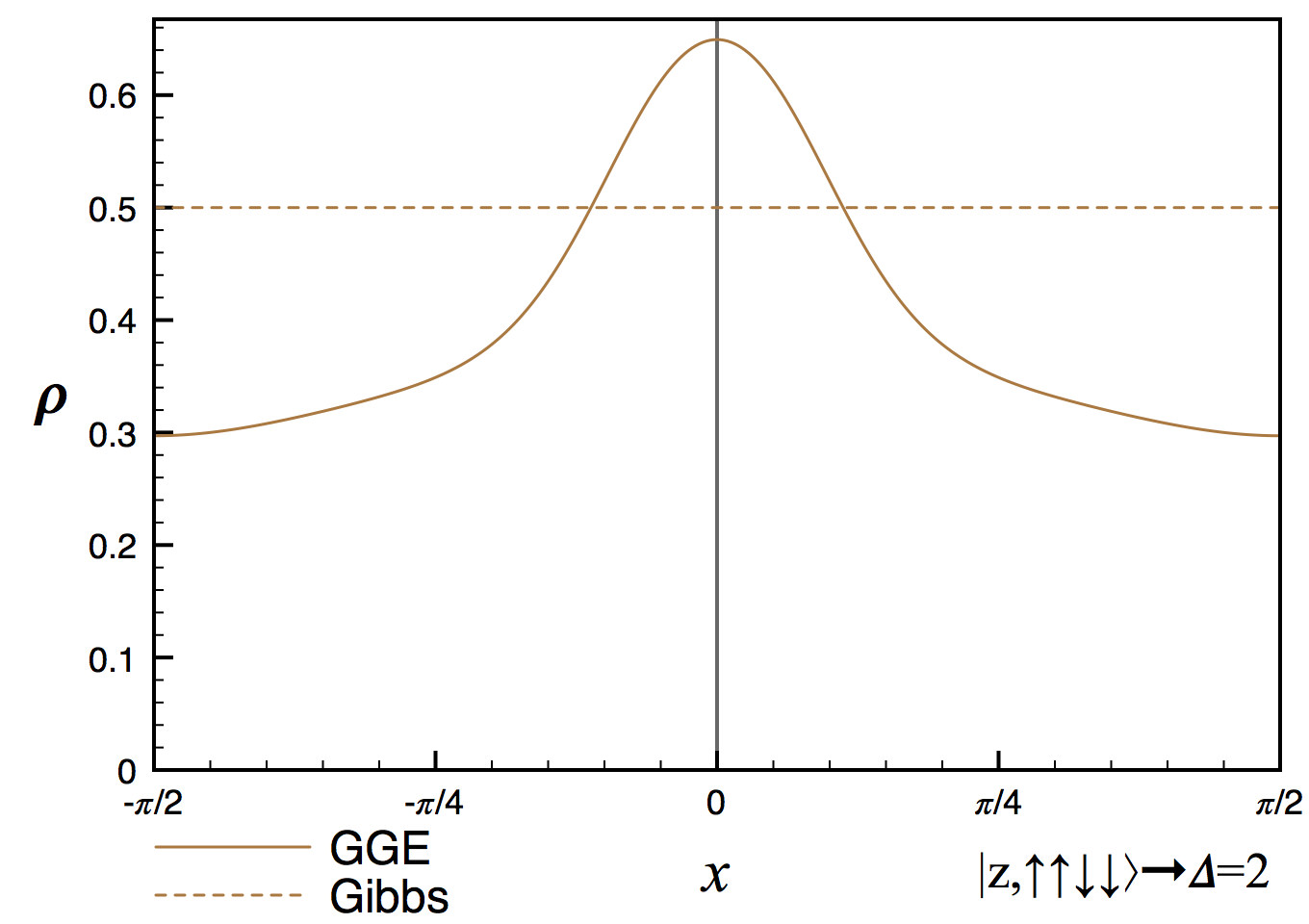}
\includegraphics[width=0.45\textwidth]{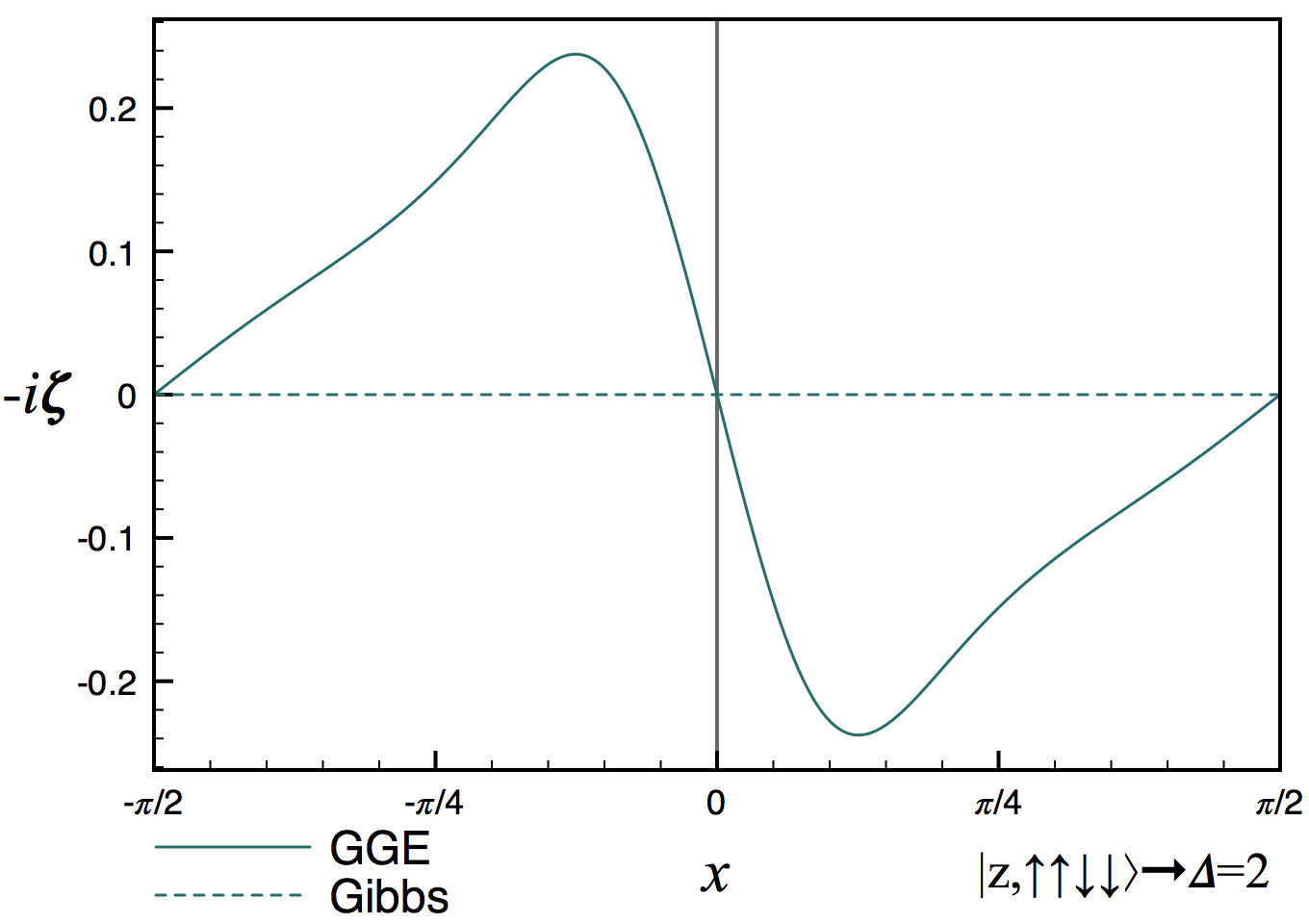}
\caption{Quench from the $s=2$ ferromagnetic domain state
  $\ket{\dots\uparrow\uparrow\downarrow\downarrow\uparrow\uparrow\downarrow\downarrow\dots}$. The effective
  temperature of the Gibbs ensemble is infinite, as one can immediately infer from
  $\braket{H^{(1)}}=2^{-L}\mathrm{Tr}[H^{(1)}]$.
}\label{fig:DW2} 
\end{figure}

Figs~\ref{fig:NS}-
\ref{fig:DW2} show results obtained from a numerical solution of the
system \fr{eq:systemh}.

In all quenches we considered, $\mathcal E(x)$ is very different from
$\mathcal E_{\rm Gibbs}(x)$. This agrees with general expectations
based on Refs~[\onlinecite{higherCL}]. A peculiar feature arising in
many quenches, see e.g. Figs~\ref{fig:NS}, \ref{fig:XNS30}, \ref{fig:XNS},
\ref{fig:XUP}, \ref{fig:MG}, is the presence of zeroes
in $\rho(x)$, which are associated with logarithmic singularities of
$\mathcal E(x)$. Ultimately such singularities are a consequence of
the long-range nature of $H_{\rm GGE}$ for these quenches, i.e. the
magnitudes of the Lagrange multipliers $\lambda_\ell$ decay very
slowly with $\ell$. Very similar singular behaviour has previously
been reported in $H_{\rm GGE}$ after quenches in models that have free
fermionic spectra. In these cases the singular behaviour was traced
back to the fact that the initial state is an eigenstate of
(generally nonlocal) conservation laws \cite{F:13a,FE_13a}.
Such a relation holds true for the XXZ chain as well: the appropriate
(nonlocal) charges are given by
\begin{multline}\label{eq:Q}
Q(k)=\frac{1}{\sinh\eta}\sum_{j=0}\frac{1}{
  j!}\Bigl[\Bigl(\frac{-2k-i\eta}{\sinh\eta}\Bigr)^j+\Bigl(\frac{-2
    k+i\eta}{\sinh\eta}\Bigr)^j\Bigr] \frac{H^{(j+1)}}{L}+\frac{\sinh\eta}{\cosh\eta-\cos(2k)}
\equiv \\
\frac{i}{\eta L}\Bigl[\tau^\prime\Bigl(-\frac{2 k}{\eta}\Bigr)\tau^{-1}\Bigl(-\frac{2 k}{\eta}\Bigr)+\tau^\prime\Bigl(2i-\frac{2 k}{\eta}\Bigr)\tau^{-1}\Bigl(2i-\frac{2 k}{\eta}\Bigr)\Bigr]+\frac{\sinh\eta}{\cosh\eta-\cos(2k)}
\, .
\end{multline}
Here $\tau$ is the transfer matrix of the XXZ model
\be
\tau(i+\lambda)=\mathrm{Tr}_{\mathcal{T}}\Bigl[\prod_j L(\lambda;\sigma_j)\Bigr]\, ,
\ee
where the L-operator $L(\lambda;\sigma_j)$ is defined in \eqref{eq:LM}
and $\mathcal T$ denotes the auxiliary space (on which the $\tau^\alpha$ act). 
The additive constant and the normalisation have been chosen in such a way that
\be
\braket{\Psi_0|Q(k)|\Psi_0}=\rho^{(1)}(k)d(k)\, ,
\ee
where $\rho^{(1)}$ is defined in \eqref{eq:rho1}. Therefore, the
``small quench limit'' of Ref~[\onlinecite{FE_13b}] corresponds to the
regime $\braket{\Psi_0|Q(k)|\Psi_0}\ll 1$. 
Using the identity 
\be\label{eq:identity}
\frac{1}{2}\sum_{j=0}\frac{1}{
  j!}\Bigl[\Bigl(\frac{-2y-i\eta}{\sinh\eta}\Bigr)^j+\Bigl(\frac{-2
    y+i\eta}{\sinh\eta}\Bigr)^j\Bigr] \Bigl(\frac{\sinh
  \eta}{2}\frac{\rm d}{\rm d x}\Bigr)^j d(x)=\pi \delta(x-y)\ ,
\qquad x,y\in\Bigl(-\frac{\pi}{2},\frac{\pi}{2}\Bigr),
\ee
one then finds
\be
\frac{H_{\rm GGE}}{L}
\sim \int_{-\frac{\pi}{2}}^{\frac{\pi}{2}}\frac{\mathrm d k}{\pi}
\mathcal E(k) \Bigl(Q(k)-\frac{\sinh\eta}{\cosh\eta-\cos(2k)}\Bigr)\, .
\ee
This is to be interpreted as the density matrices corresponding to the
operators on the two sides of the equation yielding identical local
properties. If $\mathcal E(k)$ diverges at a particular value $k_0$,
the Lagrange multiplier of the conserved charge $Q(k_0)$ is infinite,
which implies that only a subspace of the Hilbert space contributes to
the generalized Gibbs ensemble. 

In the transverse-field Ising chain the long time behaviour of
transverse correlations\cite{CEF} after a quench of the transverse
field is determined precisely by the degrees
of freedom that are almost ``frozen'' by the above mechanism. In
particular, transverse correlations decay at late times like
$t^{-3/2}$ rather than the naive expectation $t^{-1/2}$, because the
Bogoliubov modes with momenta $0$ and $\pi$ are removed by the
aforementioned projection mechanism: the initial state is an
eigenstate of the conserved charges $Q_0=\alpha^\dag_{0}\alpha_{0}$
and $Q_\pi=\alpha^\dag_{\pi}\alpha_{\pi}$. These observations suggest
the possibility that in the XXZ case zeroes in $\rho(x)$ might affect
the late time behaviour of observables in a similar fashion. 

\section{Numerical Results}\label{s:num}

In this section we present extensive numerical studies of the quench
dynamics of the XXZ chain with $\Delta>1$ by means of the
time-dependent density matrix renormalisation group (tDRMG)
\cite{tDMRG} and infinite time-evolving block decimation (iTEBD)
\cite{iTEBD} algorithms. The latter has the advantage of working
directly in infinite systems avoiding both finite size and revival
effects. However, since as well known, the main limitation to the
working of both algorithms is the fast growth of the entanglement
entropy after a global quantum quench, for any practical purpose the
two techniques are equivalent. The tDMRG computations are performed on
finite chains of $L$ spins ($L$ is taken to be even) with open
boundary conditions
\be\label{H_XXZ}
H = J \sum_{j=1}^{L-1} \left( S^{x}_{j}S^{x}_{j+1} +
S^{y}_{j}S^{y}_{j+1} + \Delta\,S^{z}_{j}S^{z}_{j+1} \right ).
\ee
In the following we will set $J=1$.
The late time behaviour of short-range correlators will then be
compared with the GGE predictions obtained from the numerical solution
of the system of nonlinear integral equations reported in
Sec.~\ref{s:nlies}. Depending on the initial state the relevant
generating function \eqref{eq:Om0} is computed either numerically or
analytically.

\subsection{Details of the tDMRG analysis}

When necessary, the algorithm initially performs a static subroutine
which selects the initial state $|\Psi_{0}\rangle$ as the ground state
of a given Hamiltonian. In the decimation process of this static
subroutine, we keep a number of states such that the energy precision
is at least of the order of $10^{-12}$. Subsequently, we perform the
evolution using the time-adapting block-decimation procedure
implemented both in tDMRG and iTEBD code. In the tDMRG code, we always
use open boundary conditions. We use the second order 
(and in some cases the fourth order) Suzuki-Trotter decomposition of the 
evolution operator with time step $dt$ which varies in the range
$[5\cdot 10^{-3}, 5\cdot 10^{-4}]$. We checked the stability of the
results with the change of $dt$ in order to be sure that no systematic
errors are introduced by time discretisation. In the tDMRG code, for
each time step, the local evolution operator is applied sequentially
on each bond starting form the left boundary of the chain and going to
the right border and coming back. We adapt in time the number of
states used to describe the reduced Hilbert space retaining at each
local step all those eigenvectors of the reduced density matrix
corresponding to eigenvalues larger than
$\lambda_{min}\in[10^{-18},10^{-20}]$, up to a maximum value $\chi_{\rm
  MAX}\in[300,1000]$ (clearly the effective maximal value used by the
algorithm strongly depends on the simulation parameters). For the
iTEBD, thanks to the invariance under two-site shift, we needs only to
apply the local evolution operator twice (on the odd and on the even
bounds, see Ref. \onlinecite{iTEBD} for details). In this algorithm the
number of state is kept fixed to $\chi_{\rm MAX}$ from the beginning
of the simulation.

\begin{figure}[t!]
\includegraphics[width=0.4\textwidth]{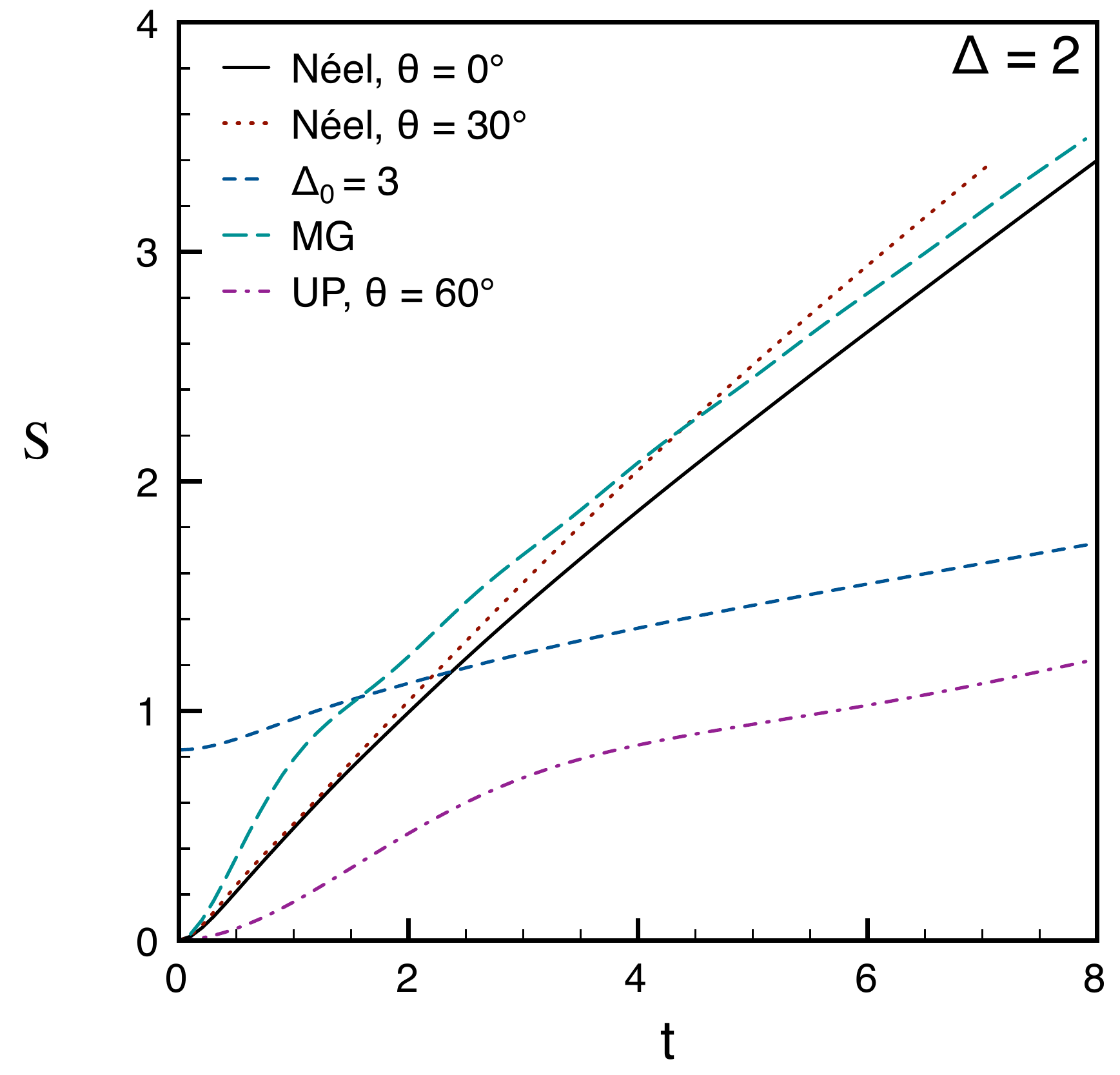}
\includegraphics[width=0.4\textwidth]{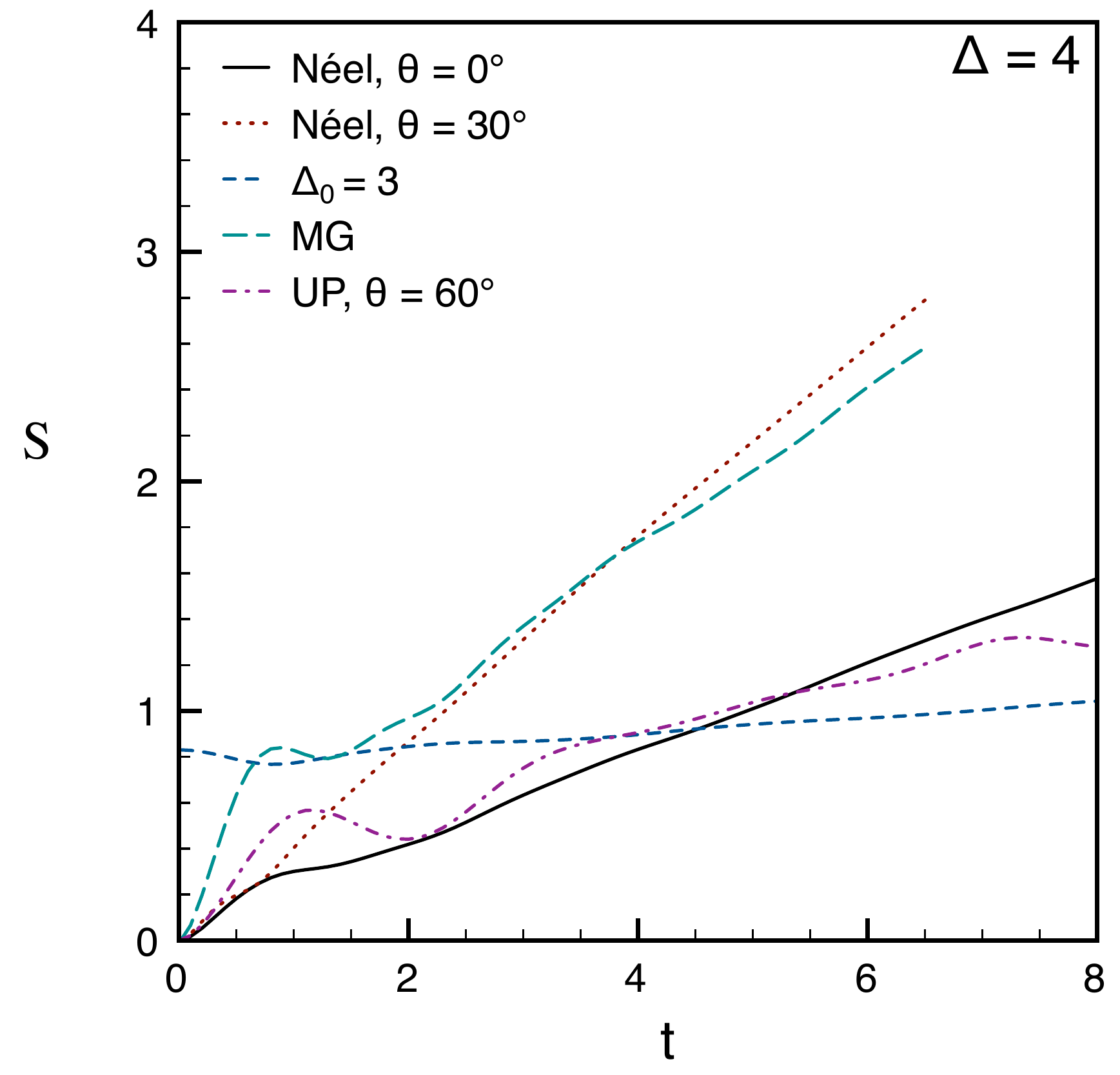}
\caption{\label{figS} Time evolution of the half-system entanglement
entropy for quenches starting from several initial states considered
in the text to XXZ chains with $\Delta=2$ (left) and $\Delta=4$
(right) respectively.
} 
\end{figure}

In order to check the GGE predictions, we focus our attention on the
following two-point spin-spin correlation functions
\be
S^{\alpha}_{j,j+\ell}=\langle\Psi_0(t)|\sigma^{\alpha}_{j}\sigma^{\alpha}_{j+\ell}|\Psi_0(t)\rangle\ ,
\quad \ell=1,2,3\ ,\ \alpha=x,y,z.
\ee
Here $\sigma^{\alpha}_{j}\equiv 2 S^{\alpha}_{j}$ are Pauli
matrices. The iTEBD algorithm operates directly in the thermodynamic limit and
hence the choice of $j$ in $S^\alpha_{j,j+\ell}$ (i.e. the location of
the first spin in the two-point function we are computing) is
irrelevant in the sense that any breaking of translational invariance
is entirely induced by the initial state and not due to finite-size effects.
This is not the case for tDMRG simulations, which are performed in
finite systems (of total even length $L$).   
Thus, in order to avoid spurious boundary effects for the largest possible time, 
we measure the correlators in the middle of the chain, i.e.
\be
F_{\alpha  1}\equiv S^\alpha_{\frac{L}{2},\frac{L}{2}+1}\ ,\
F_{\alpha  2}\equiv S^\alpha_{\frac{L}{2},\frac{L}{2}+2}\ ,\
F_{\alpha  3}\equiv S^\alpha_{\frac{L}{2}-1,\frac{L}{2}+2}\ ,\quad \alpha=x,y,z.
\label{Faj}
\ee
A list with explicit results for the expected stationary values of
\fr{Faj} is presented in Appendix \ref{app:list}. 
Even for correlators in the middle of the chain boundary effects will
start to be felt after a certain time. Such unwanted effects are
easily detected, e.g. by checking when the entanglement entropy of the
left half stops growing linearly in time. In all the plots reported in
the following only data unaffected by such boundary effects are presented.

After having under control all other sources of systematic errors
(i.e. discretisation of time and finite sizes), the only limitation of
the numerical algorithms is given by the finite number of states kept
in the decimation. Indeed, the computational complexity of the time
evolution of a quantum system on a classical computer using any
algorithm based on matrix product states (including tDMRG and iTEBD)  
is essentially set by the growth of the bipartite entanglement. 
In general for a global quantum quench, the entanglement entropy is expected to grow linearly with time \cite{cc-05}.
In Fig. \ref{figS} we report the growth of the half-system entanglement entropy with time for 
some representative initial states and evolving with the XXZ Hamiltonian for $\Delta=2$:
in all cases we have an asymptotic linear increase, but the slope
varies considerably from quench to quench. Consequently, as the
entanglement increases, we have to increase exponentially with time
the dimension $\chi$ of the reduced Hilbert space in order to
optimally control the truncation error. In spite of the adaptive
choice of $\chi$, the truncation procedure remains the main source of
error of the algorithm.

For most of the quenches studied in the following we have used both
algorithms and checked that the data are equivalent. However, for the
largest times reported, the simulations are numerically demanding and
we have chosen one of the two algorithms to avoid costing
duplications. In the main text, we will discuss the numerical data
without specifying every time the used algorithm which will be
reported only in the caption of the figures.

\subsection{Tilted N\'eel state}

\begin{figure}[ht]
\includegraphics[width=0.35\textwidth]{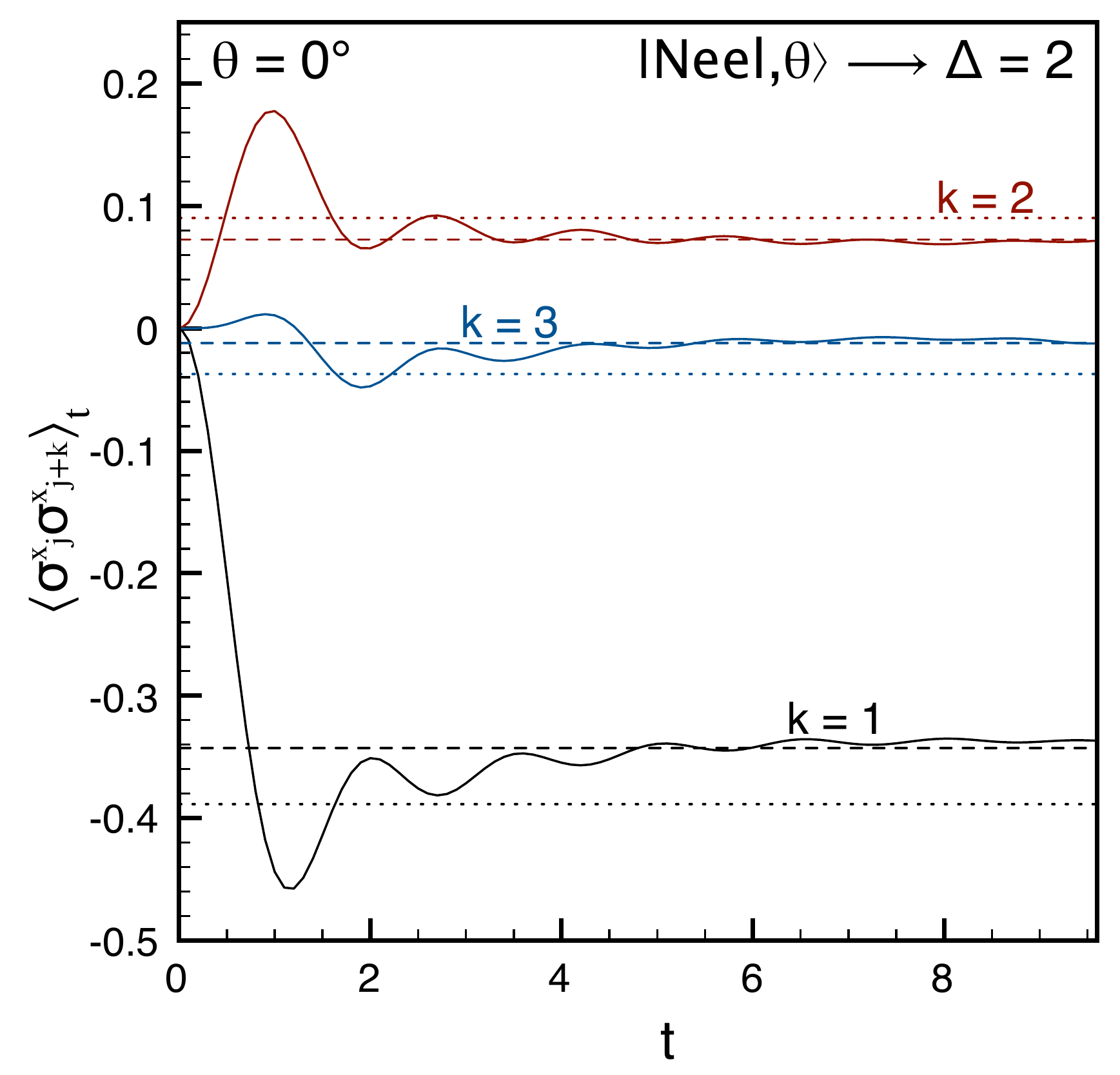}
\includegraphics[width=0.35\textwidth]{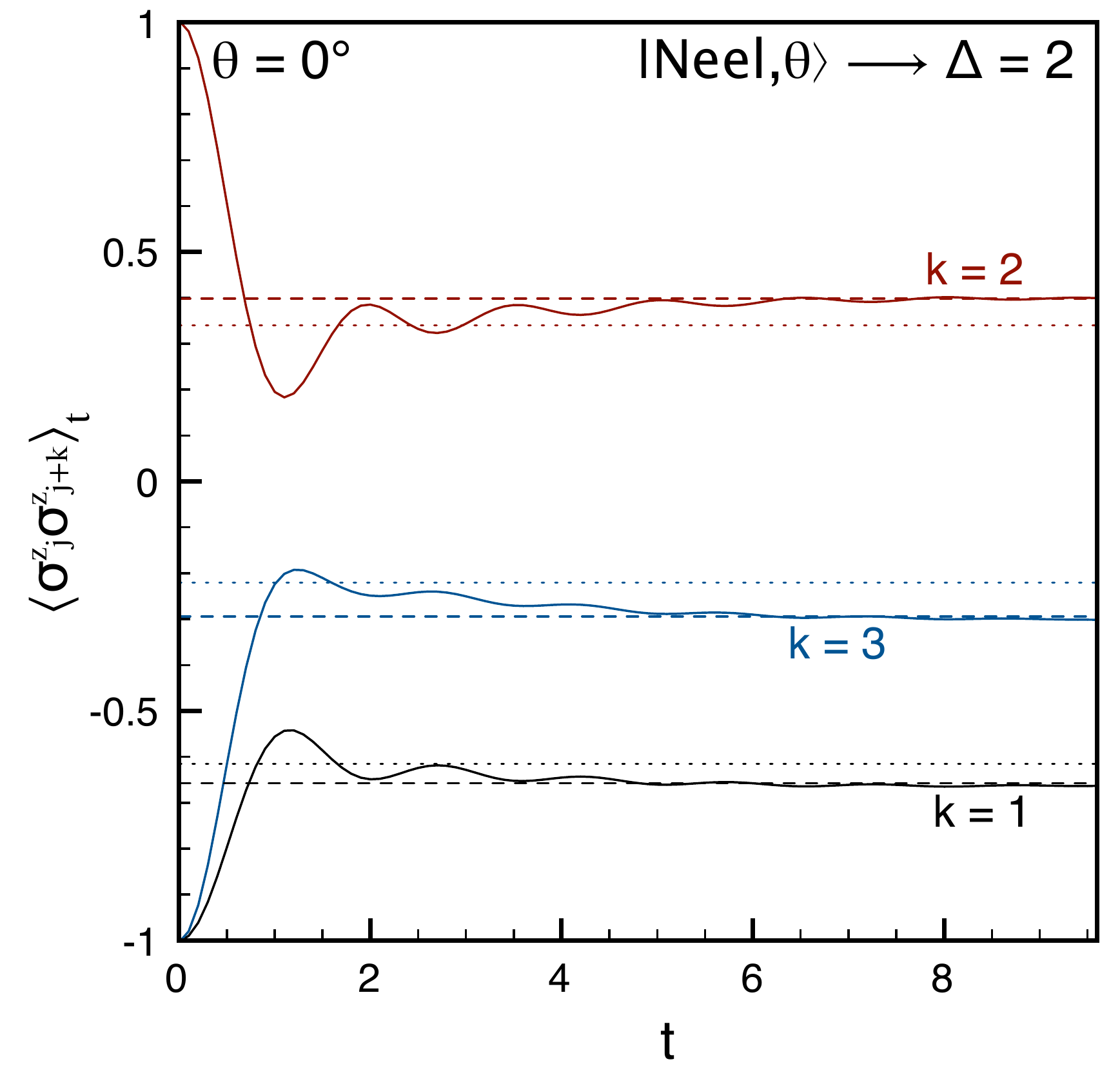}
\caption{\label{figNeel90}tDMRG results on a chain of $L=64$ sites
for a quench from a N\'eel state along the
$z$-direction (i.e. with $\theta=0^{\circ}$) to $\Delta=2$. Left panel:
transverse spin correlations ($S^x_{j,j+k}=S^y_{j,j+k}$ as the initial
state respects the U(1) spin rotational symmetry of
$H^{(1)}(\Delta)$). Right panel: longitudinal correlations.
The dashed lines indicate the GGE predictions, which are seen
to be approached fairly quickly. 
The dotted lines are the thermal expectation values (at the finite temperature given by 
the energy of the initial state) which are well separated from the GGE.}
\end{figure}

\begin{figure}[ht]
\includegraphics[width=0.35\textwidth]{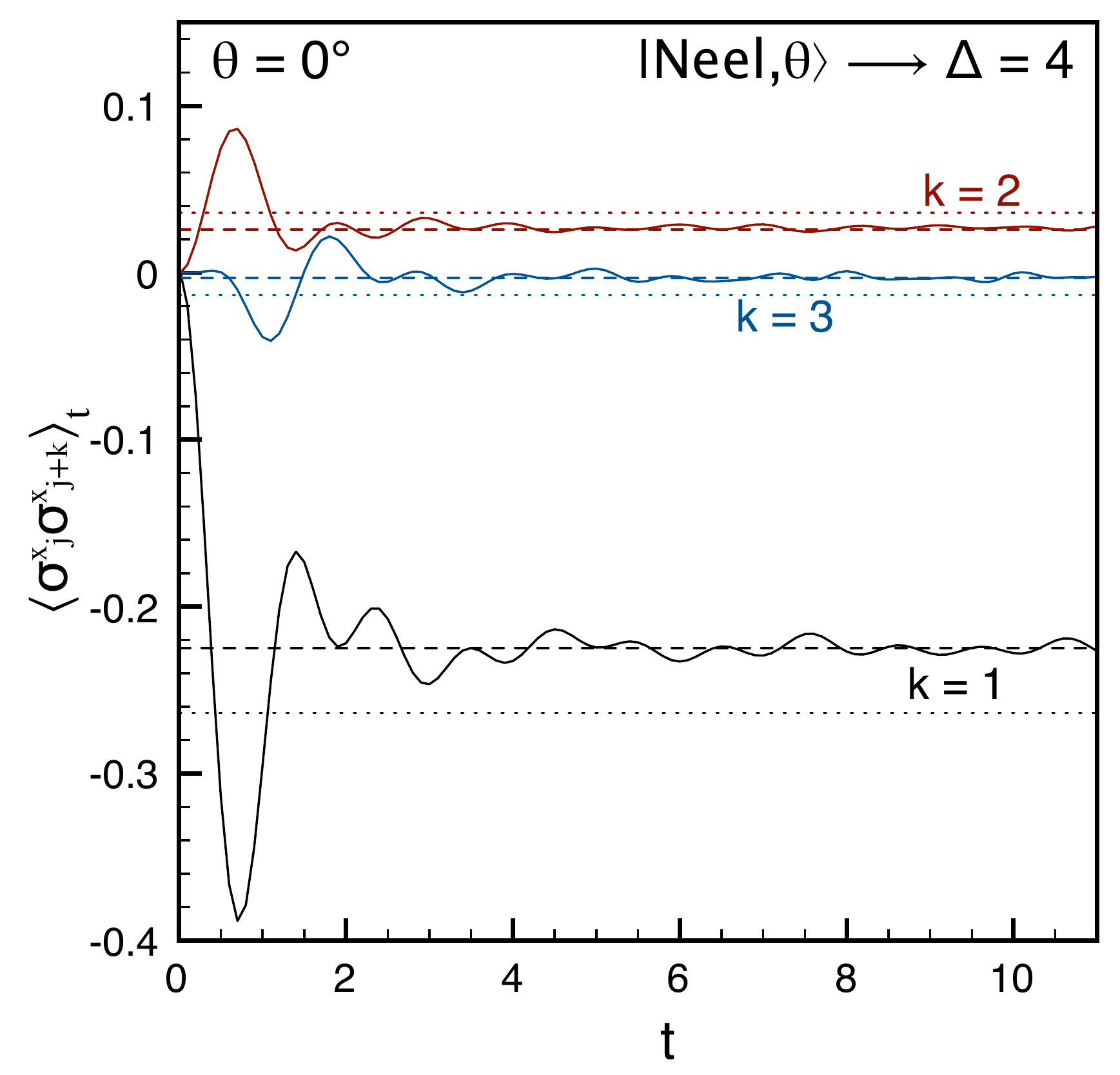}
\includegraphics[width=0.35\textwidth]{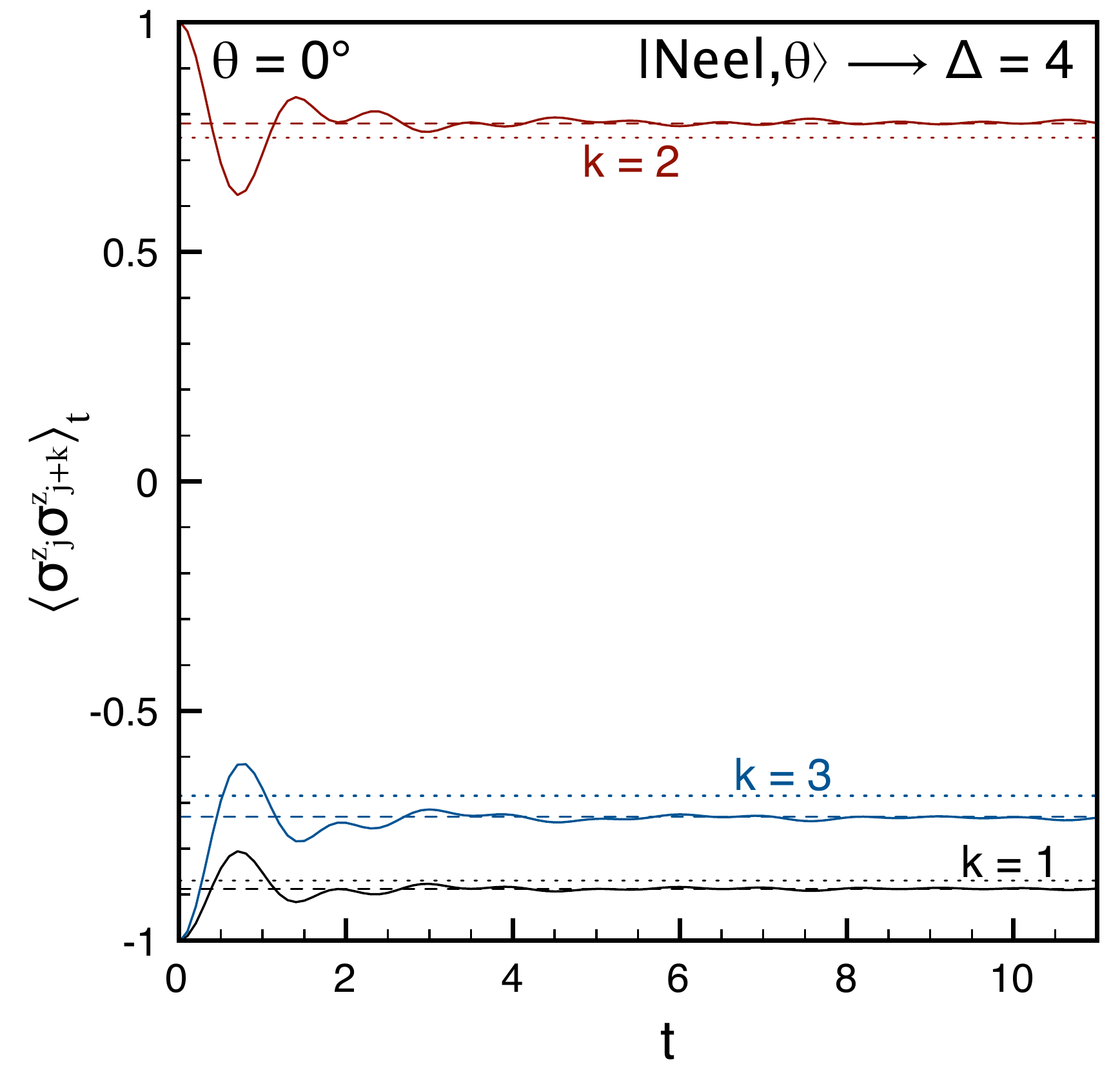}
\caption{Same as Fig.~\ref{figNeel90} but with $\Delta=4$.}
\label{figNeel90b}
\end{figure}

We first consider the evolution from a N\'eel state pointing in an
arbitrary direction in the $xz$ plane, i.e. from the initial state  
\be
\ket{\theta;\nearrow\swarrow\dots}=e^{i\theta\sum_j
  S^y_j}\ket{\uparrow\downarrow\dots}.
\ee
The N\'eel state in z-direction ($\theta=0^\circ$) respects the U(1)
symmetry of the Hamiltonian and leads to isotropic correlations in the
transverse directions, i.e. $S^x_{j,j+k}=S^y_{j,j+k}$. Results for
quenches from this state to $H^{(1)}(\Delta=2)$ and $H^{(1)}(\Delta=4)$ are
presented in Figs~\ref{figNeel90} and \ref{figNeel90b}
respectively. 

\begin{figure}[ht]
\includegraphics[width=0.32\textwidth]{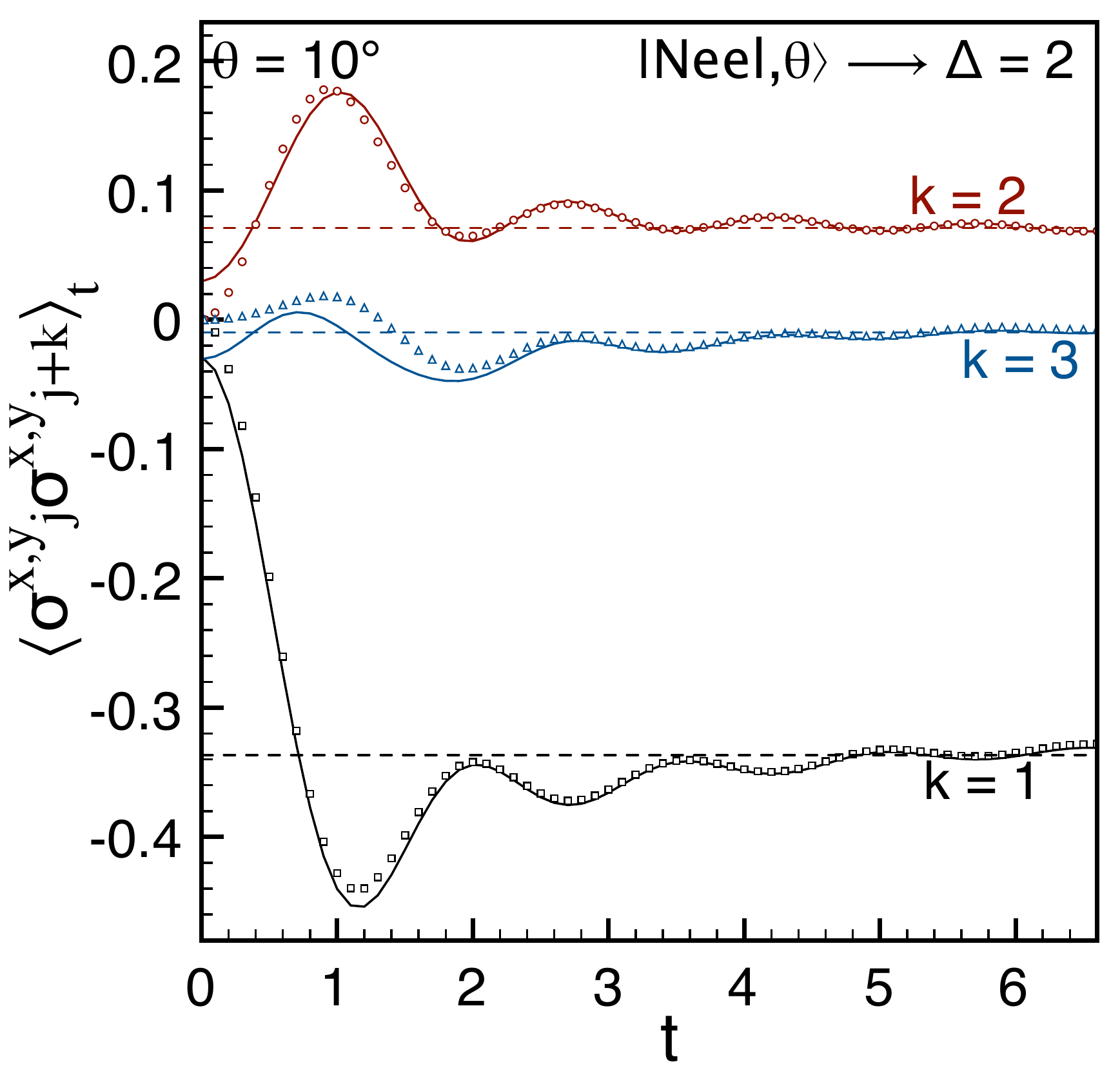}
\includegraphics[width=0.32\textwidth]{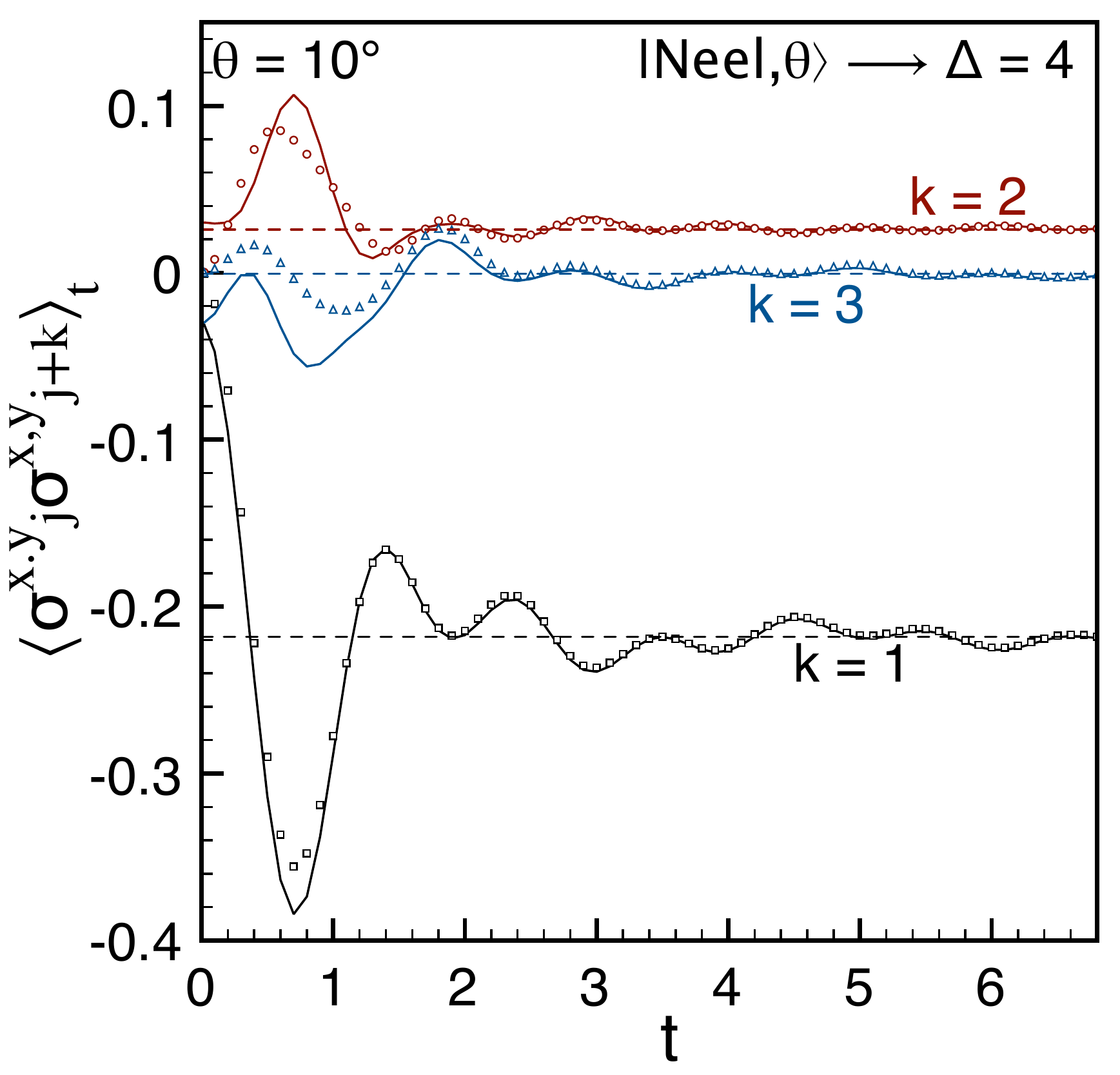}
\includegraphics[width=0.32\textwidth]{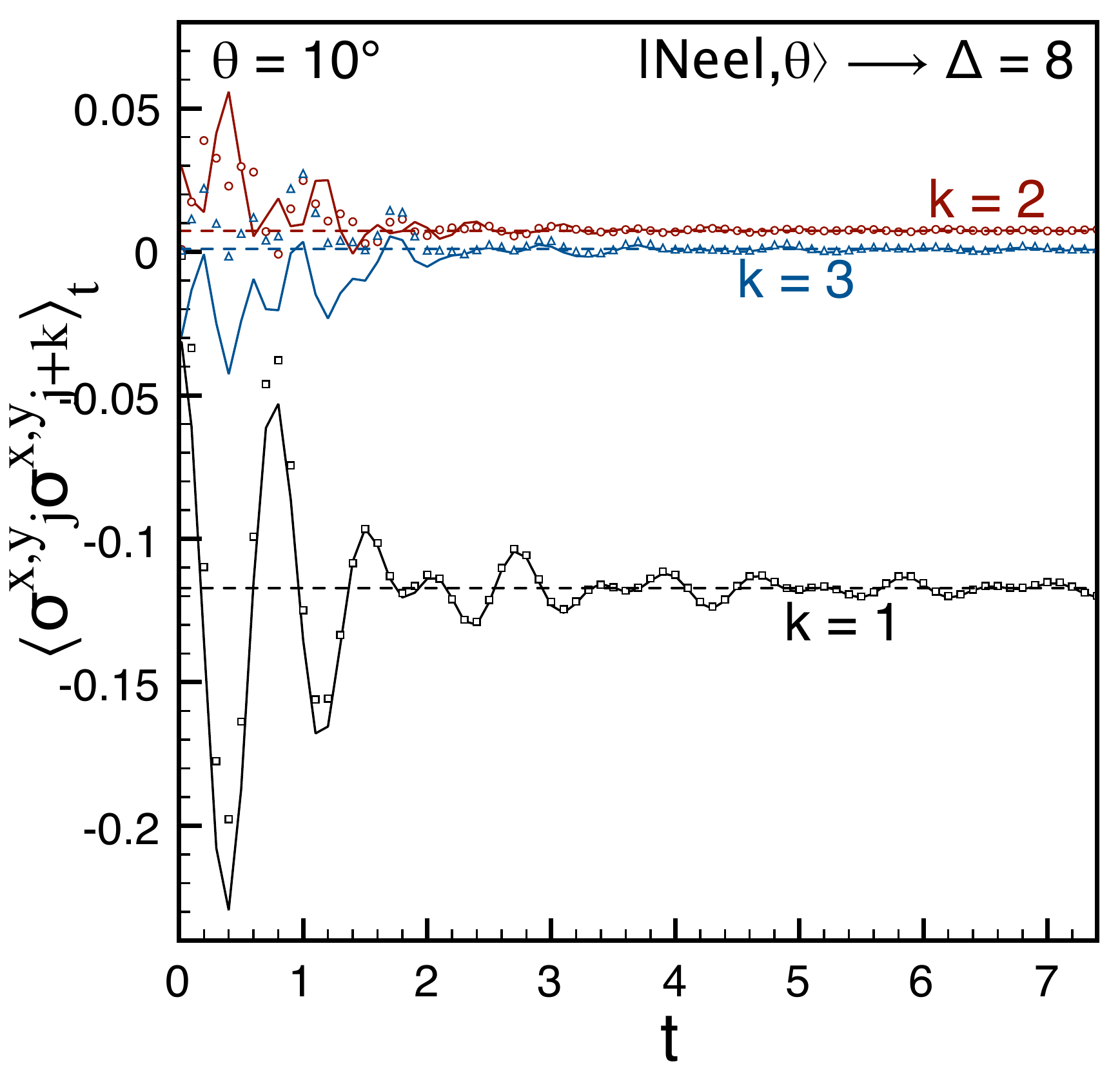}\\
\includegraphics[width=0.32\textwidth]{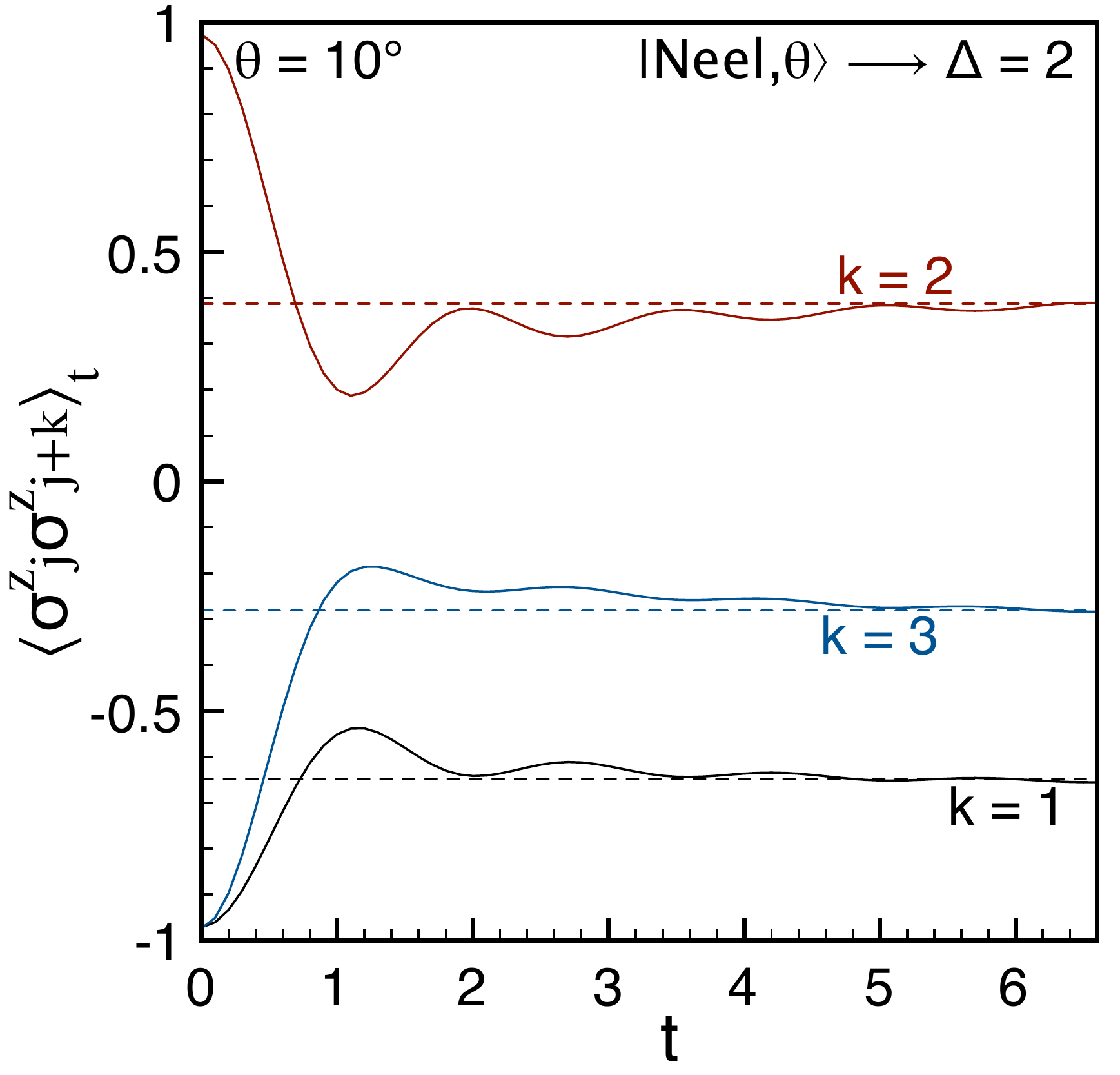}
\includegraphics[width=0.32\textwidth]{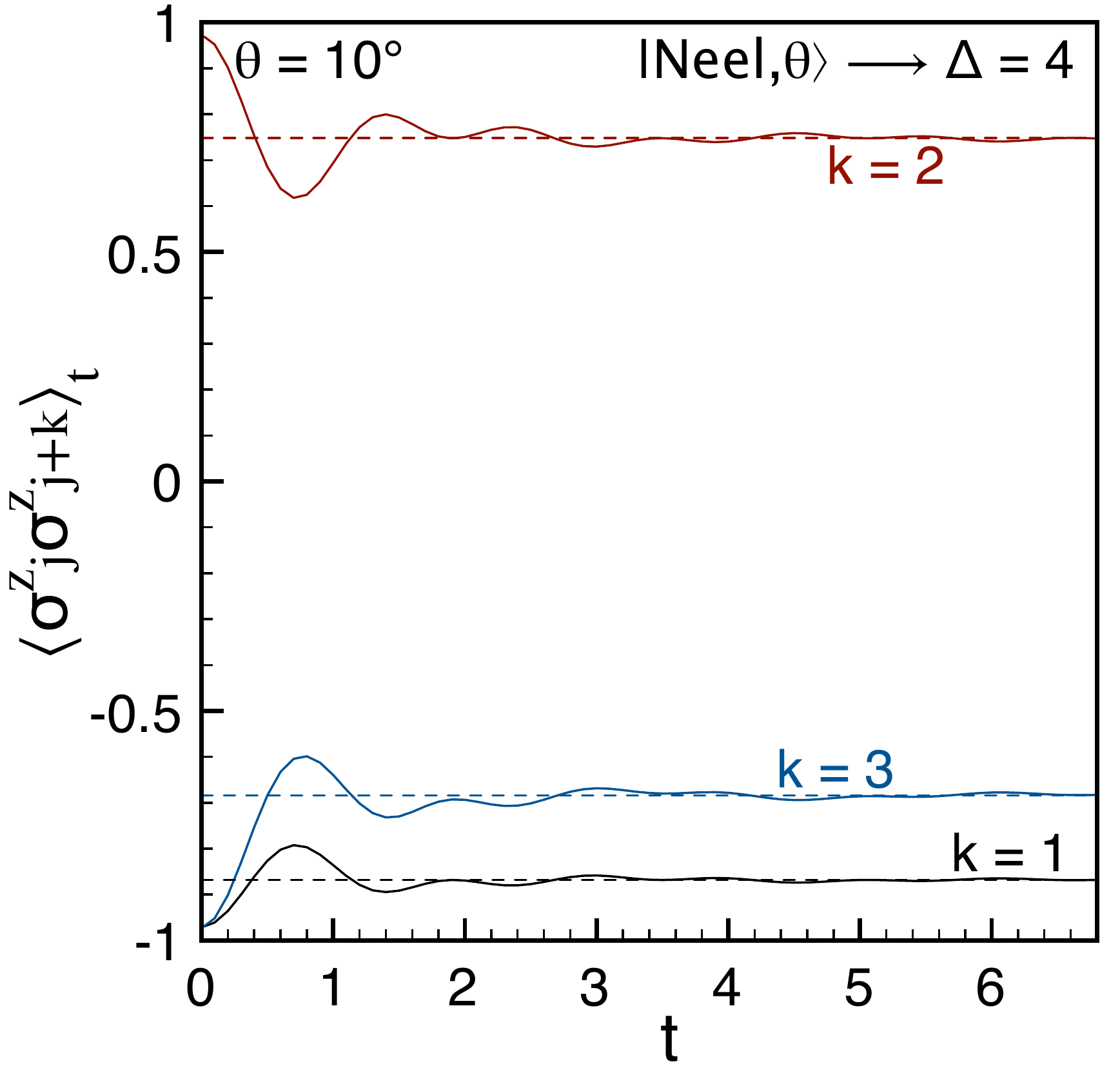}
\includegraphics[width=0.32\textwidth]{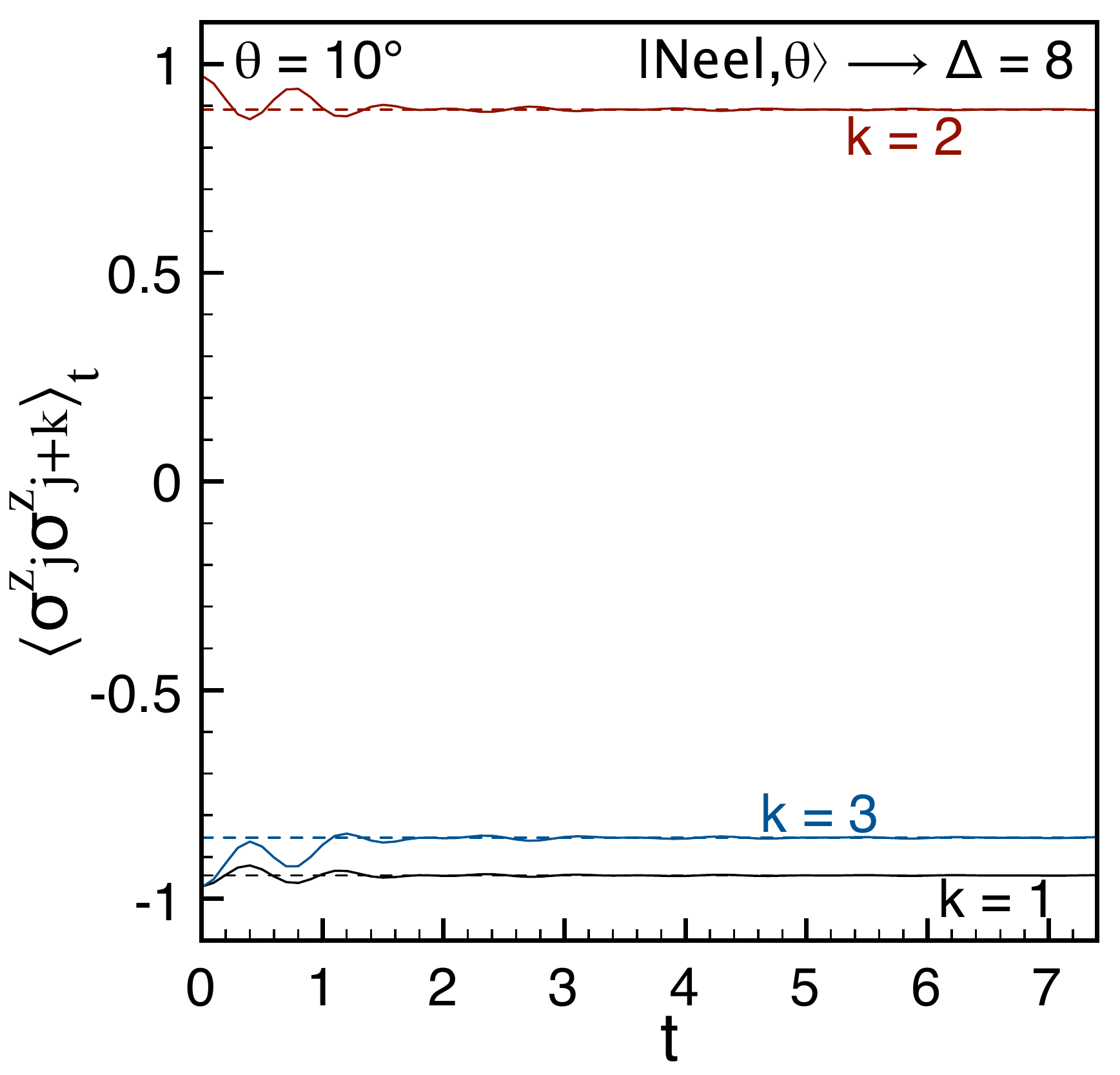}
\caption{\label{figNeel80}Quench from a N\'eel state with
  $\theta=10^{\circ}$ to $\Delta=2,4,8$ (from left to right). Top row:
transverse correlations $\langle\sigma^x_j\sigma^x_{j+k}\rangle_t$
(solid lines) and $\langle\sigma^y_j\sigma^y_{j+k}\rangle_t$ (symbols)
for distances $k=1,2,3$. The rotational
symmetry in the $xy$ plane is restored at $t\approx 2$.
Bottom row: longitudinal correlations for distances $k=1,2,3$.
All correlators approach the GGE predictions (dashed lines) at late
times.}  
\end{figure}

\begin{figure}[ht]
\includegraphics[width=0.32\textwidth]{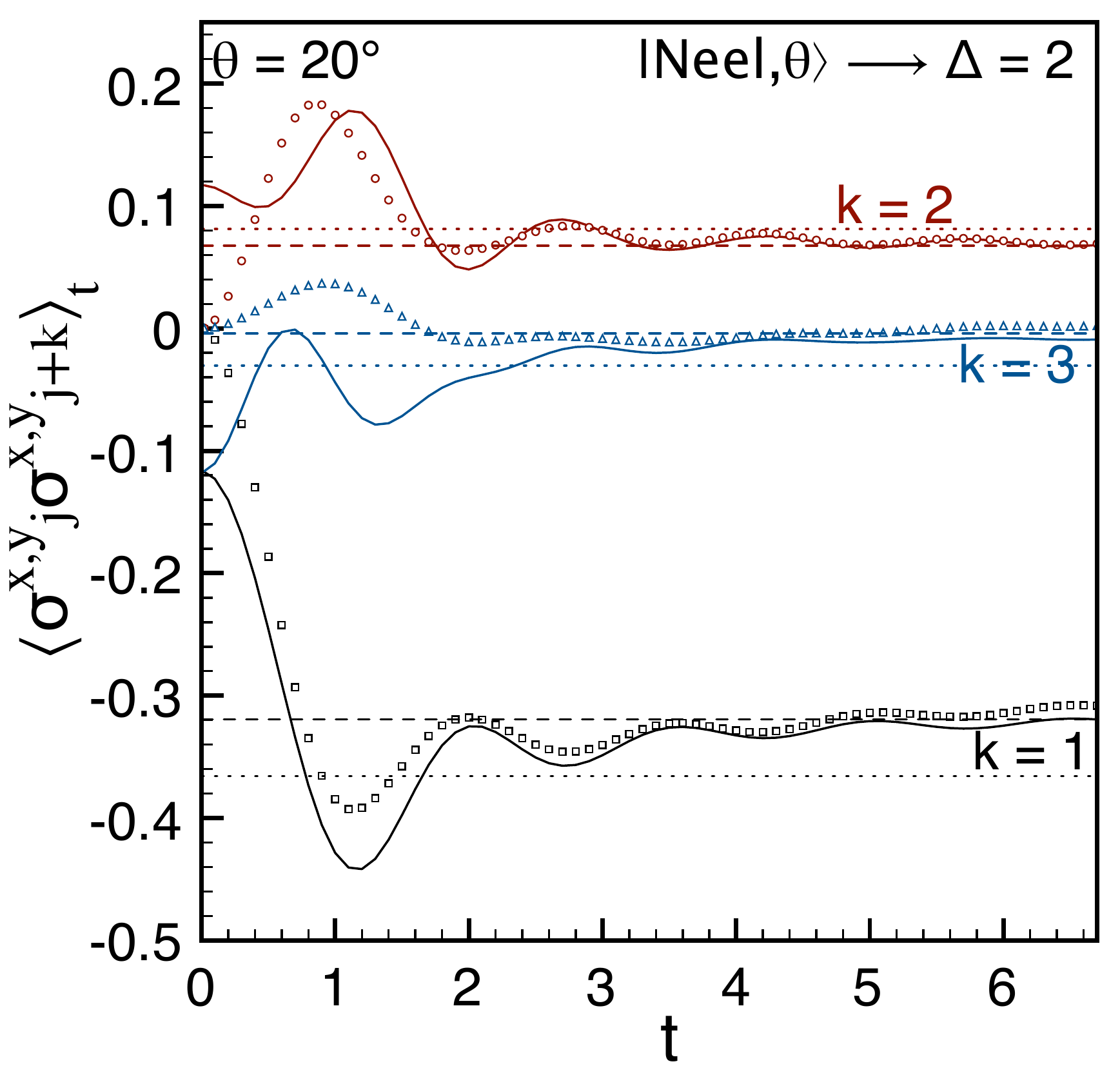}
\includegraphics[width=0.32\textwidth]{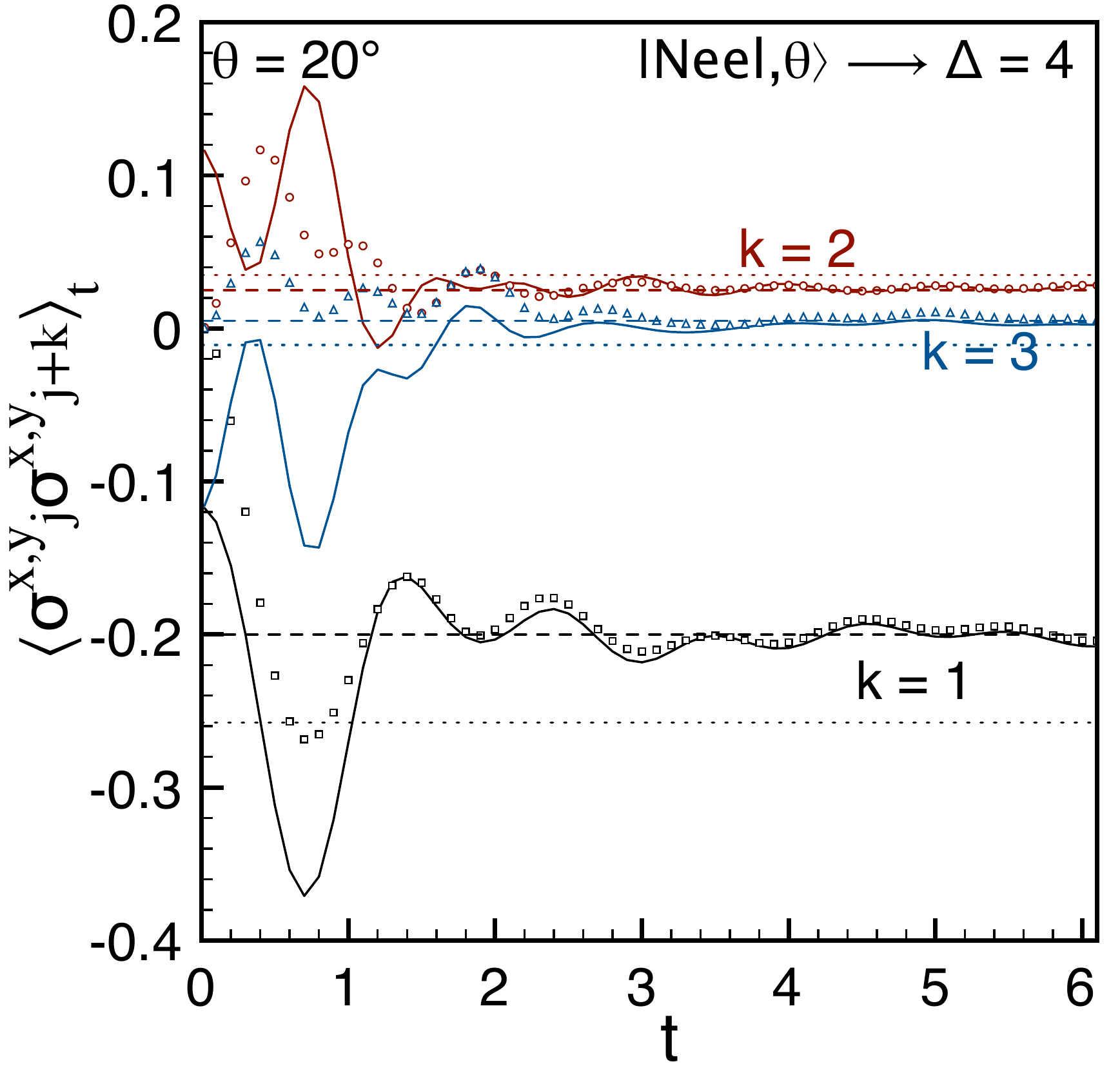}
\includegraphics[width=0.32\textwidth]{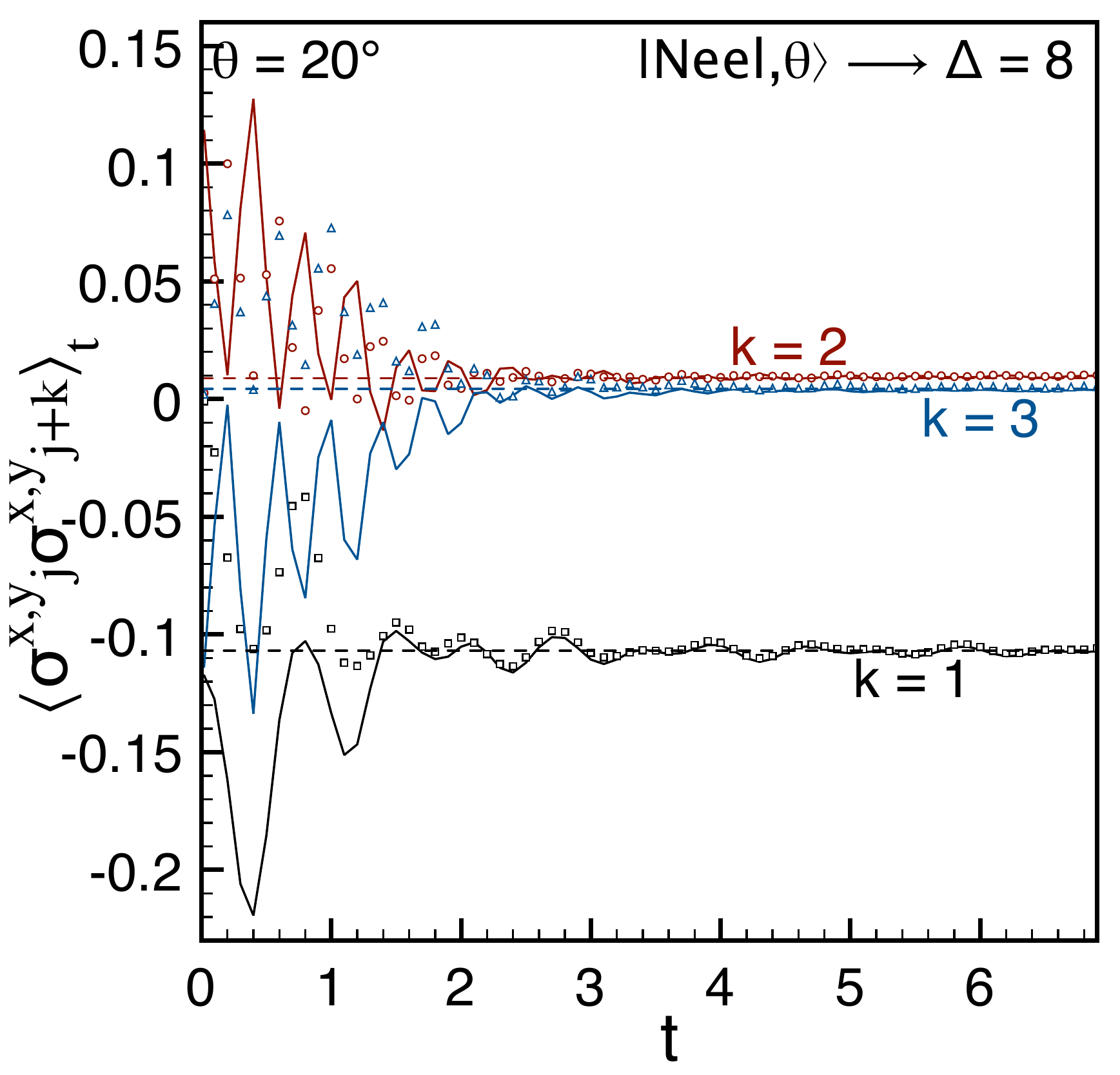}\\
\caption{\label{figNeel70}
Same as Fig. \ref{figNeel80}, but with initial state 
with $\theta=20^{\circ}$.
}
\end{figure}
\begin{figure}[ht]
\includegraphics[width=0.32\textwidth]{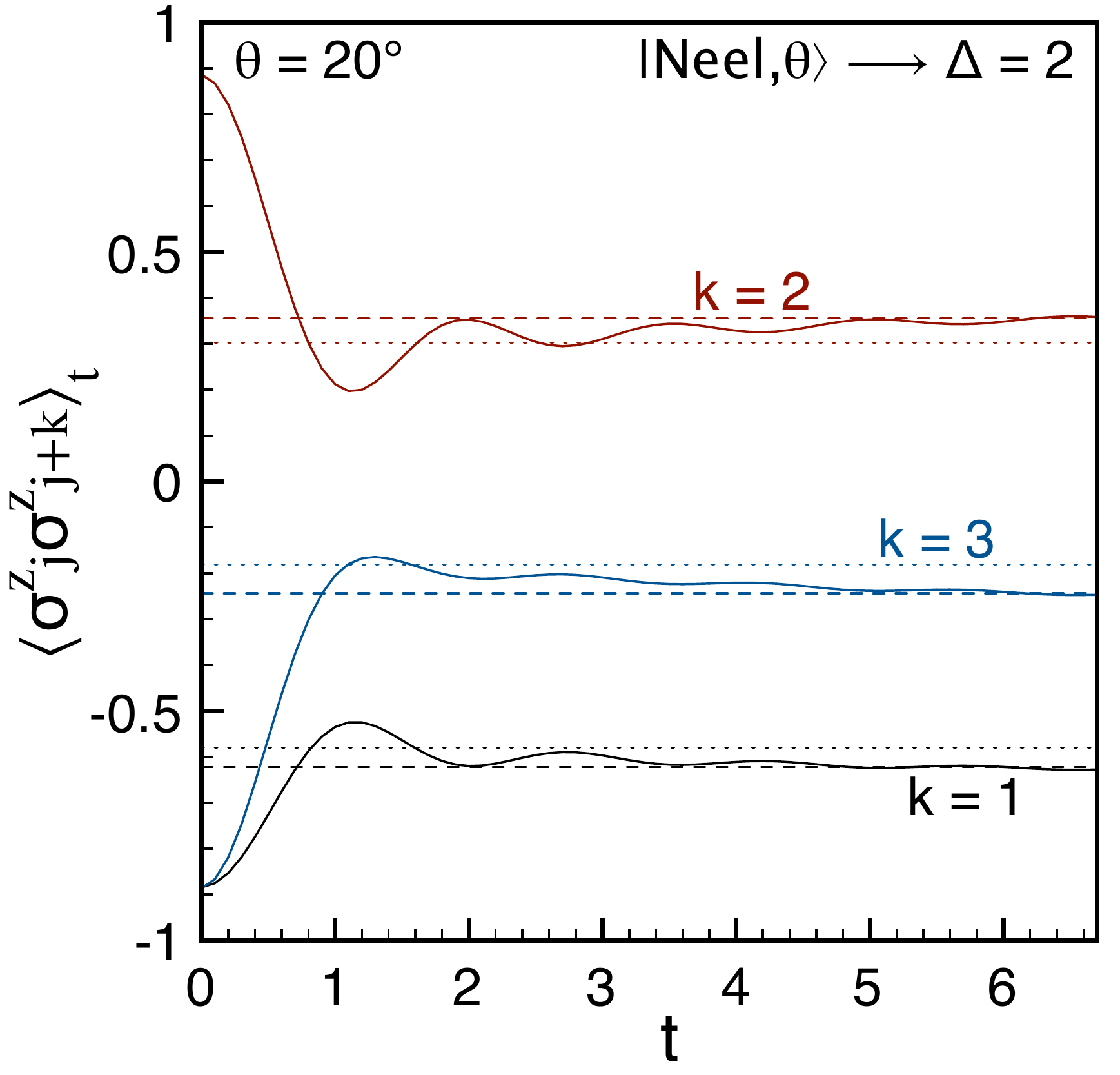}
\includegraphics[width=0.32\textwidth]{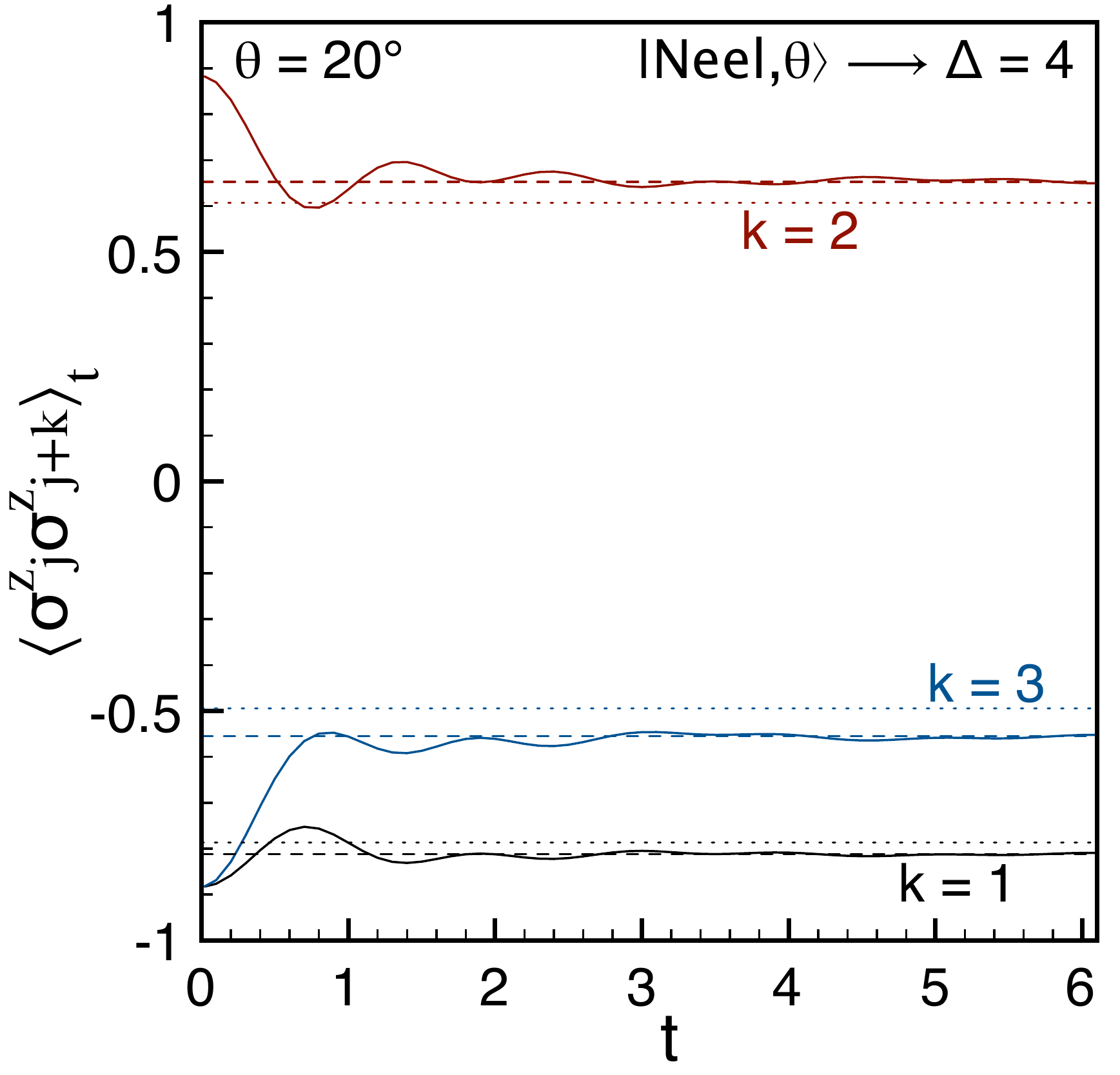}
\includegraphics[width=0.32\textwidth]{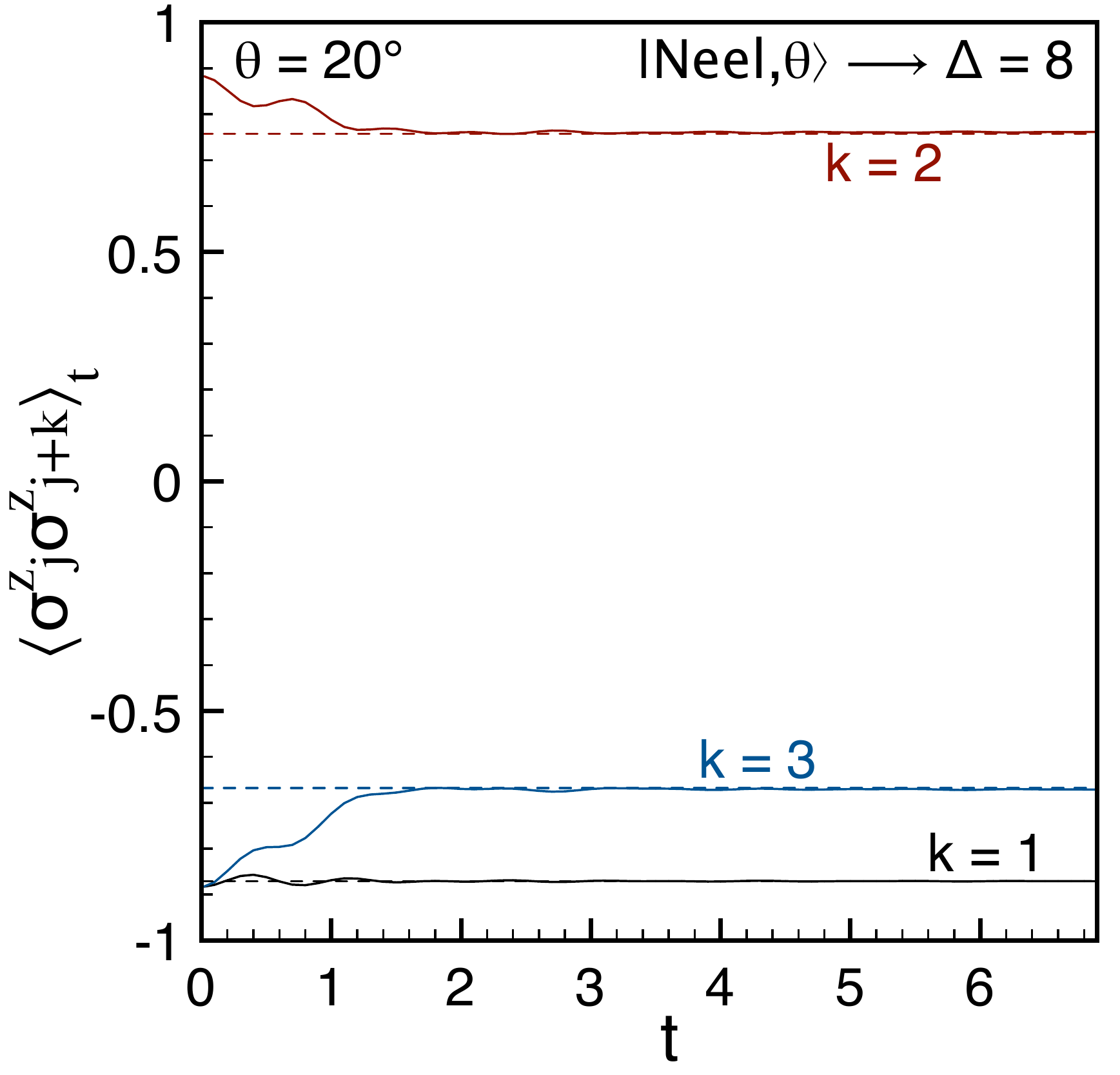}
\caption{\label{figNeel70b}
Same as Fig. \ref{figNeel80}, but with initial state 
with $\theta=20^{\circ}$.
}
\end{figure}

\begin{figure}[ht]
\includegraphics[width=0.32\textwidth]{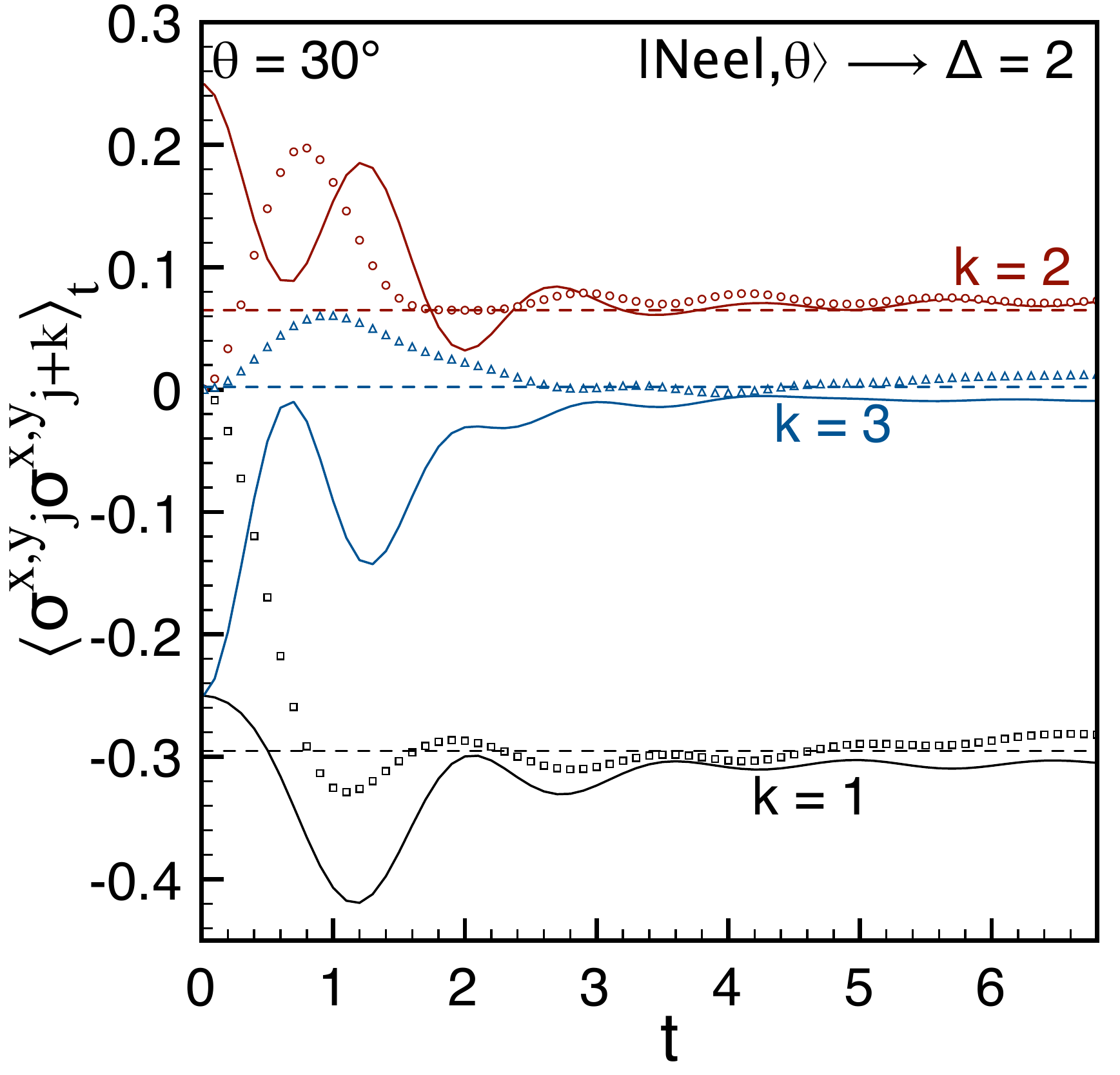}
\includegraphics[width=0.32\textwidth]{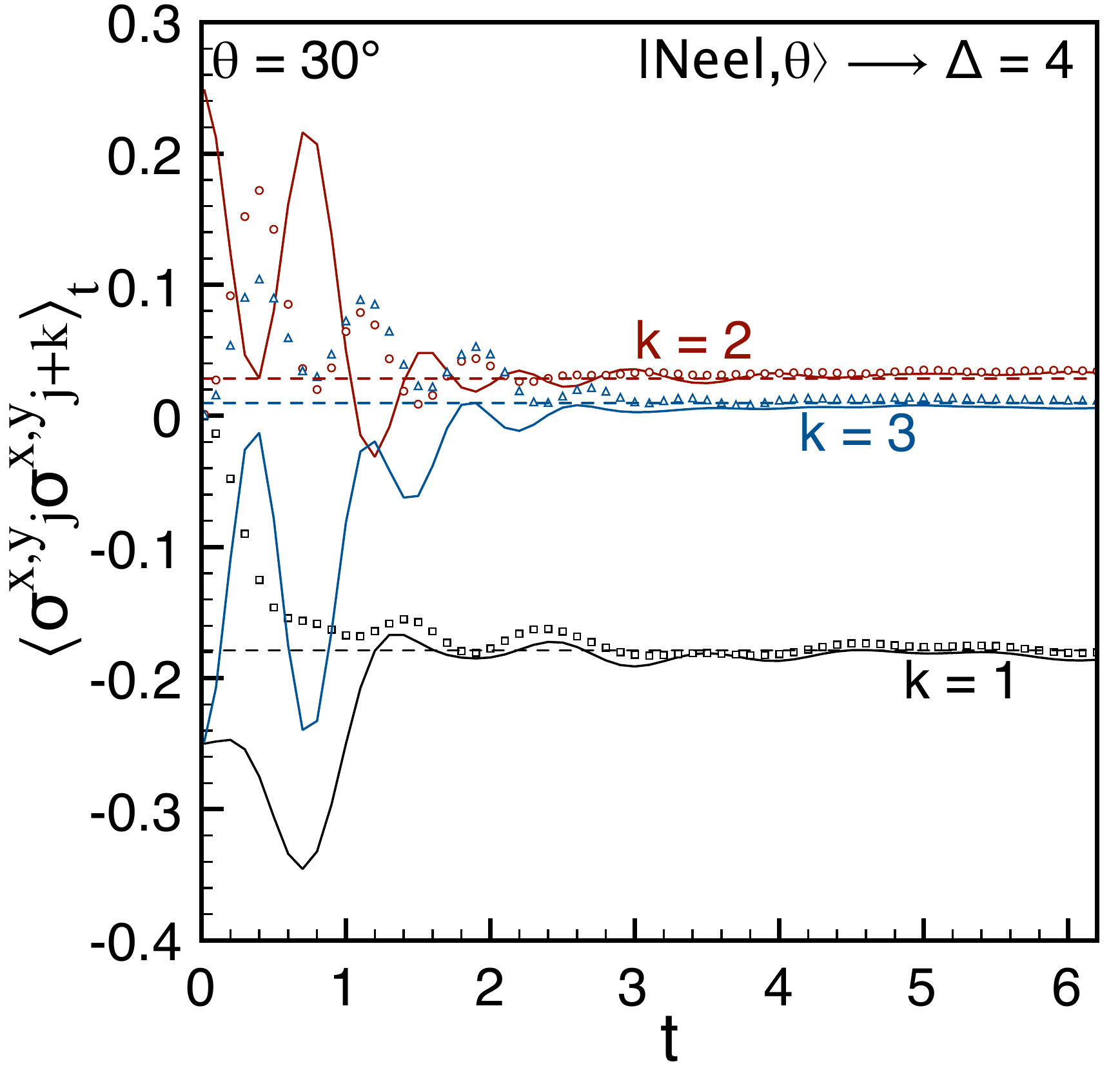}
\includegraphics[width=0.32\textwidth]{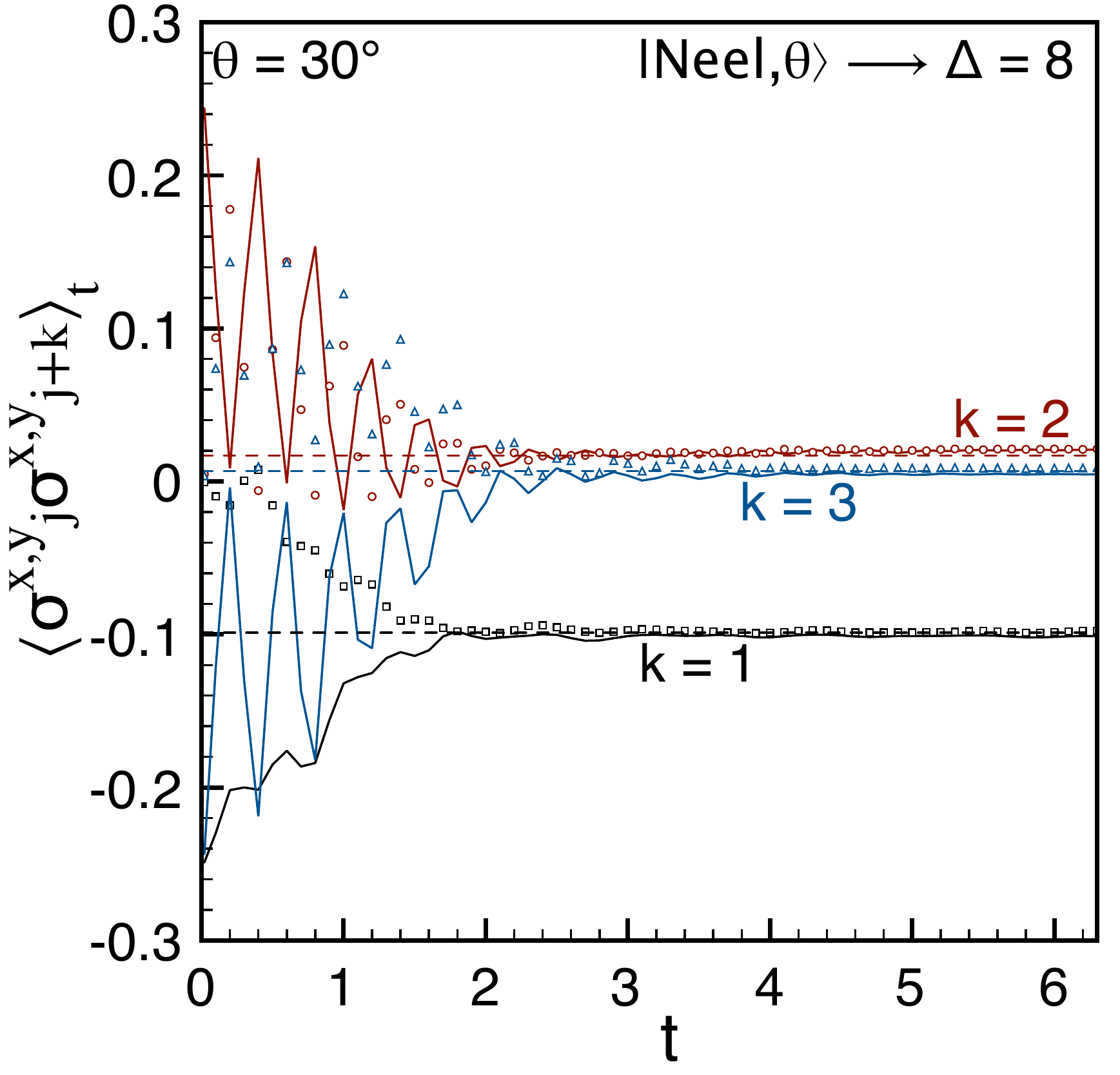}\\
\includegraphics[width=0.32\textwidth]{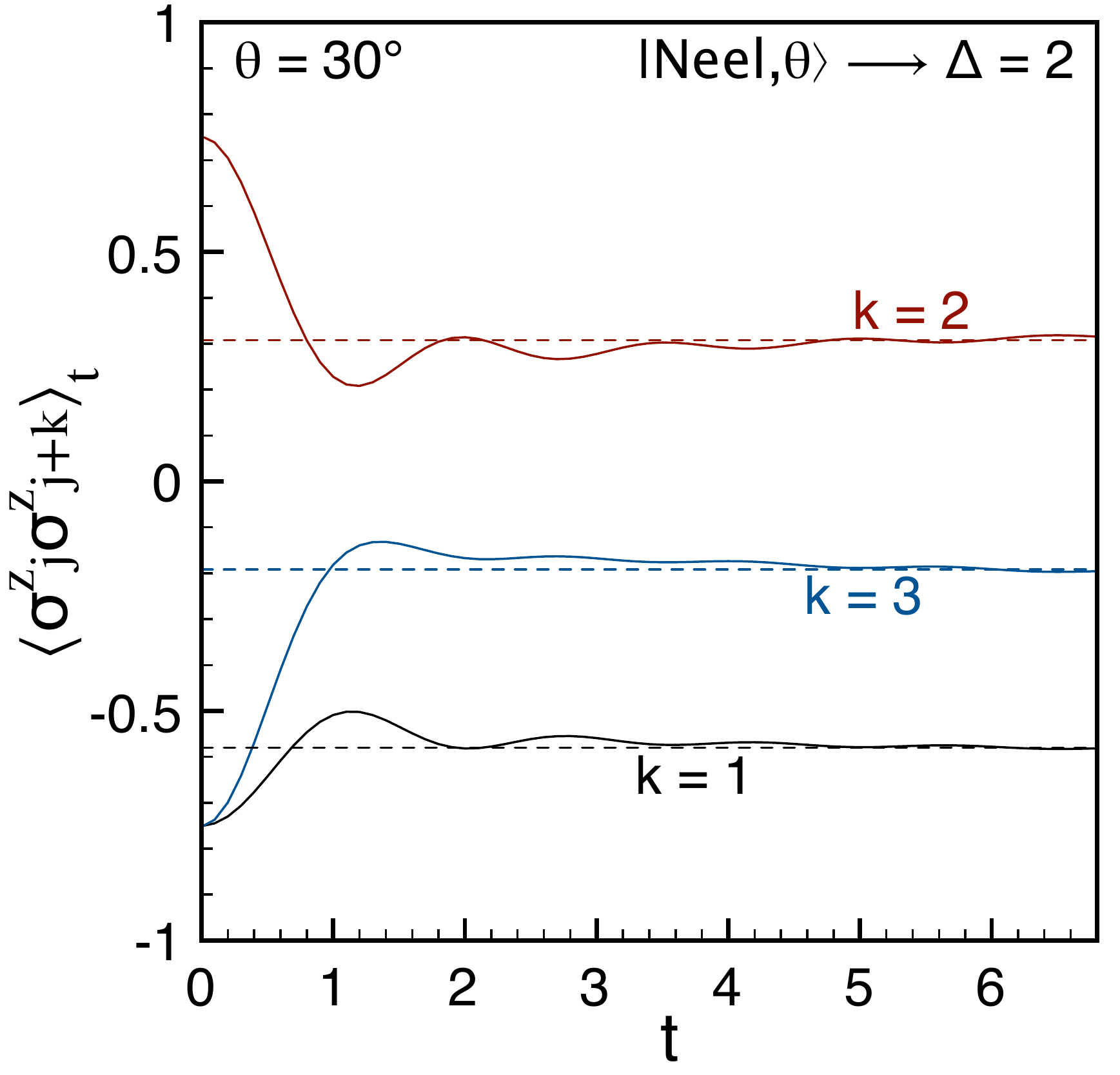}
\includegraphics[width=0.32\textwidth]{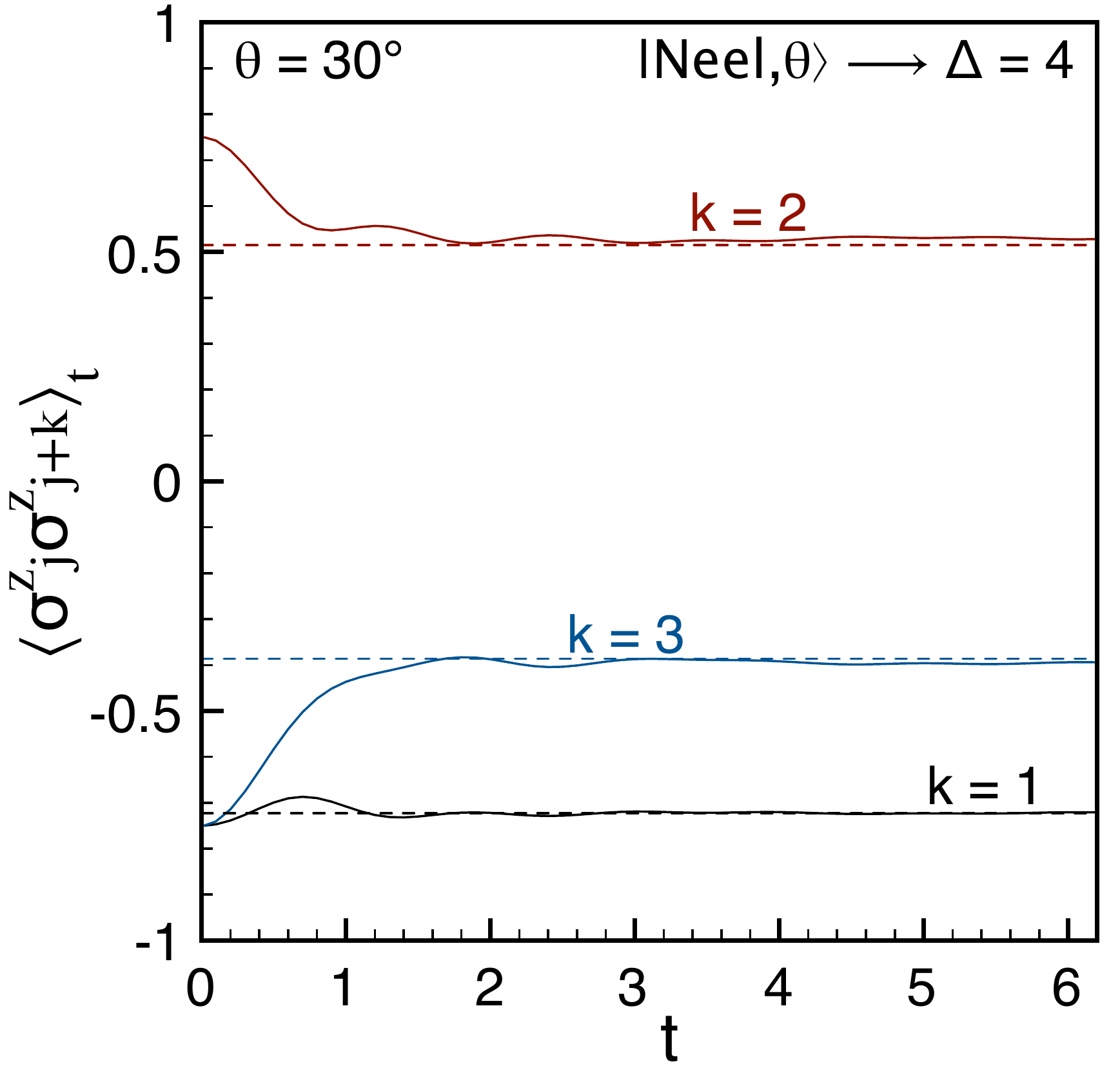}
\includegraphics[width=0.32\textwidth]{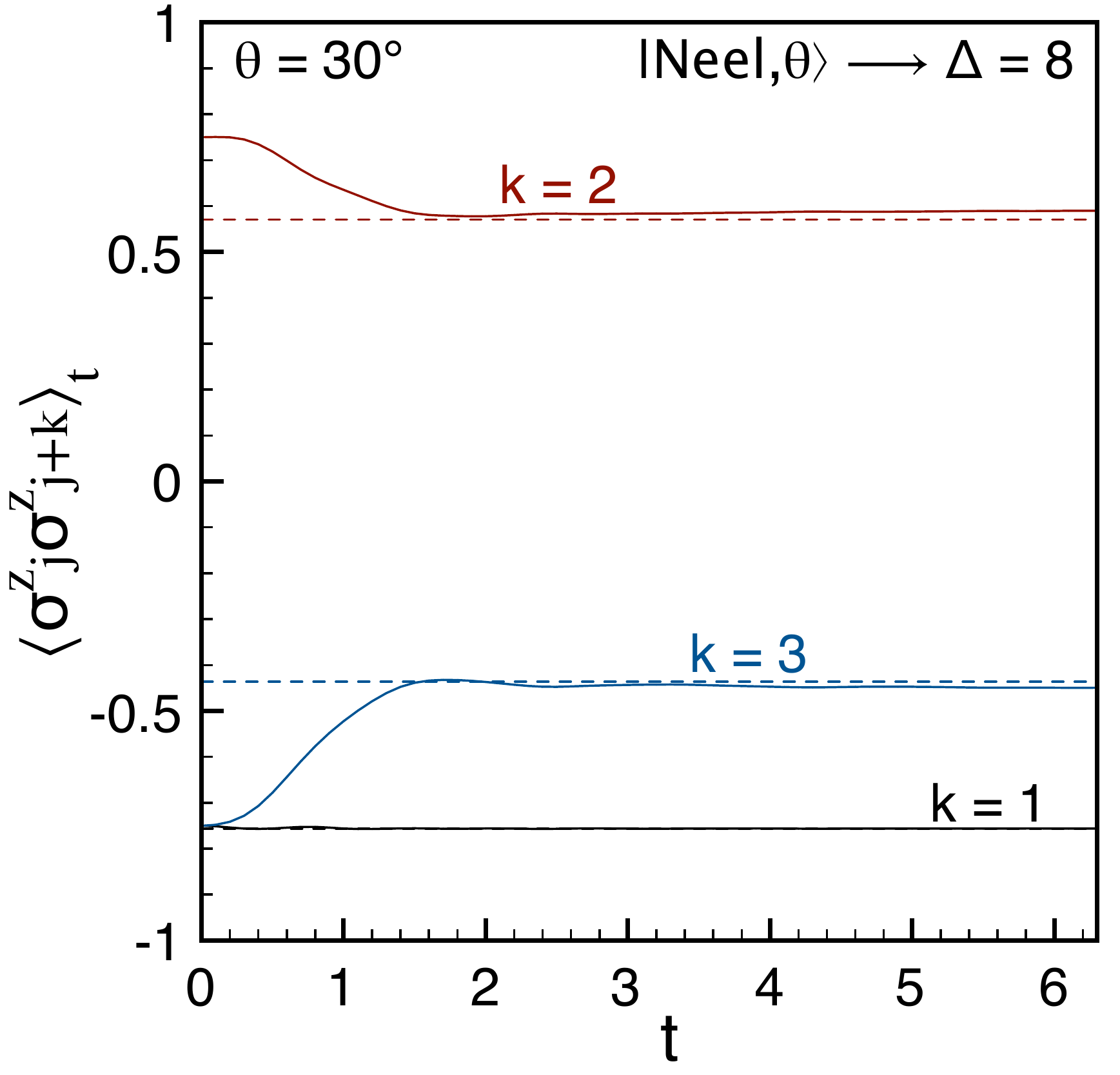}
\caption{\label{figNeel60}The same as Fig. \ref{figNeel80}, but with initial state 
with $\theta=30^{\circ}$. } 
\end{figure}
We observe that all correlation functions appear to relax to
time-independent values, which are compatible with the predictions of
the GGE.
The quench originating from the N\'eel state has been thoroughly
analysed previously, and our results are in perfect agreement with
those reported in Ref.~[\onlinecite{bpgda-10}]. In particular, the
oscillatory behaviour during relaxation reflects the presence of
multiple frequencies, with the principal frequency  proportional to
the anisotropy\cite{bpgda-10} $\Delta$. Hence, the larger the value of
$\Delta$, the easier it is to observe the relaxation because the
oscillations around the asymptotic value are faster. 
In the figures we also report the Gibbs values at temperatures fixed by the 
initial state energies. 
It is evident that, in some cases, these values are well separated from the 
GGE ones and those are the ideal candidates to distinguish the two ensembles 
in real experiments.

Next we consider quenches from N\'eel states where the order parameter
points along an arbitrary direction, a situation which to the best of
our knowledge has not been previously considered in the
literature. This case presents a 
very interesting difference compared to the N\'eel state in z-direction:
for any nonzero tilt $\theta$ the initial state breaks the rotational
symmetry in the $xy$ plane  of the XXZ Hamiltonian. This means that
transverse correlations in the x and y directions are no longer
required to be equal by symmetry, and at short times they are indeed
generically quite different. On the other hand, in the GGE the U(1)
symmetry is restored. It is therefore important to understand on what
time scales the symmetry restoration occurs. In
Figs~\ref{figNeel80}, \ref{figNeel70}, \ref{figNeel70b}, and \ref{figNeel60} we report
results for quenches from tilted N\'eel states at angles
$\theta= 10^{\circ}$, $\theta= 20^{\circ}$, and  $\theta= 30^{\circ}$ respectively.
In all cases the transverse correlations are seen to relax in an
oscillatory manner to stationary values compatible with restoration
of the spin-rotational symmetry around the z-axis. Like in the
$\theta=0$ case, the oscillations are irregular (indeed even more
irregular than before), which indicates the presence of multiple
frequencies. The principal frequency again appears to be proportional
to the anisotropy $\Delta$. As a result it is easier to observe the
relaxation for large $\Delta$, because the oscillations around the
asymptotic value are faster. In fact, for $\Delta=2$, the correlations
do not look particularly stationary even at the latest times
accessible to us, because they oscillate around their asymptotic
values with a very large period.

Another interesting issue is the influence of the \emph{strength} of
the U(1) symmetry breaking in the initial state: clearly increasing
$\theta$ leads to a stronger breaking of the symmetry, and the naive
expectation would be that this results in a slower relaxation to a
stationary regime. Interestingly, this expectation is not entirely borne
out by the numerical results: a comparison of Figs \ref{figNeel80},
\ref{figNeel70}, \ref{figNeel70b}, and \ref{figNeel60} indicates that the symmetry is
restored (in the sense that $S^x_{j,j+k}$ becomes approximately equal
to $S^y_{j,j+k}$) on a time scale that appears to not be strongly
$\theta$-dependent. From a computational point of view, decreasing the
values of $\theta$ leads to an increase in the required computational
resources, because the entanglement entropy grows more quickly
(cf. Fig. \ref{figS}). This makes the simulations increasingly
difficult for initial states aligned closer to the $\hat x$ axis. 

\subsection{Majumdar-Ghosh dimer product state}
We now turn to time evolution starting in the Majumdar-Ghosh ground state 
\be
\displaystyle|{\rm MG}\rangle=\prod_{j=1}^{L/2}
\frac{|\uparrow\rangle_{2j-1}\otimes|\downarrow\rangle_{2j}-
|\downarrow\rangle_{2j-1}\otimes|\uparrow\rangle_{2j}}{2}.
\label{MG}
\ee
\begin{figure}[ht]
\includegraphics[width=0.32\textwidth]{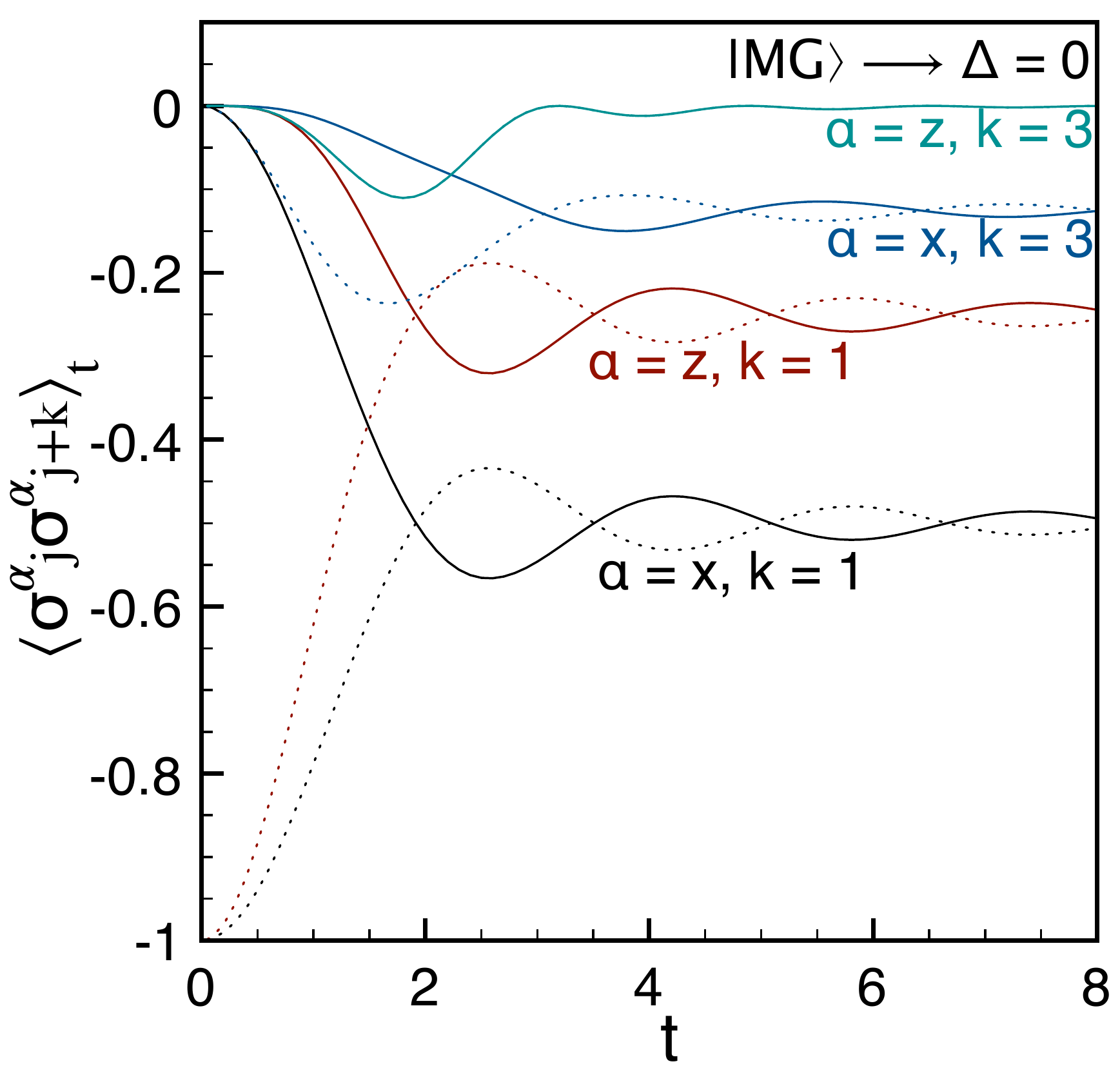}
\includegraphics[width=0.32\textwidth]{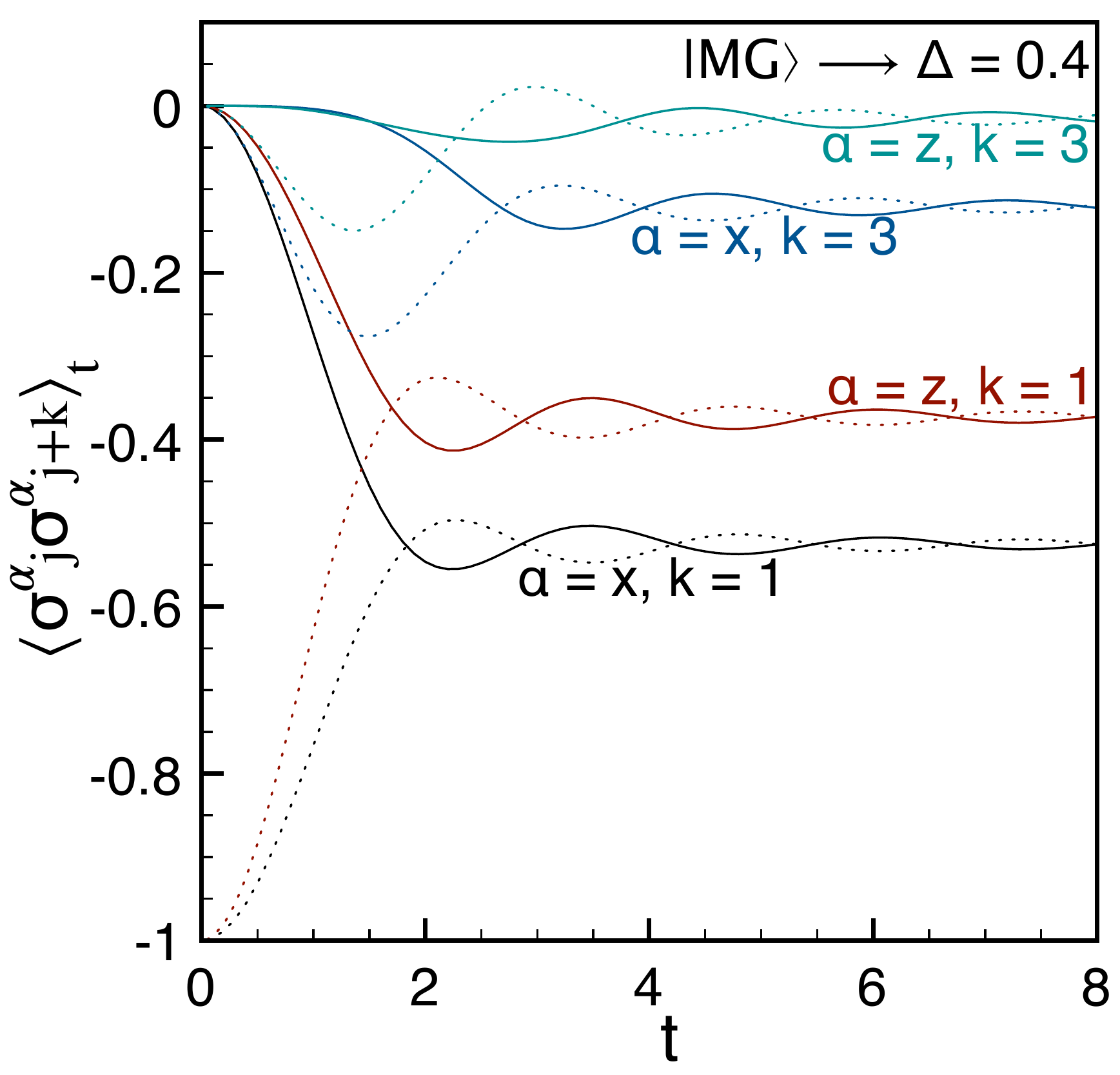}
\includegraphics[width=0.32\textwidth]{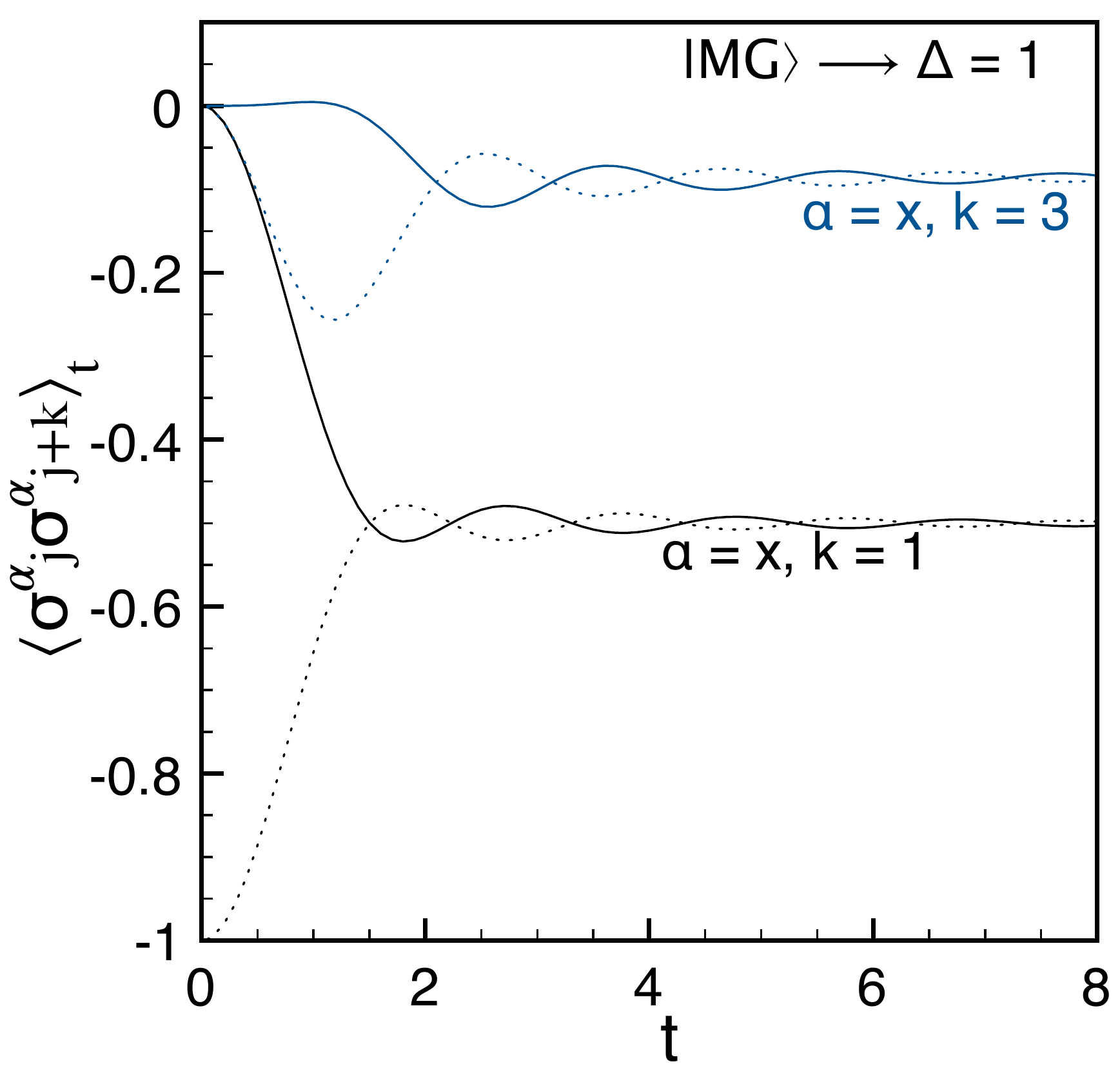}
\caption{\label{figMGtoD} 
Quench from the Majumdar Ghosh state to different $\Delta\geq 0$. 
We focus on correlators at distances $1$ and $3$, as nearest-neighbour
correlations are insensitive to translational symmetry breaking and
the data approach the GGE values quite rapidly. Solid (dotted) lines
correspond to $j=L/2$ and $j=L/2-1$ respectively. For small values of
$\Delta$ relaxation to stationary values is observed. For small values
of $\Delta>1$ the observed relaxation is compatible with the
predictions of the GGE (dashed lines). For larger values of $\Delta$
relaxation does not occur on the accessible time scales.
} 
\end{figure}
\begin{figure}[ht]
\includegraphics[width=0.32\textwidth]{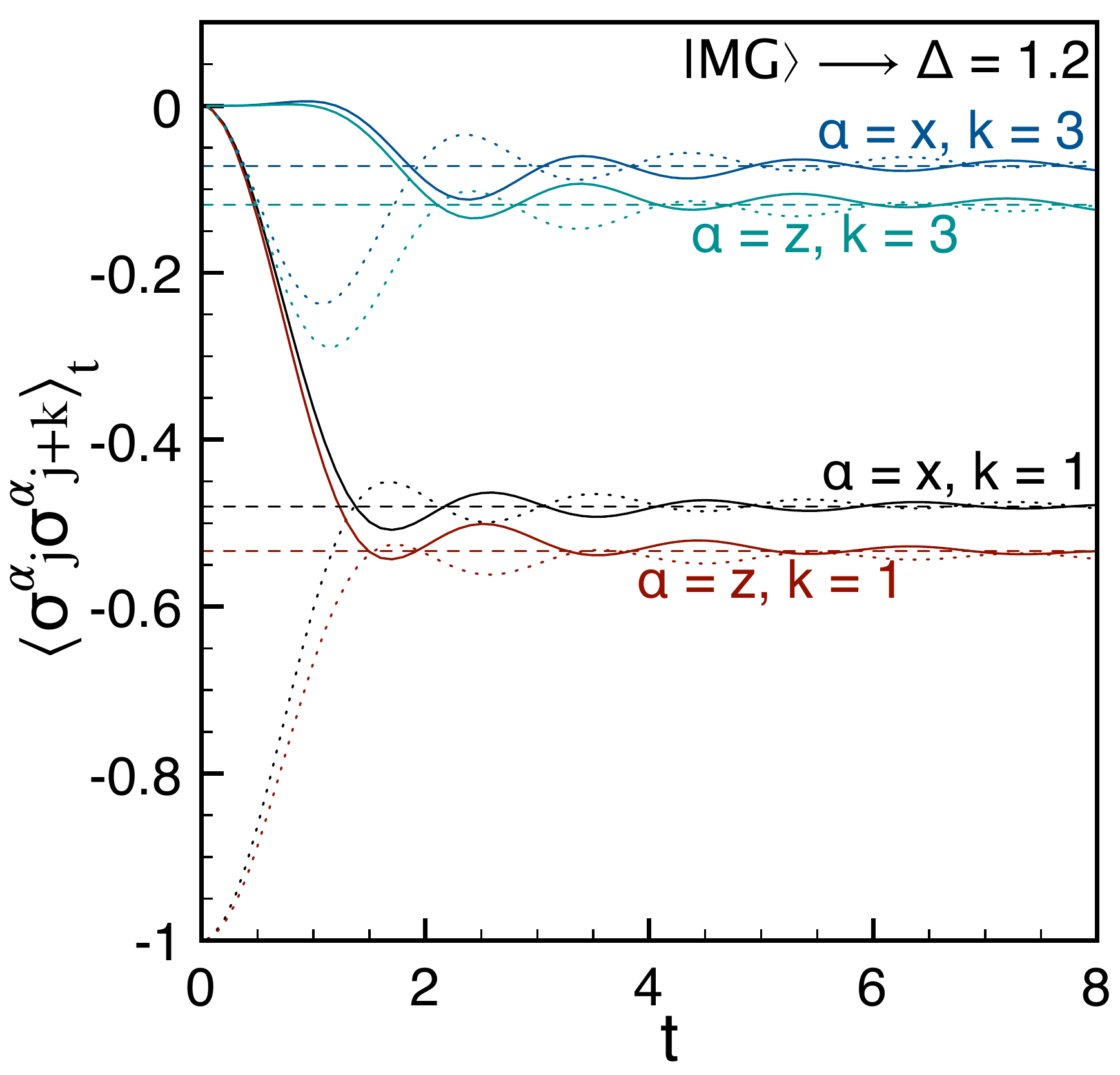}
\includegraphics[width=0.32\textwidth]{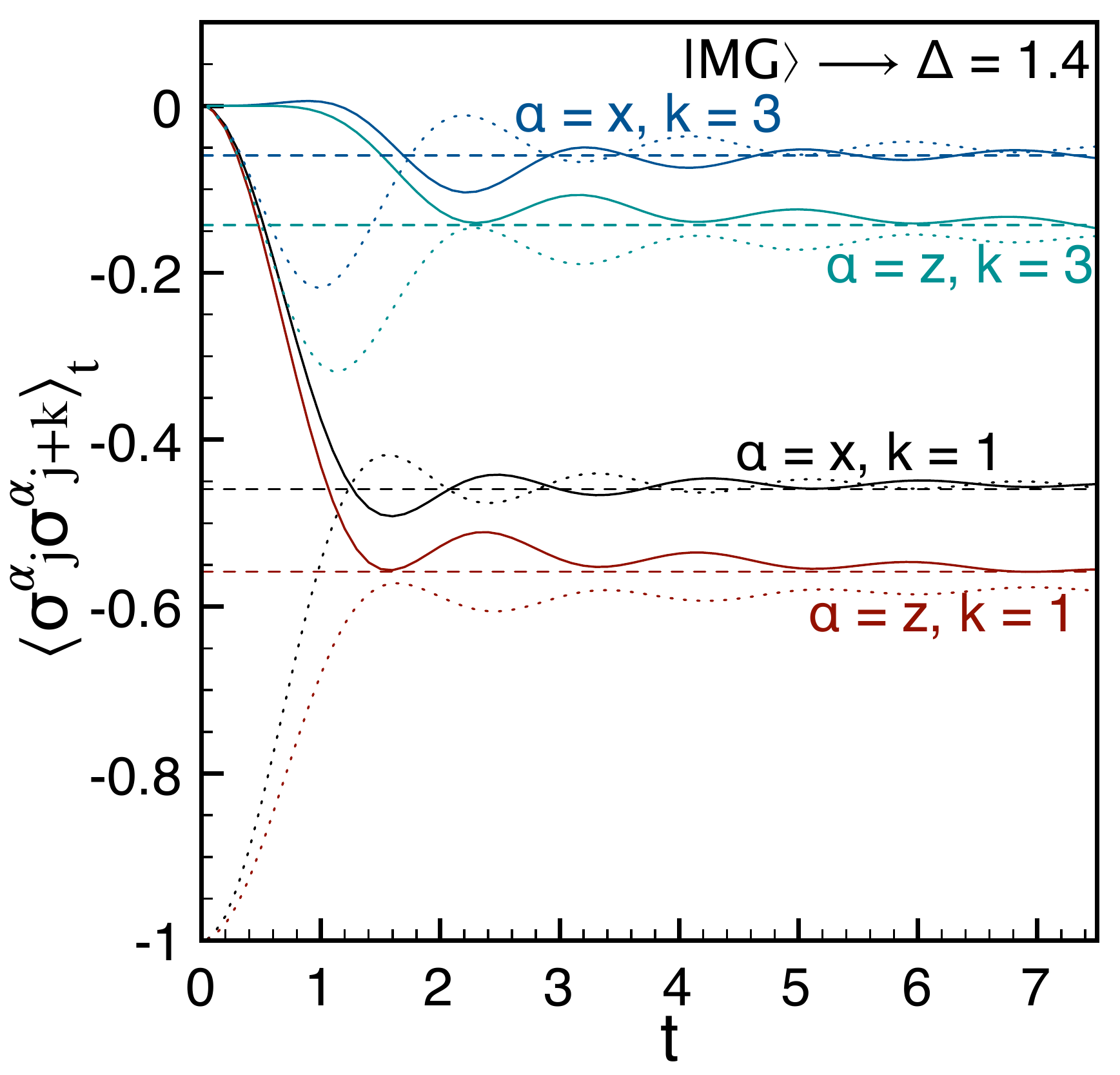}
\includegraphics[width=0.32\textwidth]{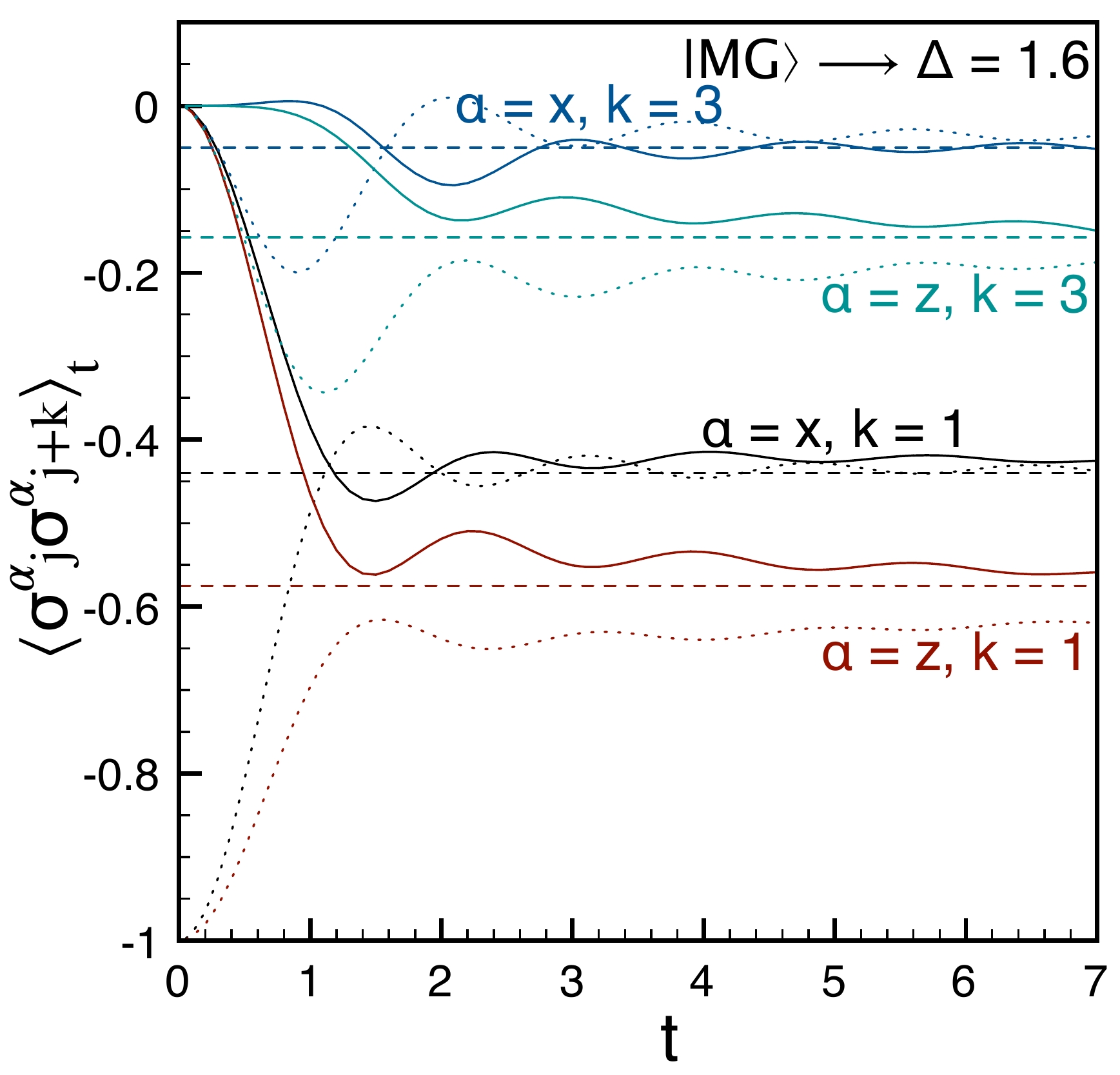}
\caption{Same as Fig.~\ref{figMGtoD} with $\Delta=1.2,1.4,1.6$.}
\label{figMGtoDb} 
\end{figure}
\begin{figure}[ht]
\includegraphics[width=0.32\textwidth]{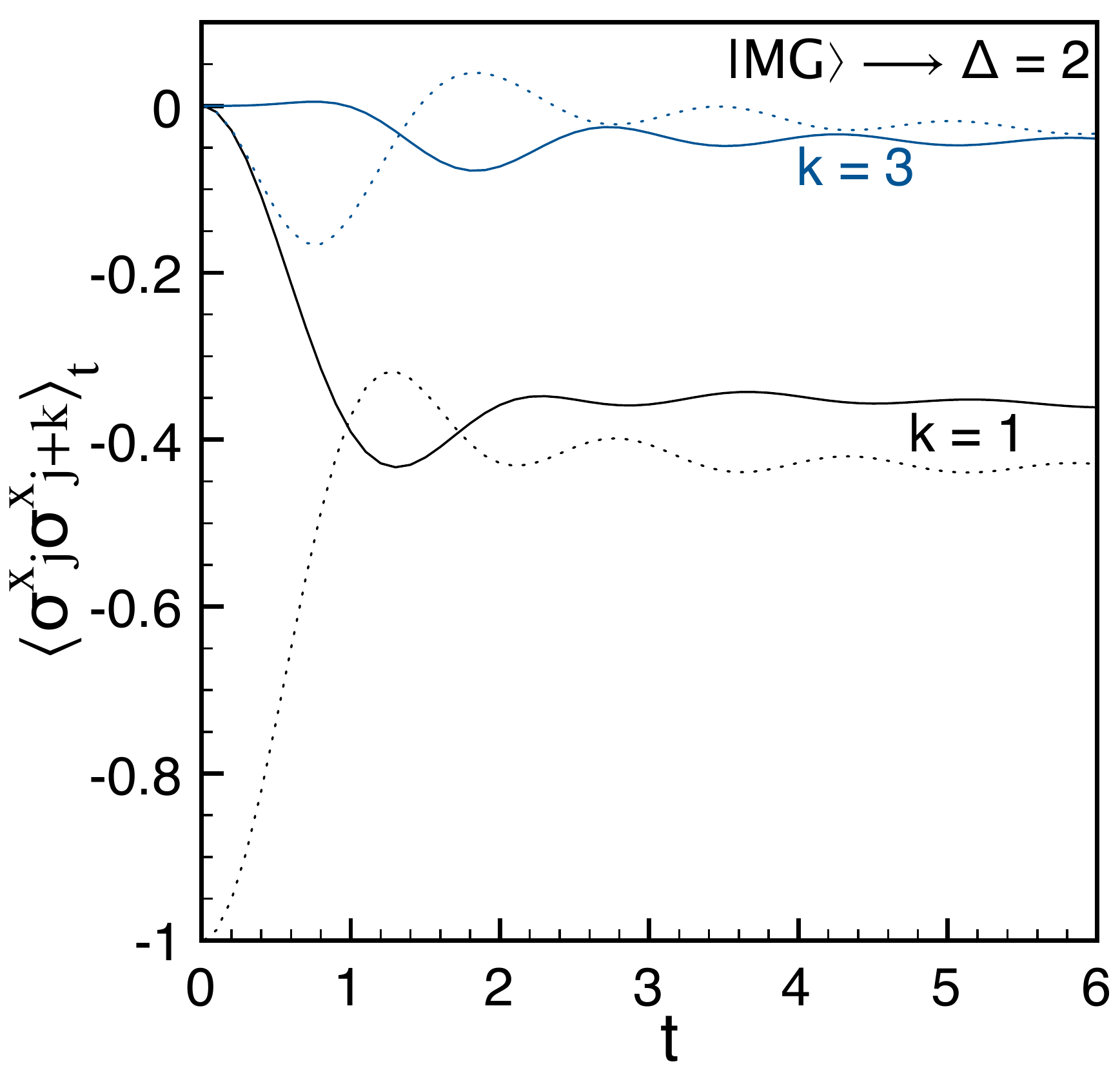}
\includegraphics[width=0.32\textwidth]{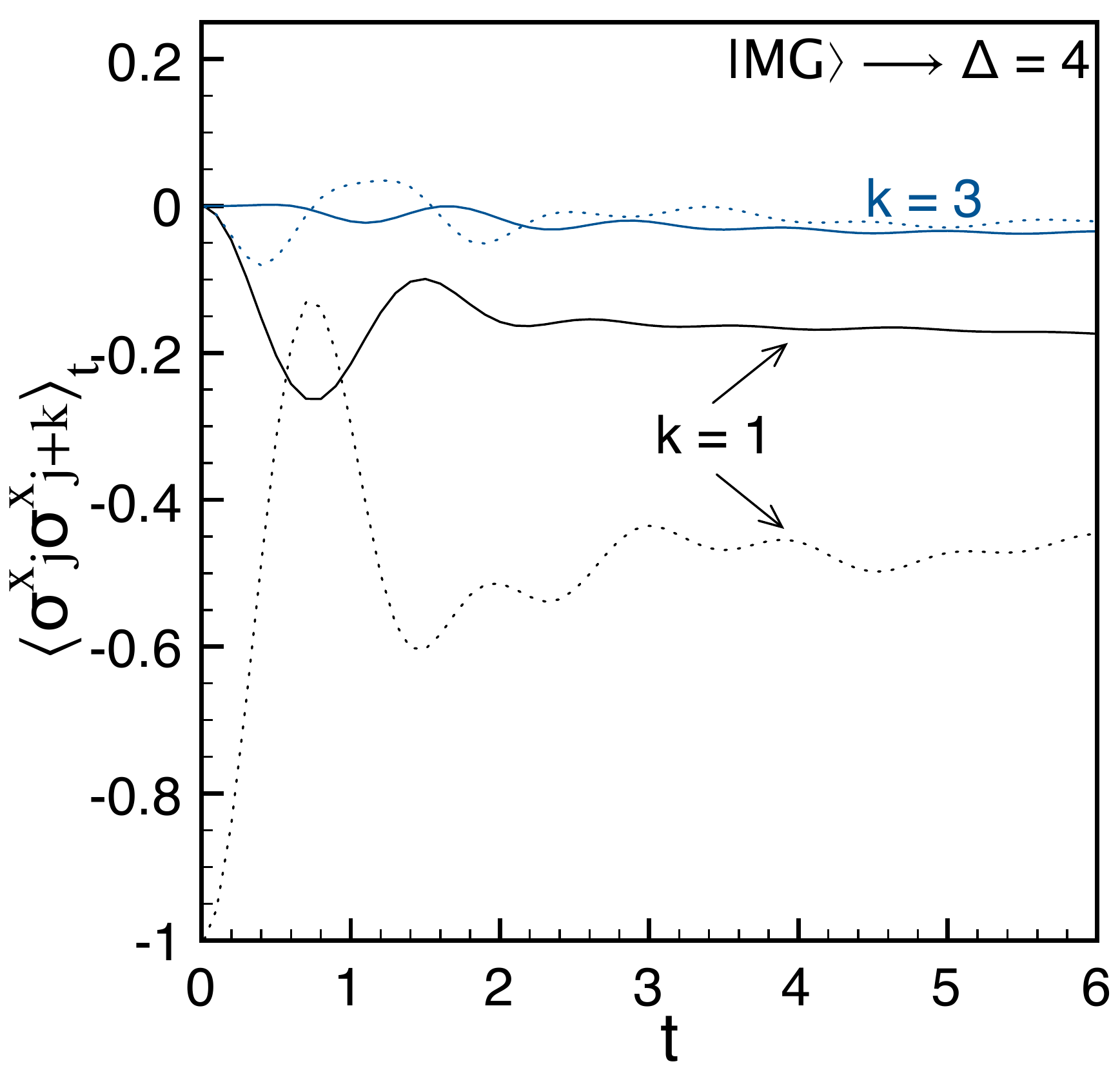}
\includegraphics[width=0.32\textwidth]{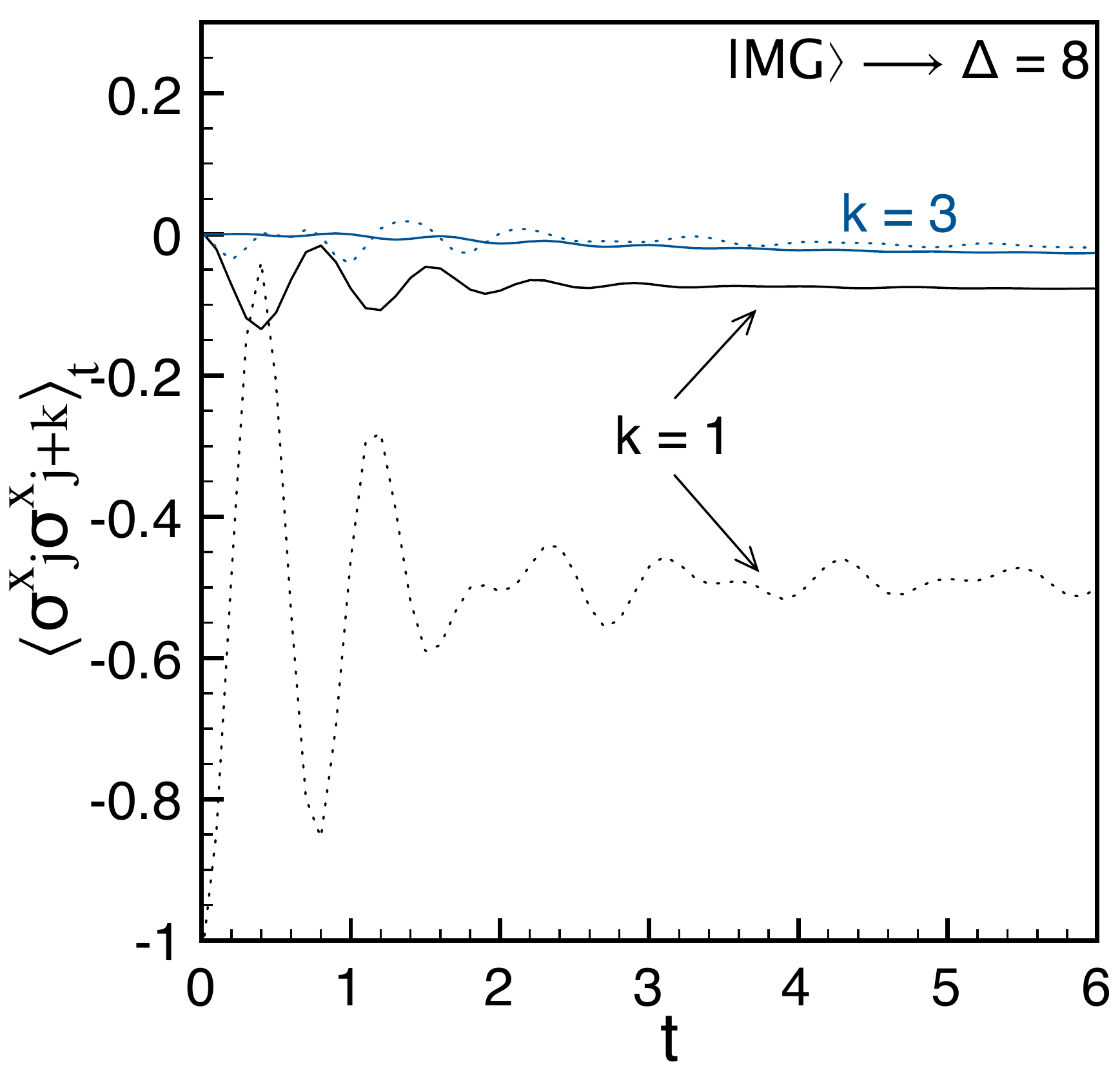}\\
\includegraphics[width=0.32\textwidth]{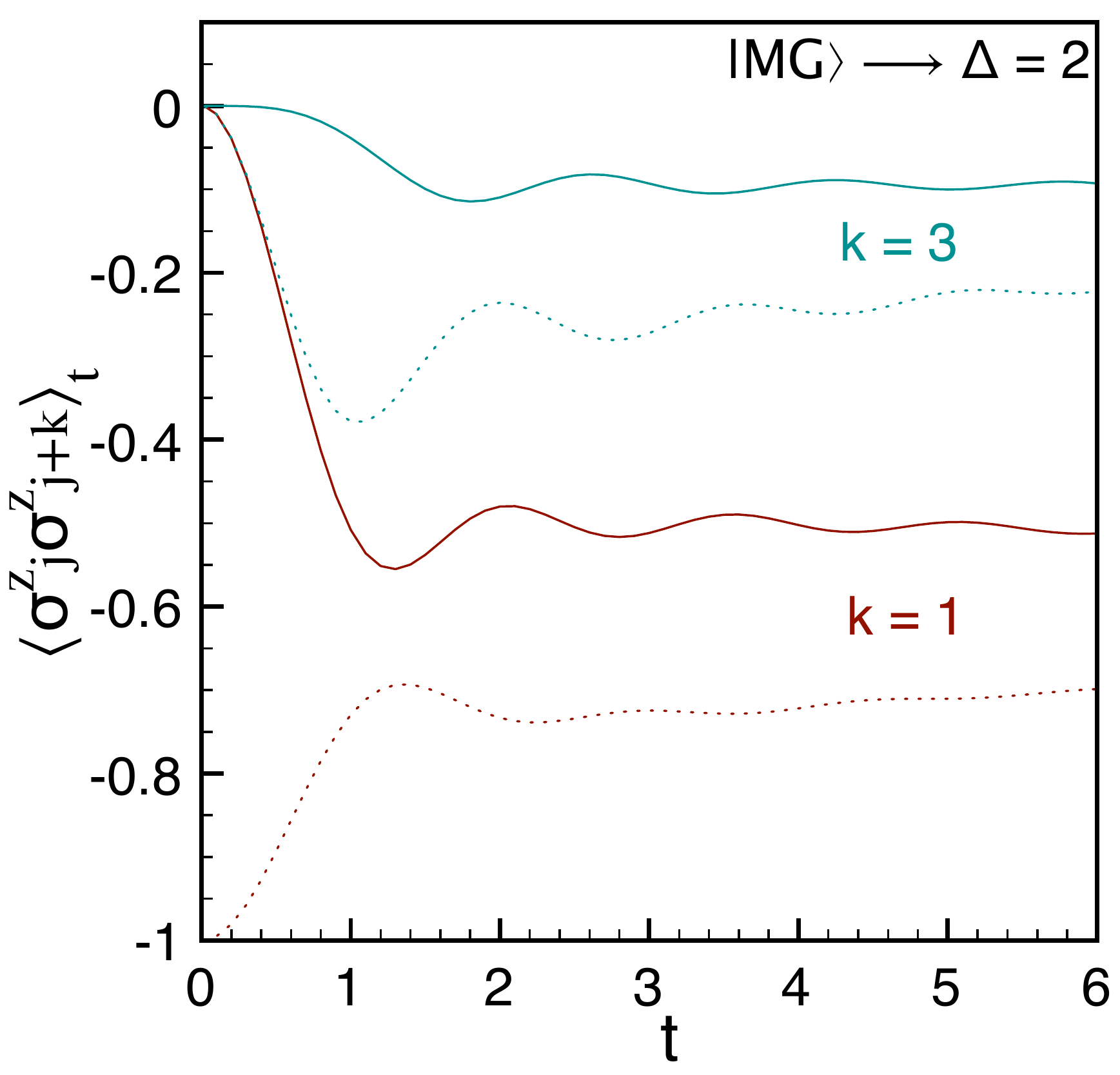}
\includegraphics[width=0.32\textwidth]{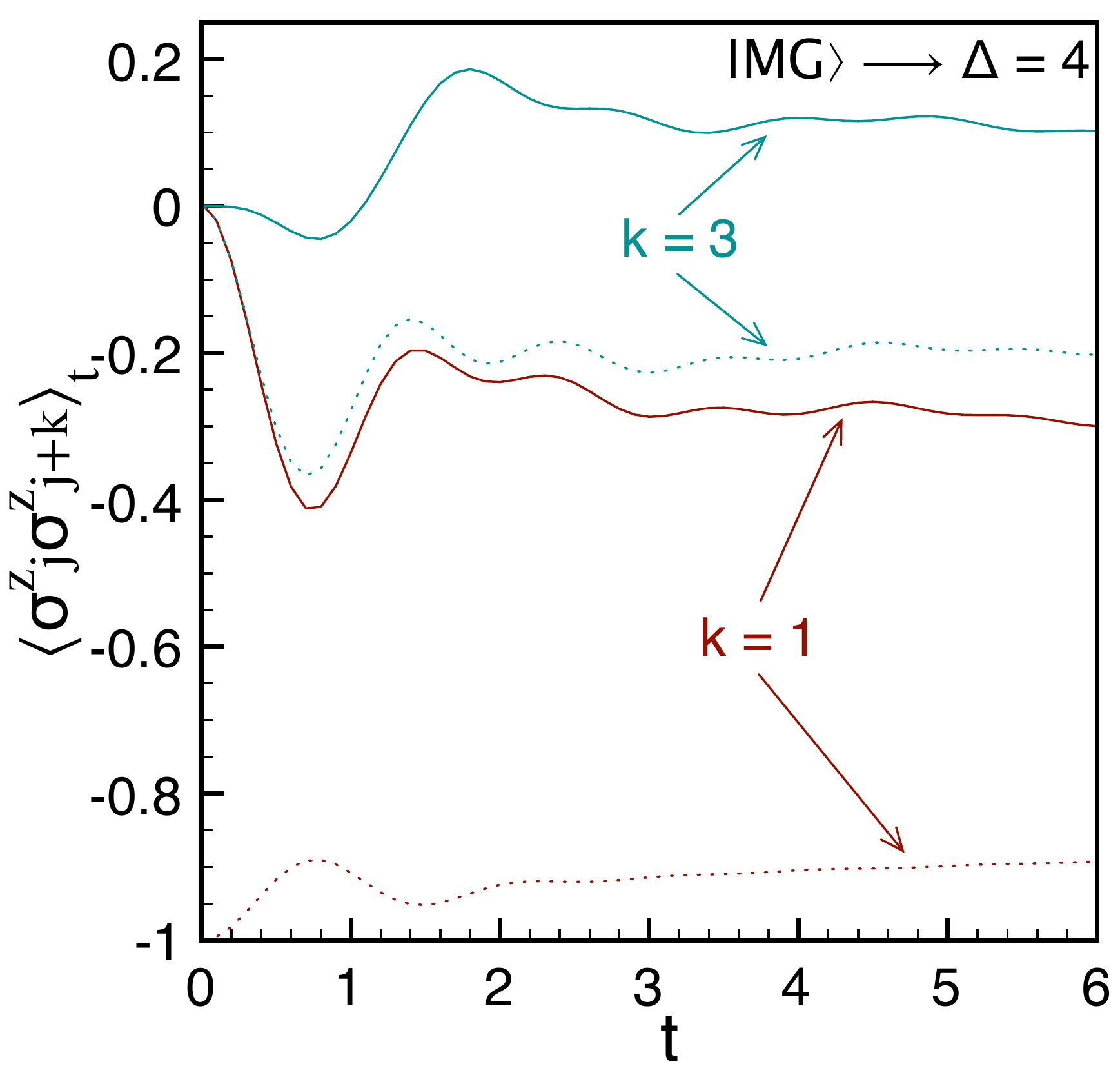}
\includegraphics[width=0.32\textwidth]{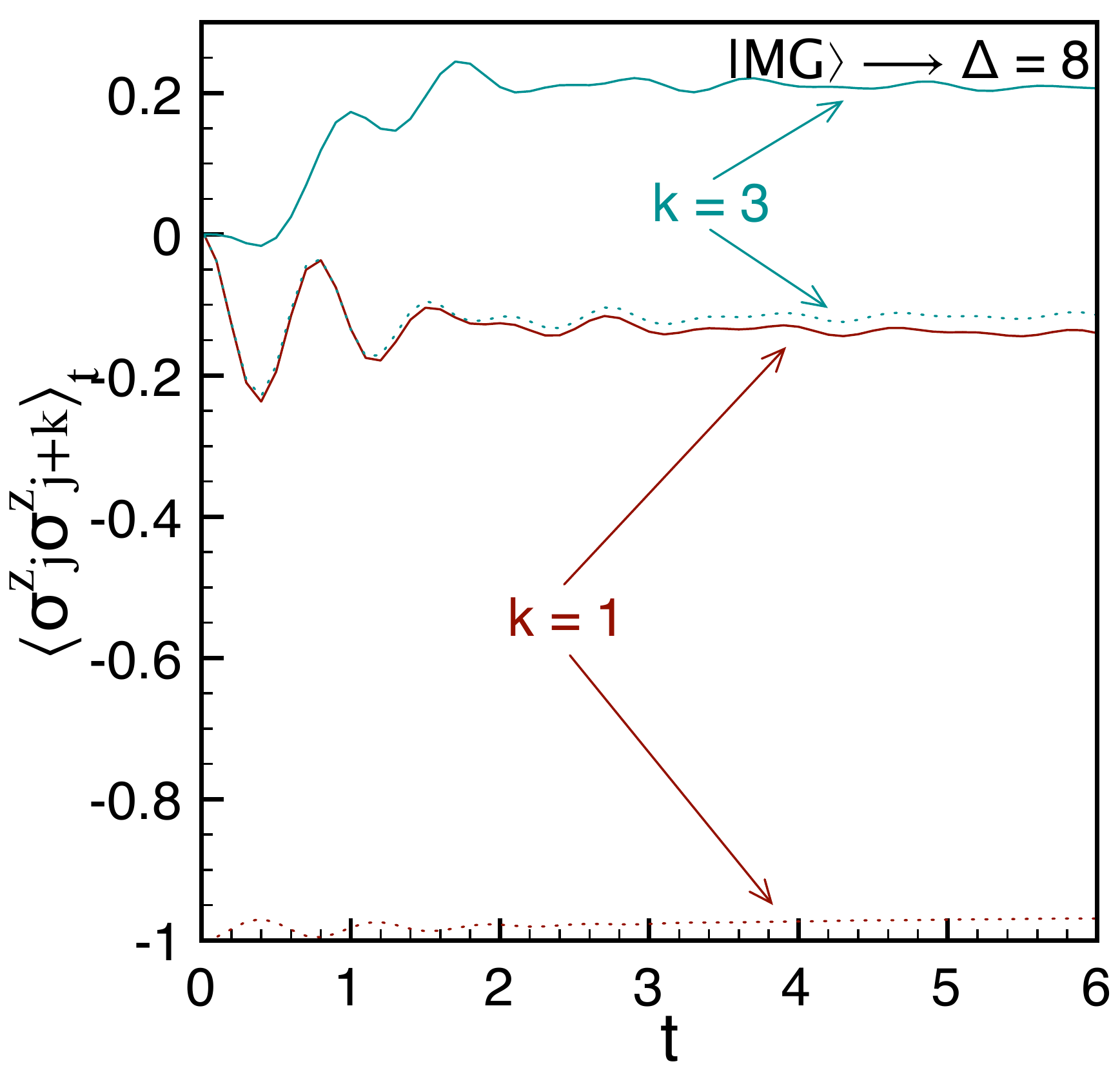}
\caption{Same as Fig.~\ref{figMGtoD} with $\Delta=2,4,8$. 
Transverse and longitudinal correlations are shown separately for the
sake of clarity.}
\label{figMGtoDc} 
\end{figure}
This quench exhibits very interesting physical features, but is
also quite demanding numerically (for $\Delta$ larger than $\simeq
1.4$), because of the fast growth of the entanglement entropy 
(cf. Fig. \ref{figS}, where the entanglement growth is comparable with
the N\'eel state with $\theta=30^\circ$). The state \fr{MG} breaks
translational invariance, while the GGE is translationally
invariant. This implies that translational symmetry should get
restored. In order to analyse this symmetry restoration, we compute
correlators with even and odd parities using tDRMG, i.e.
$\langle{\rm MG}(t)|\sigma^{\alpha}_{L/2}\sigma^{\alpha}_{L/2+k}|{\rm MG}(t)\rangle$ and 
$\langle{\rm MG}(t)|\sigma^{\alpha}_{L/2-1}\sigma^{\alpha}_{L/2-1+k}|{\rm MG}(t)\rangle$ 
for $\alpha=x,z$ and $k=1,3$. We note that it is sufficient to
consider transverse correlations in $x$ direction, as the initial
state is U(1) invariant. Furthermore, as $|{\rm MG}\rangle$ is
invariant under translations by two sites, next-nearest neighbour
correlators are insensitive to the breaking of translational symmetry.

In Figs~\ref{figMGtoD}, \ref{figMGtoDb} and \ref{figMGtoDc} we show
the time evolution of transverse and longitudinal correlations at
distances $1$ and $3$ for quenches to the Heisenberg chain with
several values of $\Delta$. We include results for $\Delta\leq 1$ in
order to elucidate the general trend of the $\Delta$-dependence. The
$\Delta=0$ case is exactly solvable by free-fermion methods \cite{mau-prep}, and our
numerical results agree perfectly with the analytical results in this
case. At late times all correlations relax in an oscillatory manner to
stationary values given by the appropriate GGE. The qualitative
behaviour of correlation functions is essentially unchanged for
anisotropies smaller than $\Delta=1$, but GGE predictions for the
stationary values are not yet available. Increasing $\Delta$ further,
the time evolution is seen to become less regular, involving several
oscillation frequencies. The curves for parity even and odd correlators
cease to be symmetric around the stationary value and the relaxation
is observed to slow down. The results for $\Delta=1.2,1.4, 1.6$ are
visibly compatible with relaxation to the GGE predictions (dashed lines). 
For $\Delta=2,4,8$ no relaxation is observed on the accessible time
scales. This strongly suggests a relaxation time that grows with
increasing $\Delta$. In fact, one can show that in the limit
$\Delta\to\infty$ the relaxation time diverges \cite{mau-prep}.

\subsection{Tilted ferromagnetic state}
In this case the initial state is
\be
\ket{\theta;\nearrow\nearrow\dots}=
e^{i\theta\sum_j S^y_j}\ket{\uparrow\uparrow\dots}\ .
\ee
\begin{figure}[ht]
\includegraphics[width=0.32\textwidth]{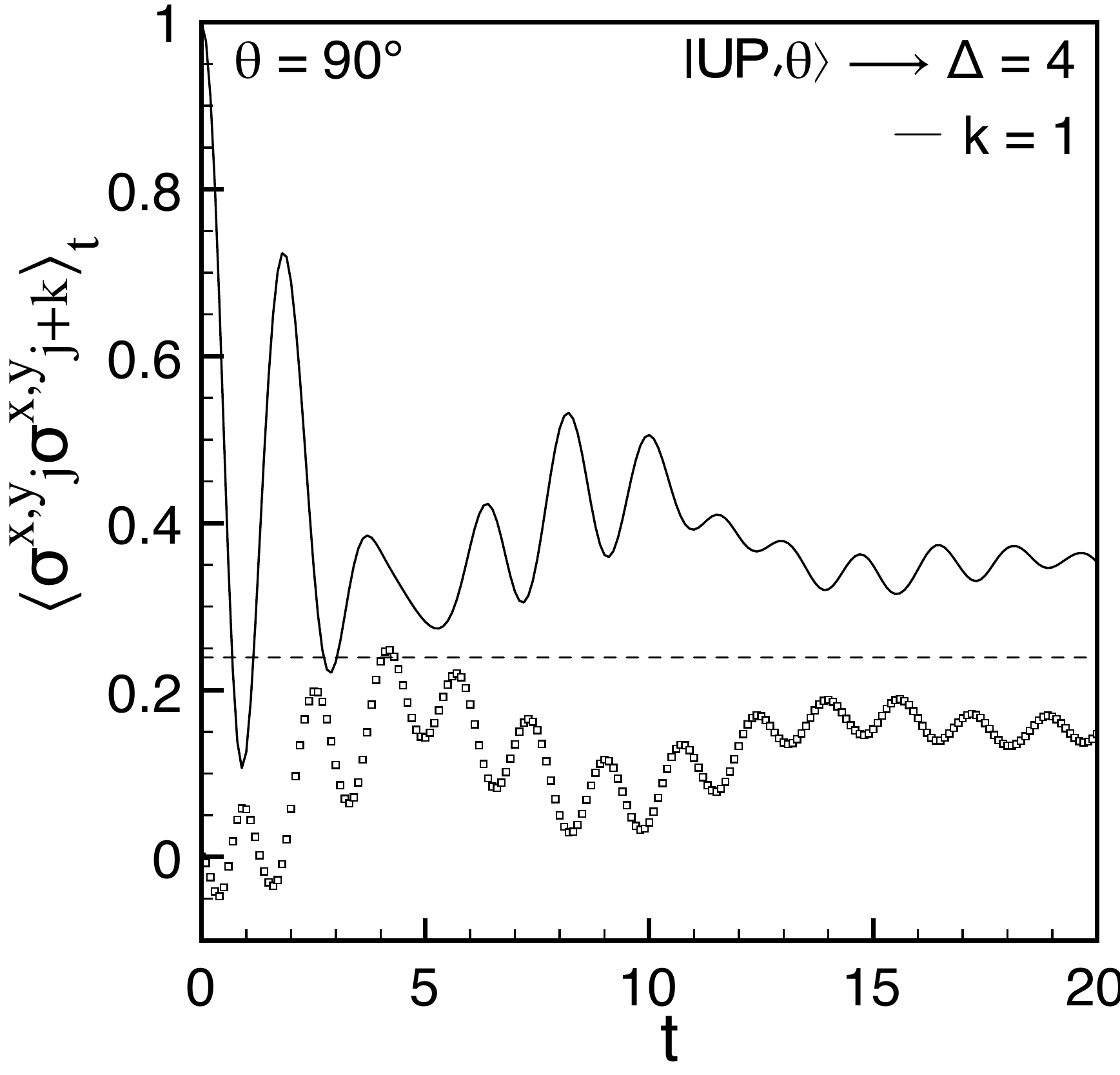}
\includegraphics[width=0.32\textwidth]{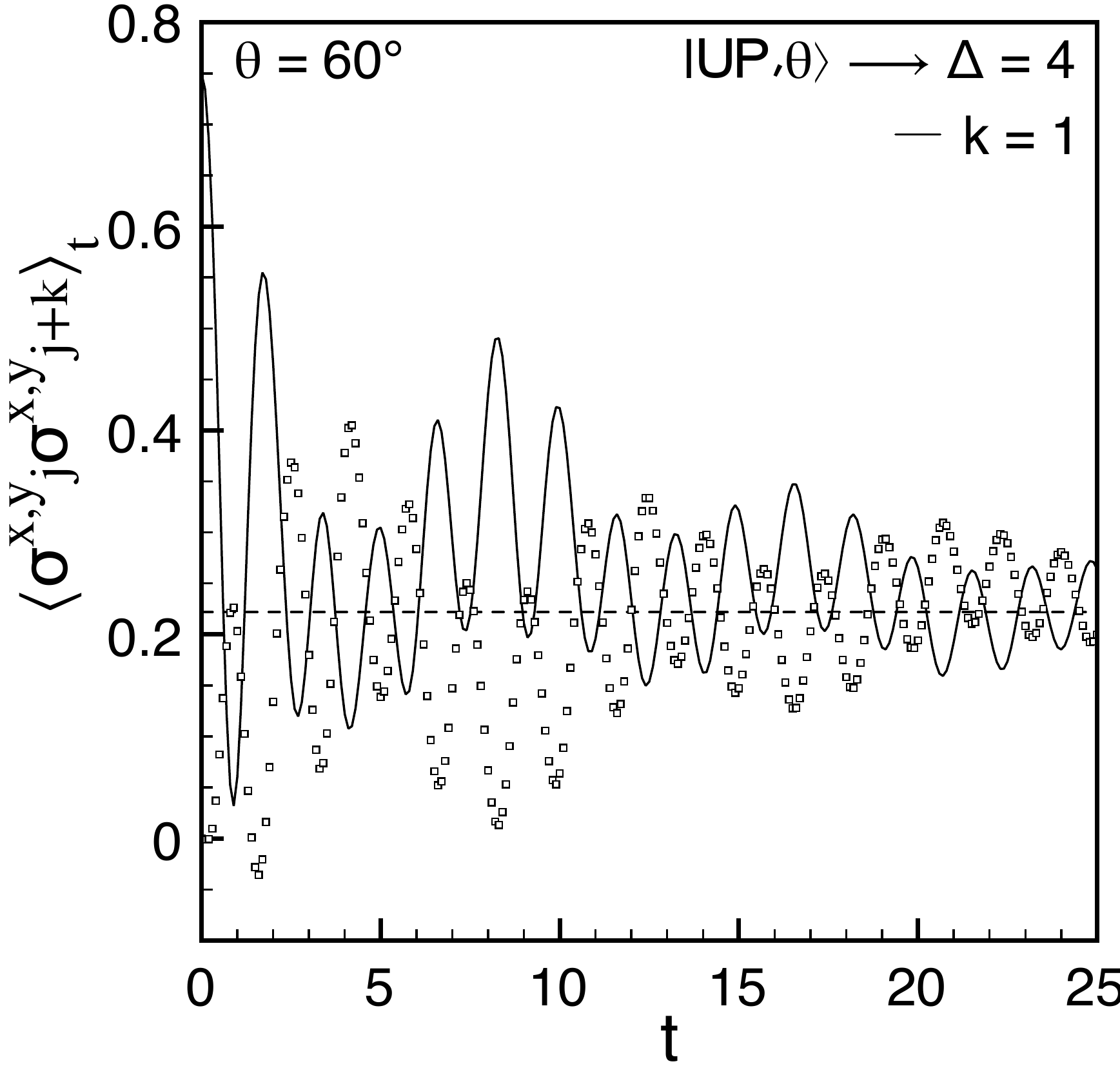}
\includegraphics[width=0.32\textwidth]{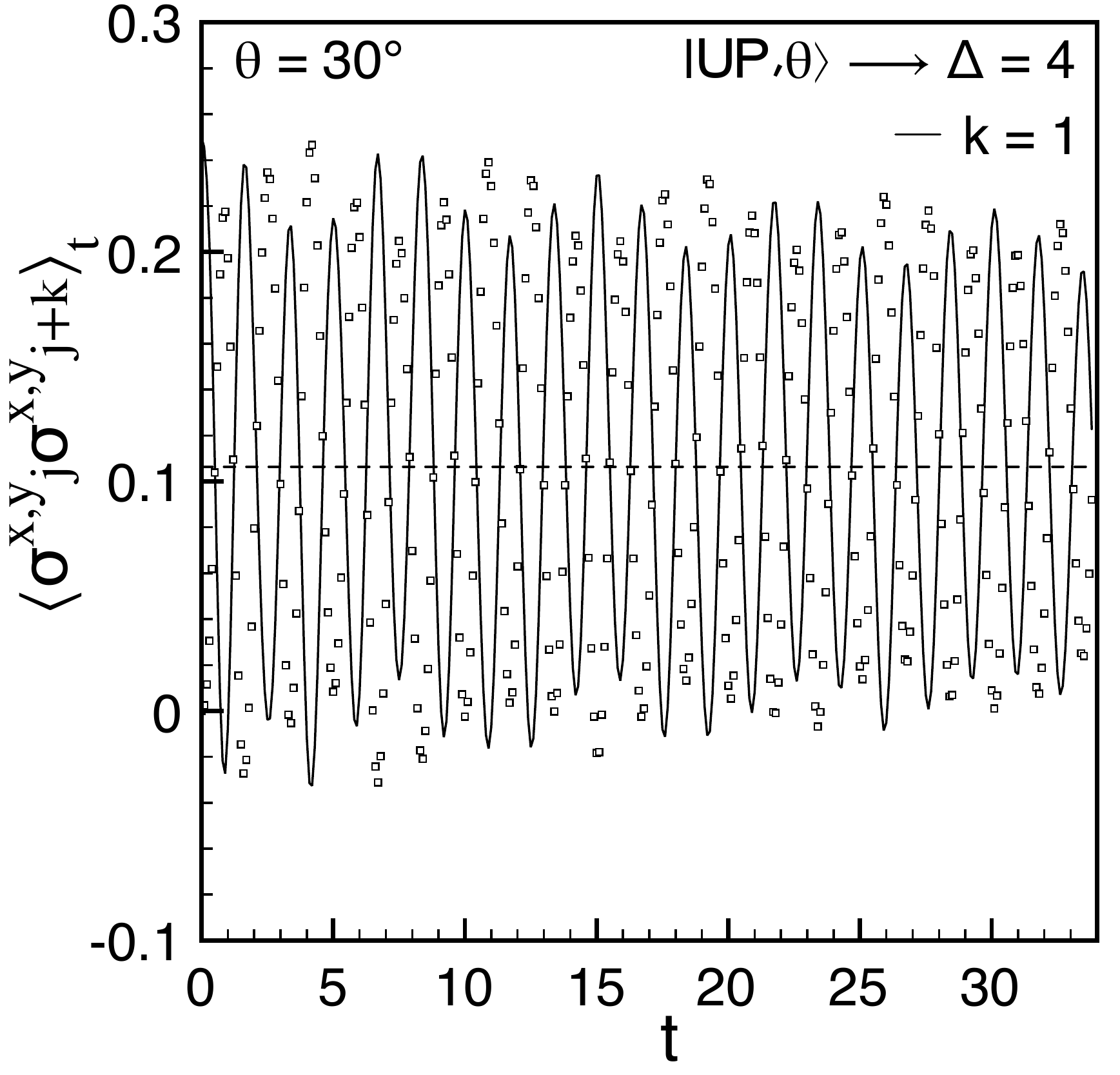}
\includegraphics[width=0.32\textwidth]{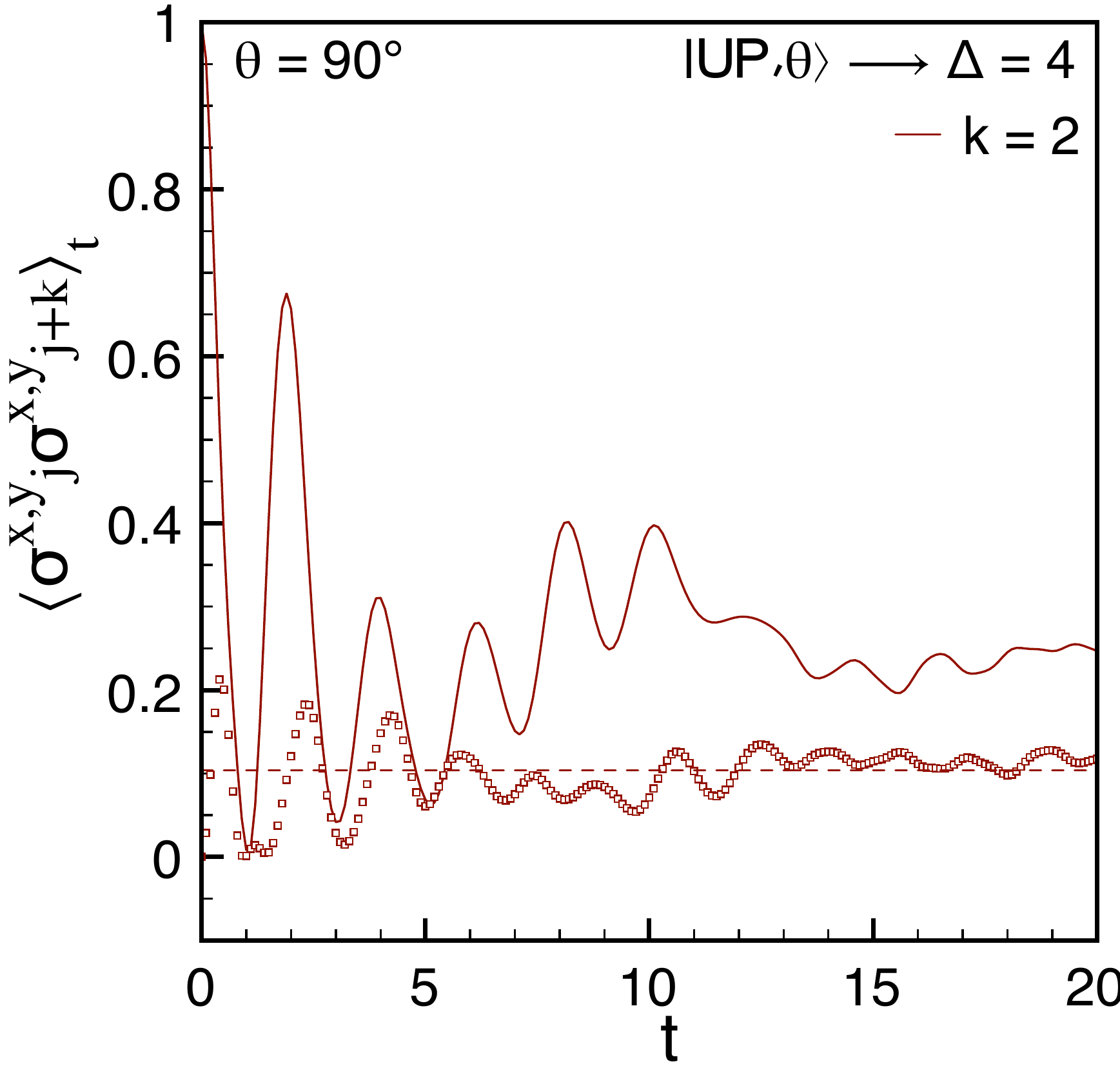}
\includegraphics[width=0.32\textwidth]{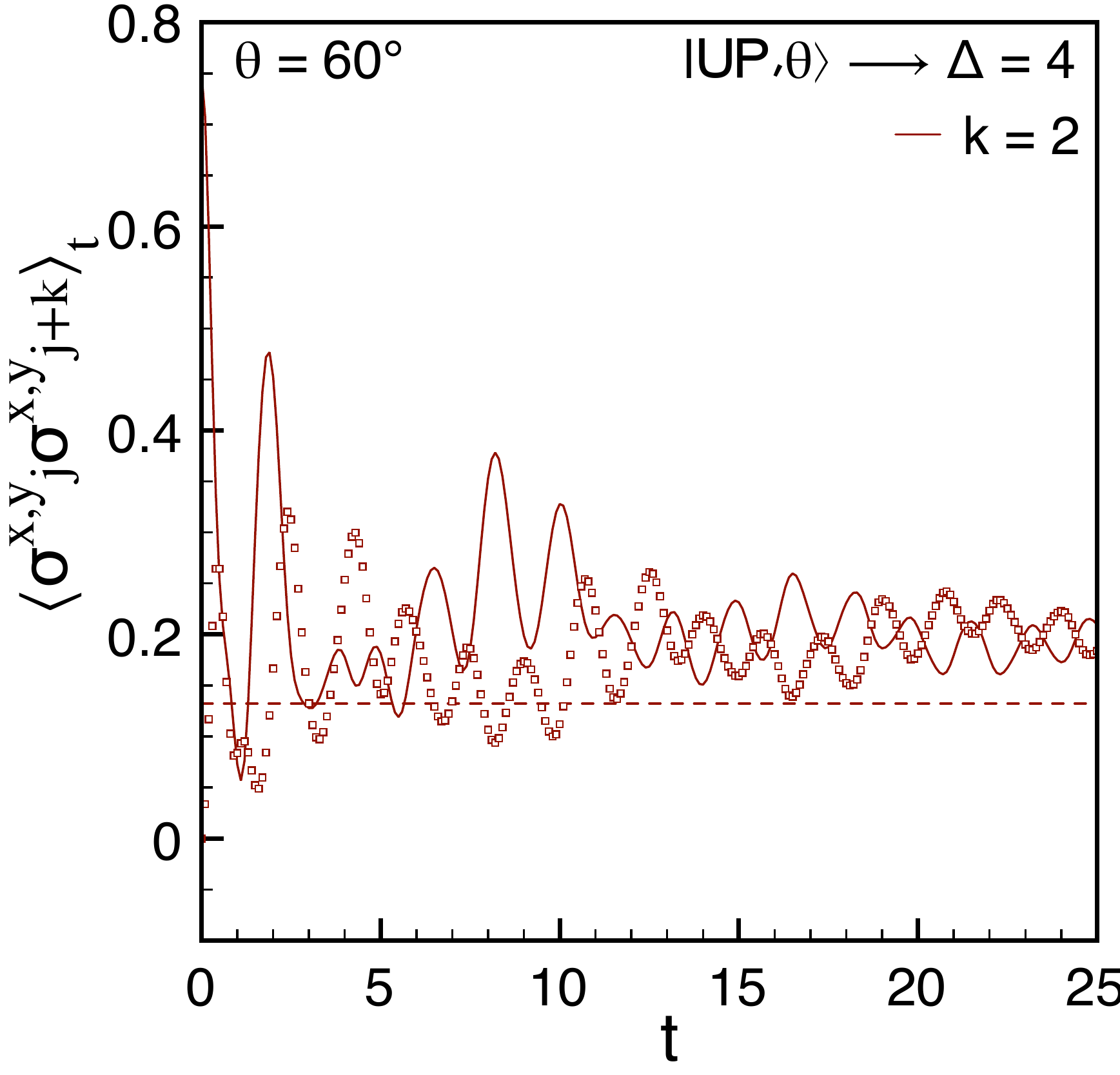}
\includegraphics[width=0.32\textwidth]{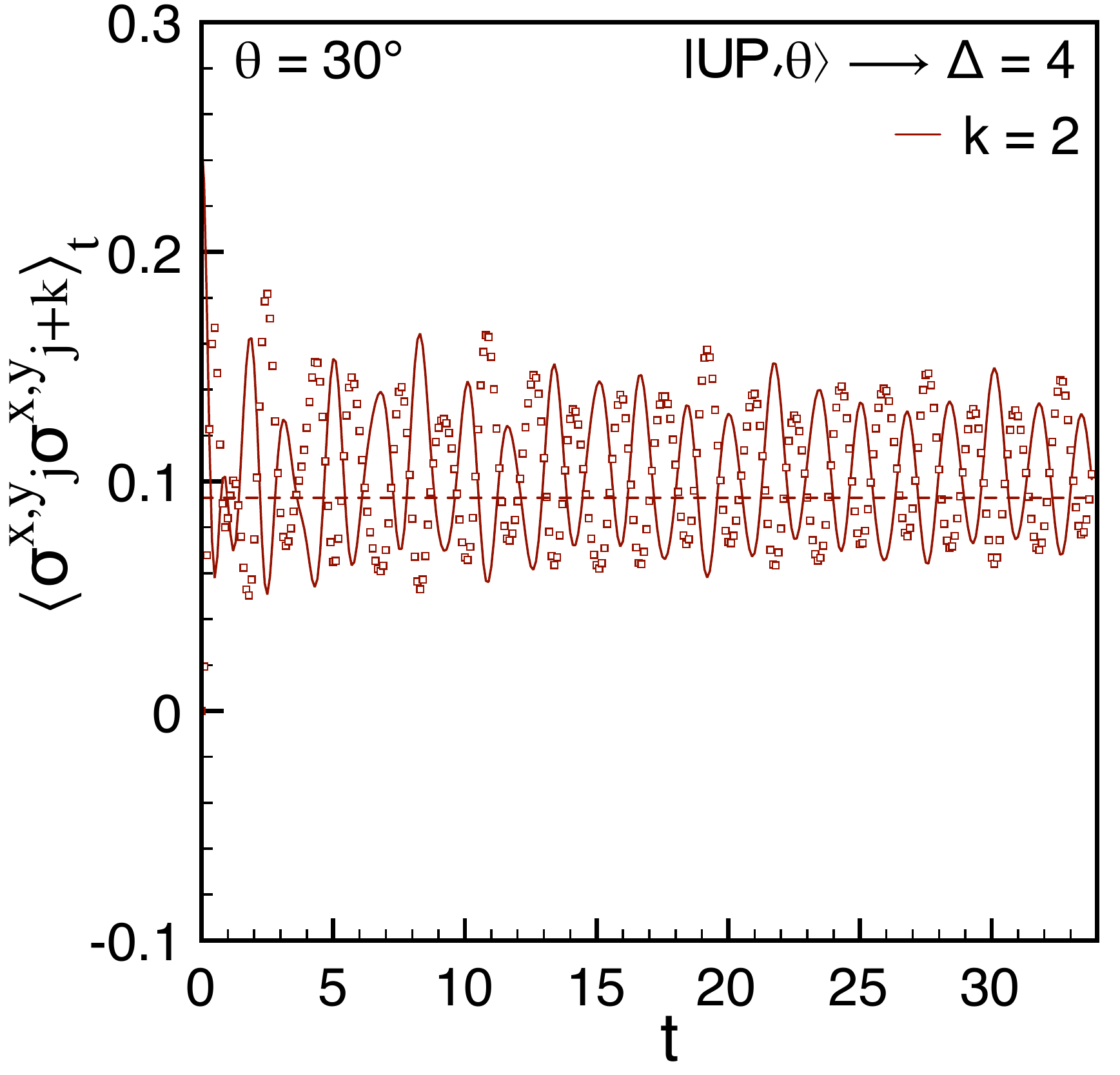}
\includegraphics[width=0.32\textwidth]{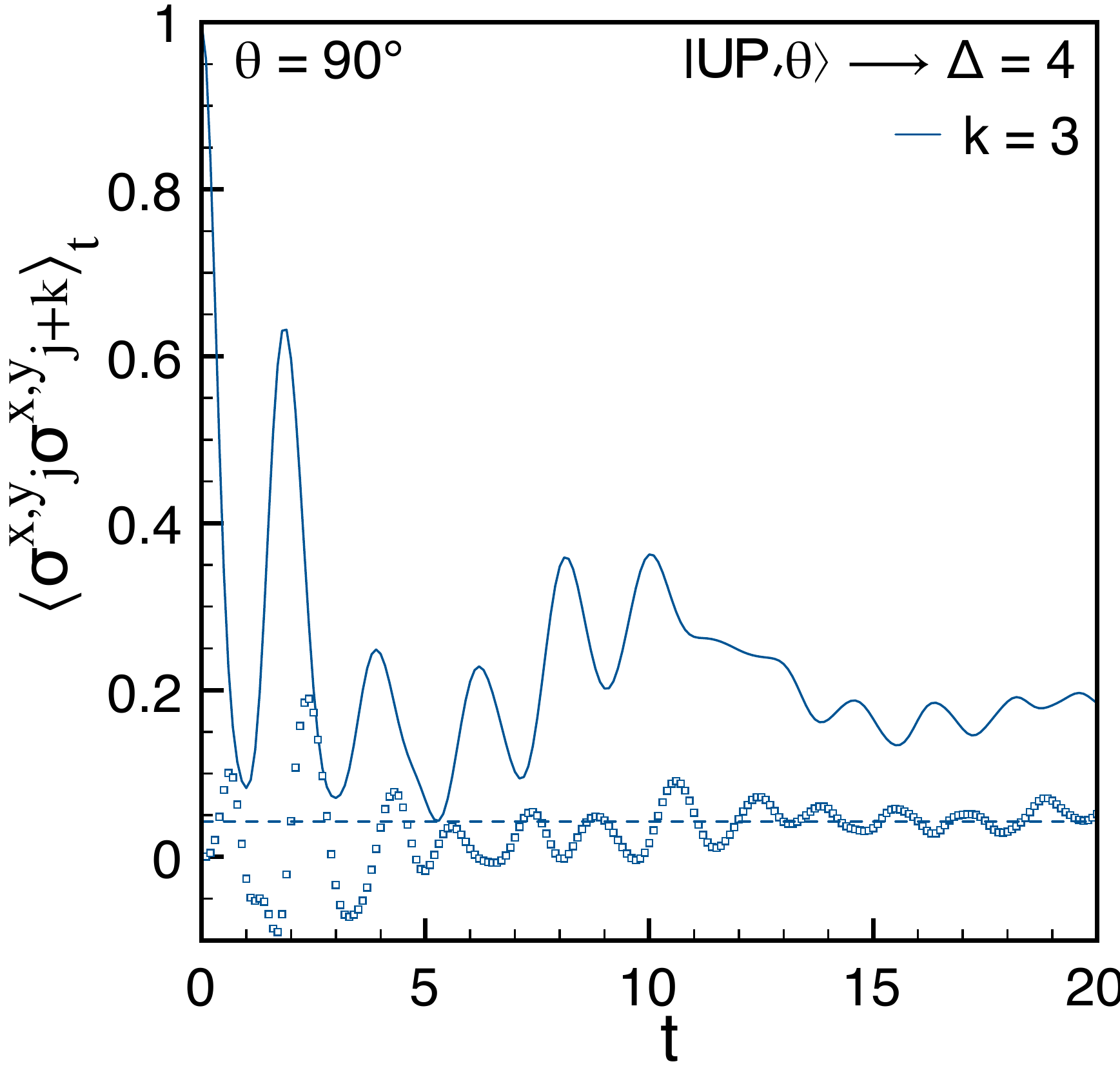}
\includegraphics[width=0.32\textwidth]{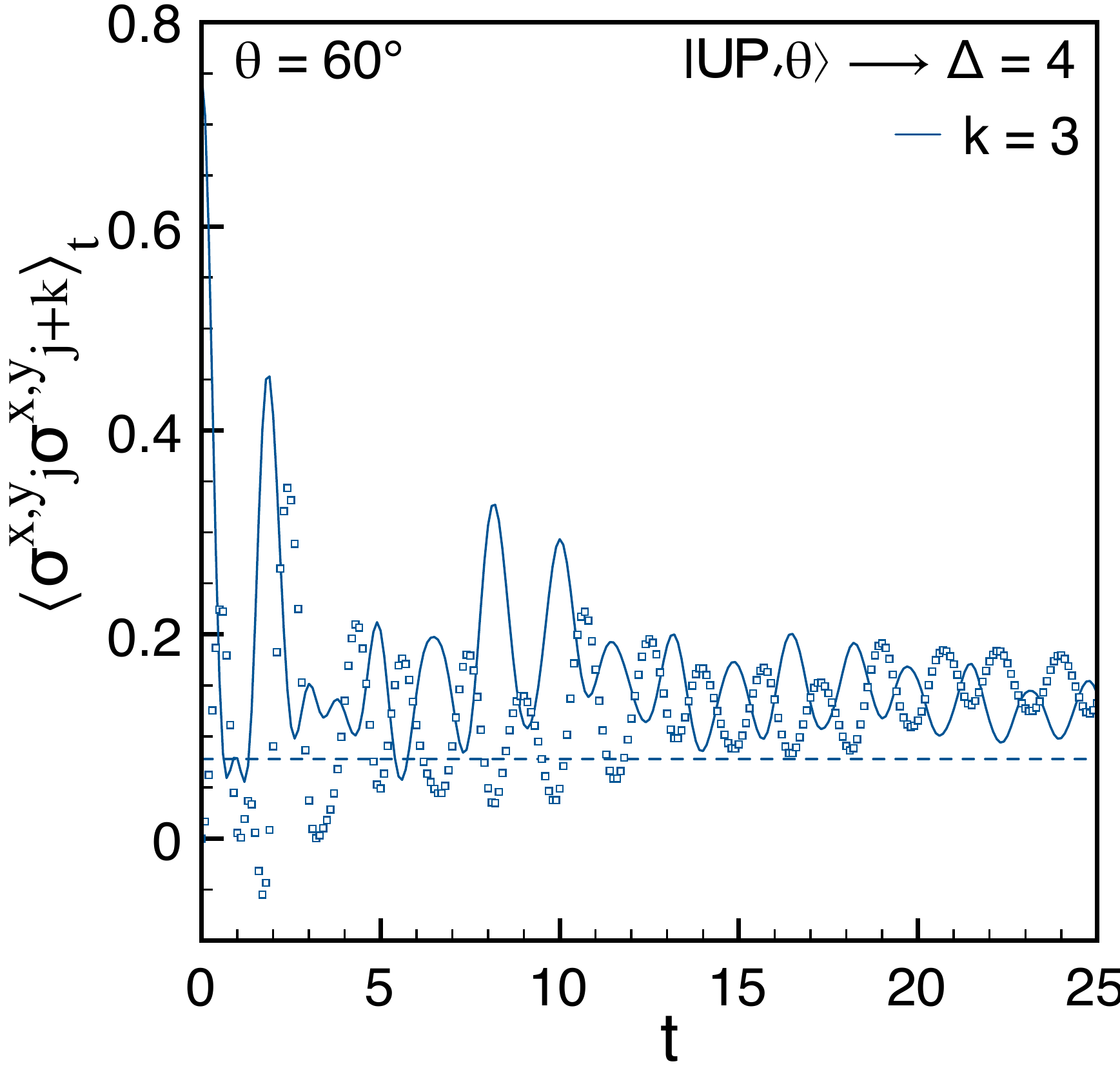}
\includegraphics[width=0.32\textwidth]{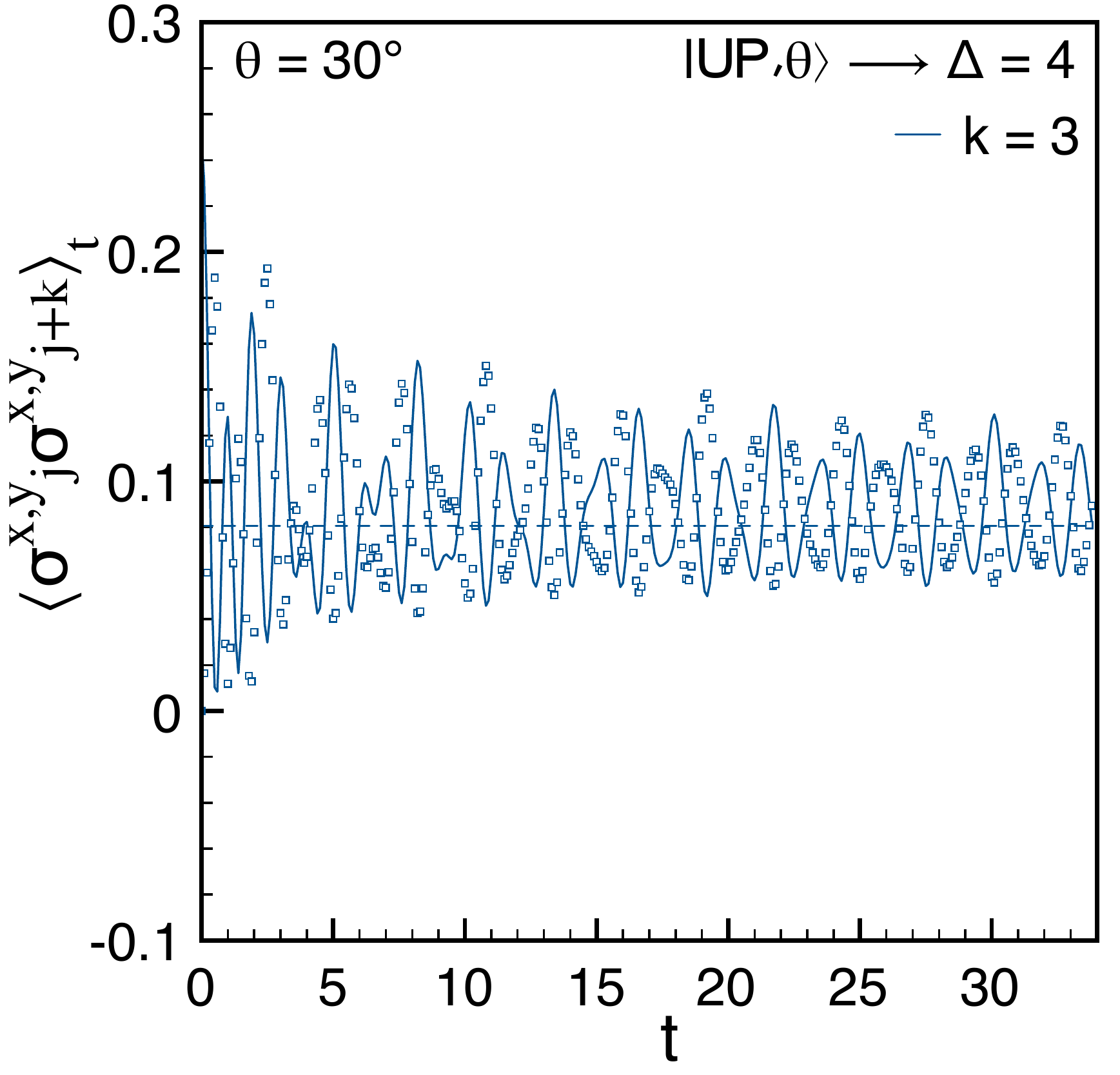}
\caption{\label{figUP1}iTEBD results for transverse
correlations after a quench from the ferromagnetic state with
$\theta=90^{\circ},60^{\circ}, 30^{\circ}$ (from left to right column)
to $\Delta=4$. Solid lines (symbols) correspond to $S^x_{j,j+k}$
($S^y_{j,j+k}$).
} 
\end{figure}
\begin{figure}[ht]
\includegraphics[width=0.32\textwidth]{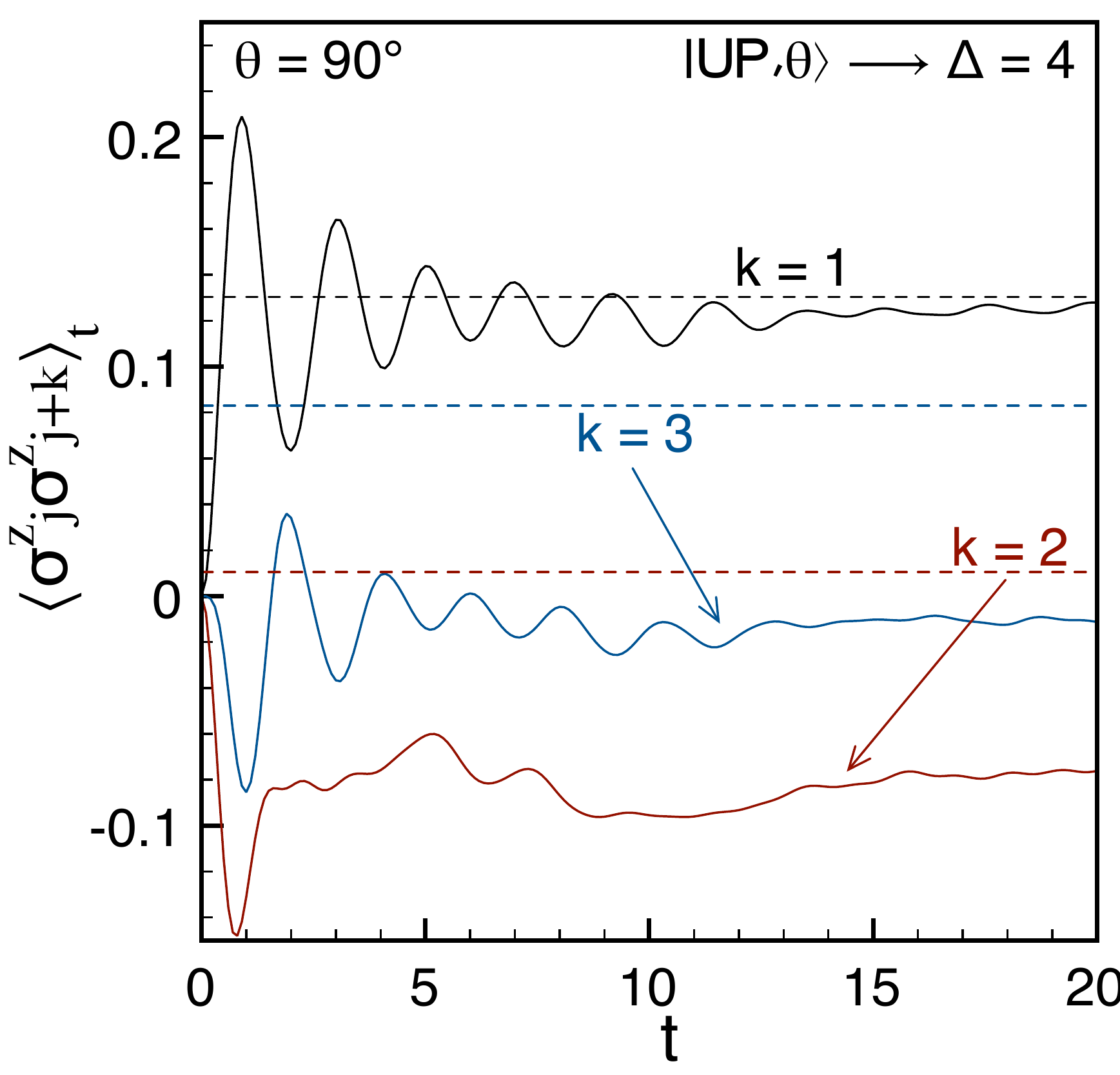}
\includegraphics[width=0.32\textwidth]{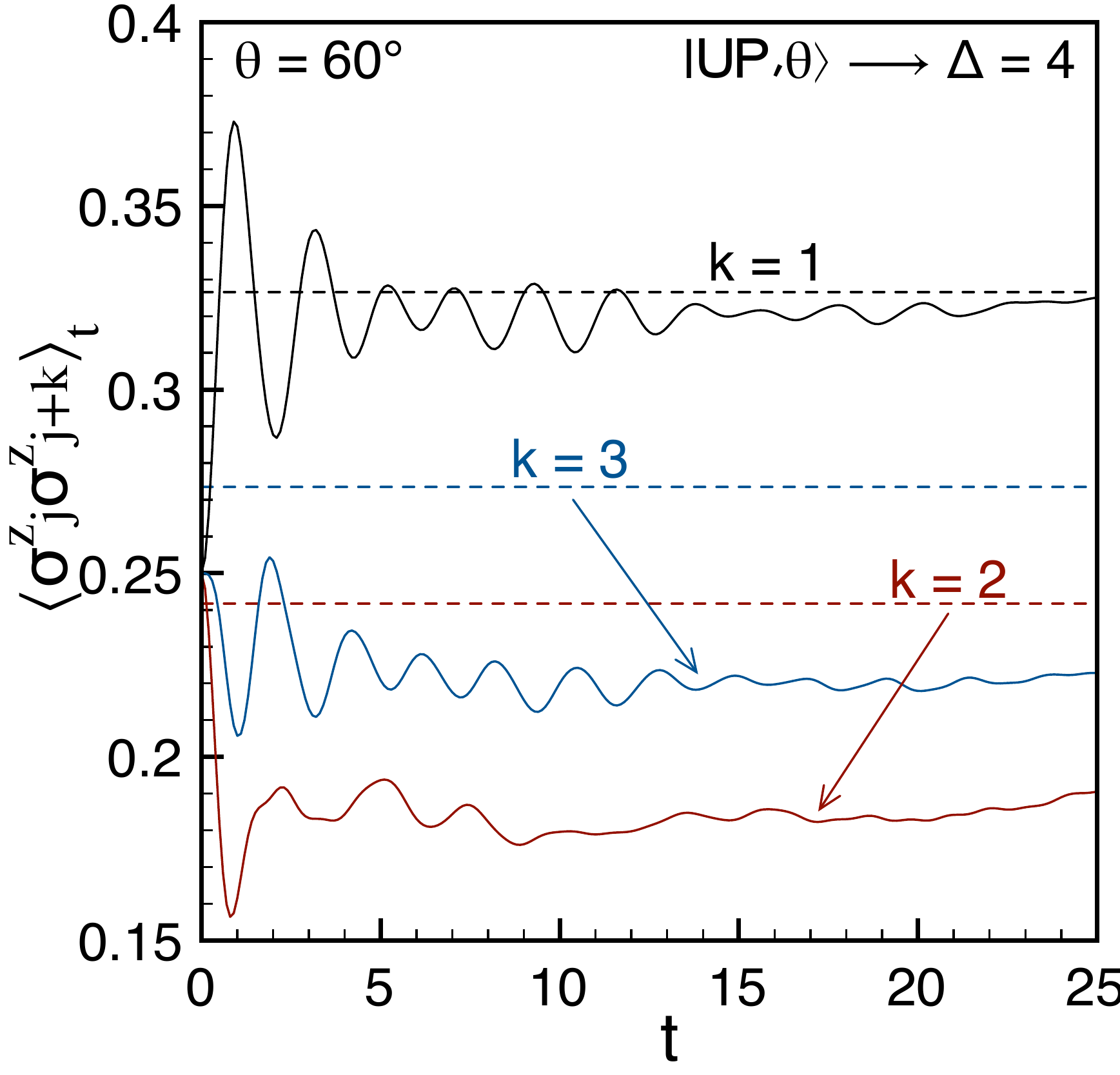}
\includegraphics[width=0.32\textwidth]{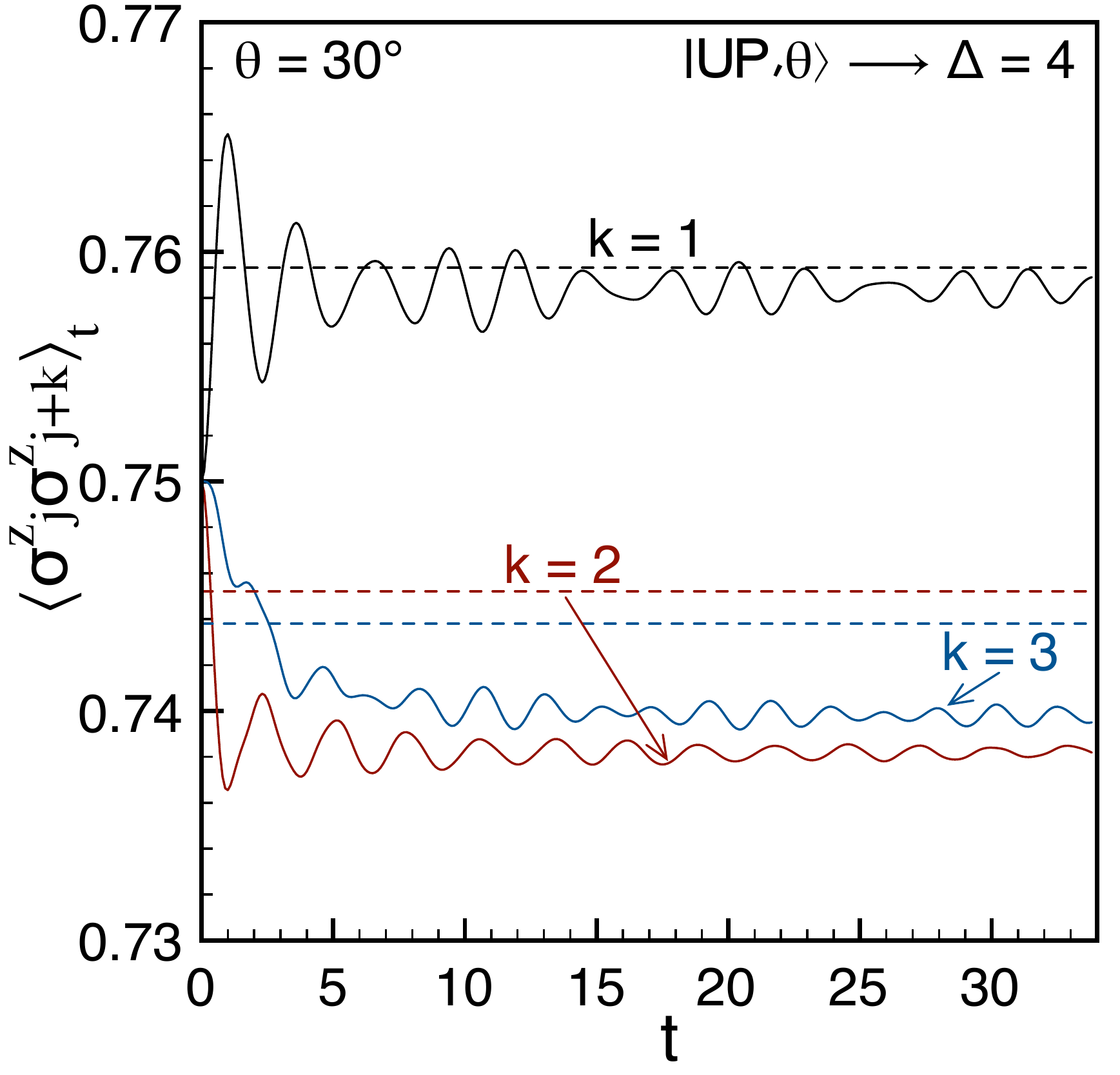}
\caption{\label{figUP4}Same as Fig.~\ref{figUP1} for longitudinal
  correlations. 
} 
\end{figure}
The tilted ferromagnetic states are similar to the tilted N\'eel
states in that they generally break the U(1) symmetry of rotations
around the z-axis of the XXZ Hamiltonian. However, the quench is more
complicated than in the N\'eel case for the following reason. 
The ferromagnetic state along the z-axis ($\theta=0$) does not break
the U(1) symmetry, but in fact is an eigenstate of the XXZ
Hamiltonian. As a result spin-spin correlation function are
time-independent in this case. When we approach $\theta=0^{\circ}$ from above,
the breaking of symmetry becomes unimportant, while at the same time
it becomes increasingly difficult to observe relaxational behaviour in
the accessible time window. Furthermore, a ferromagnetic state in an
arbitrary direction is an exact eigenstate of the isotropic ($\Delta=1$) 
Hamiltonian. Concomitantly the relaxation time diverges for quenches
from general tilted ferromagnetic states when $\Delta$ is close to $1$.
As a result of the aforementioned complications, the relaxation times
are always extraordinarily large and even though the growth of the
entanglement entropy is considerably slower than for the other initial
states we have considered (cf. Fig. \ref{figS}), which allows us to
explore larger time windows, relaxation to the GGE is not observed during  
the accessible times.

In Figs~\ref{figUP1}-\ref{figUP4} we show iTEBD data for quenches
from tilted ferromagnetic states with $\theta=90^{\circ}, 60^{\circ},
30^{\circ}$ to an XXZ chain with $\Delta=4$. The various correlators
are seen to exhibit irregular and non-monotonic oscillations. The
symmetry in the $xy$ plane is clearly not restored. The observed
oscillatory behaviour in the nearest-neighbour correlations occurs
around values that are broadly compatible with the GGE prediction.
Conversely, correlators at distance $2$ and $3$ appear to relax, but
to values that are quite distant from the GGE predictions. Our
interpretation of the data is that in all cases the time scale for
relaxation is too large to be accessible by numerical simulations. 

\begin{figure}[t]
\includegraphics[width=0.35\textwidth]{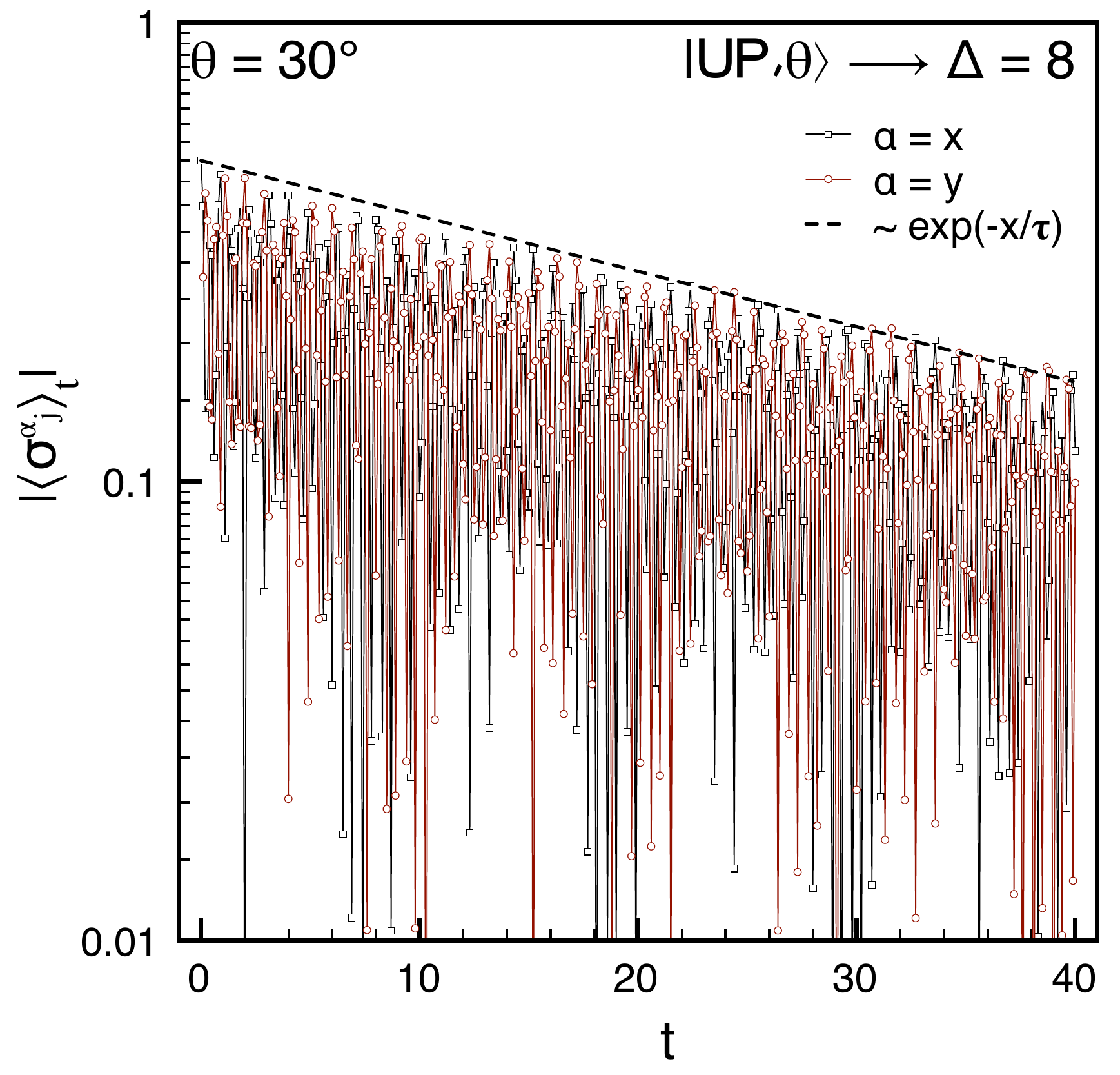}
\includegraphics[width=0.35\textwidth]{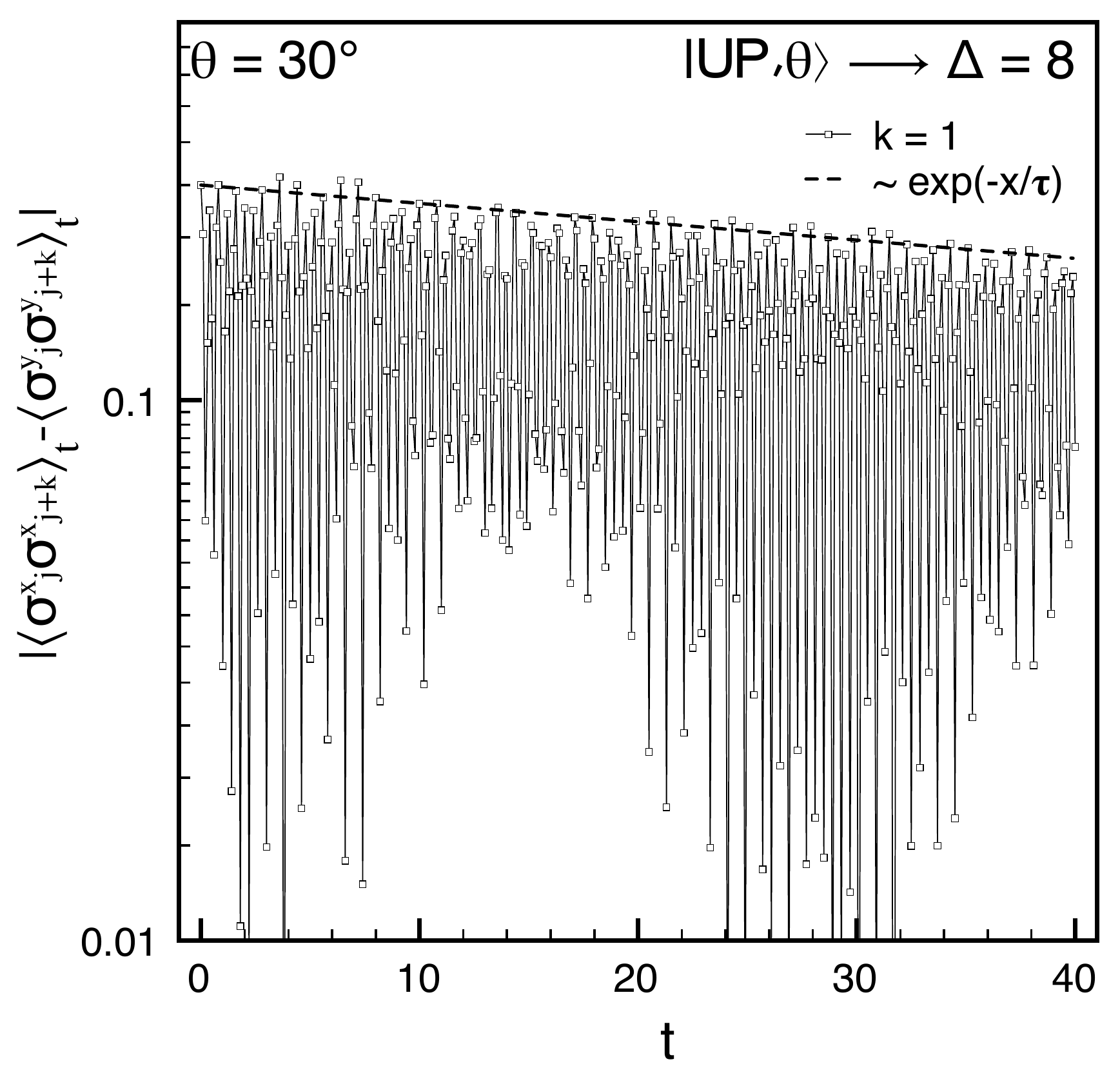}\\
\includegraphics[width=0.35\textwidth]{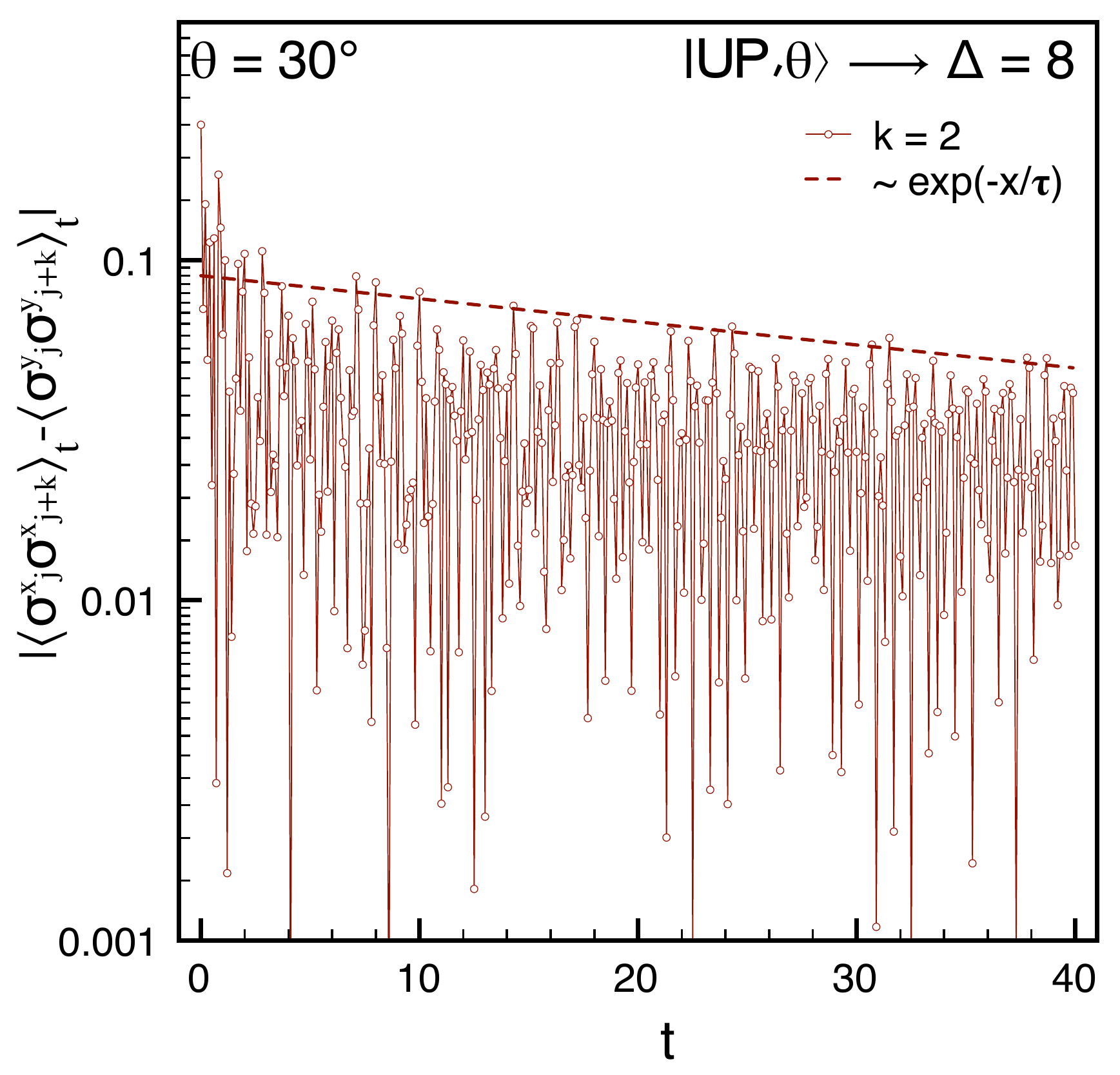}
\includegraphics[width=0.35\textwidth]{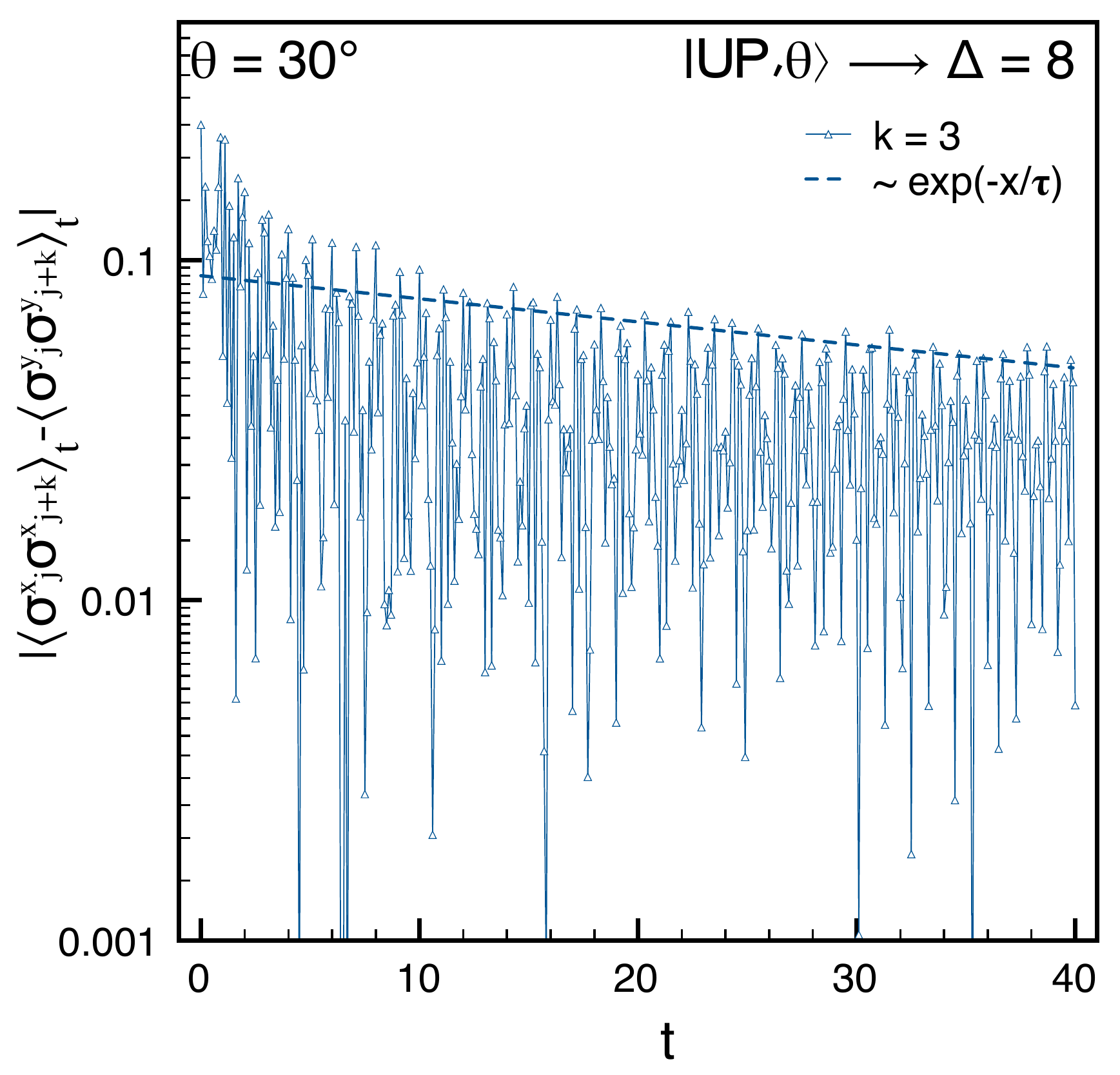}
\caption{\label{figUPLOG}
iTEBD results for quenches from the ferromagnetic state with $\theta=60^{\circ}$  to $\Delta=8$. 
The rightmost panel is the local correlator $\langle\sigma^{x,y}_j\rangle$.
The other three panels report the (absolute value of the) difference 
$|\langle \sigma^{x}_{j}\sigma^x_{j+k}\rangle-\langle\sigma^{y}_{j}\sigma^y_{j+k}\rangle|$ with $k=1,2,3$
(going from left to right).  
} 
\end{figure}
In order to lend credence to this interpretation, we have analyzed
the single-spin expectation values $\langle\sigma^{x,y}_j\rangle$ for
a quench from a ferromagnetic state with $\theta=30^{\circ}$  to $\Delta=8$. 
The results are shown in Fig. \ref{figUPLOG}. The data is compatible
with exponential relaxation to the expected GGE value zero, but with a
very large relaxation time. In order to see whether there is any
evidence for restoration of spin-rotational symmetry around the
z-axis, it is useful to plot the differences between transverse correlators 
$|\langle
\sigma^{x}_{j}\sigma^x_{j+k}\rangle-\langle\sigma^{y}_{j}\sigma^y_{j+k}\rangle|$
for distances $k=1,2,3$ as functions of time. The data are compatible
with a very slow exponential decay to zero, indicating symmetry
restoration at vary late times. A naive fit of the maxima for the
difference at $k=1$ gives a time scale $\tau\sim130$, which implies
that in order to observe the true asymptotic value with a precision of
$0.01$, we should roughly run the simulation up to $t\sim300$ which is
clearly beyond our capability. For smaller values of $\Delta$, the
relaxation times increases because we are getting closer to the
isotropic point $\Delta=1$, where relaxation is absent. Consequently,
 an analysis like in Fig. \ref{figUPLOG} becomes even more
difficult, but we are confident that the same qualitative scenario is
valid. The upshot is as follows: we believe that quenches starting
from tilted ferromagnetic states are characterised by very large
relaxation times. This prevents us from checking the GGE predictions.

\subsection{Interaction quenches}

\begin{figure}[t]
\includegraphics[width=0.35\textwidth]{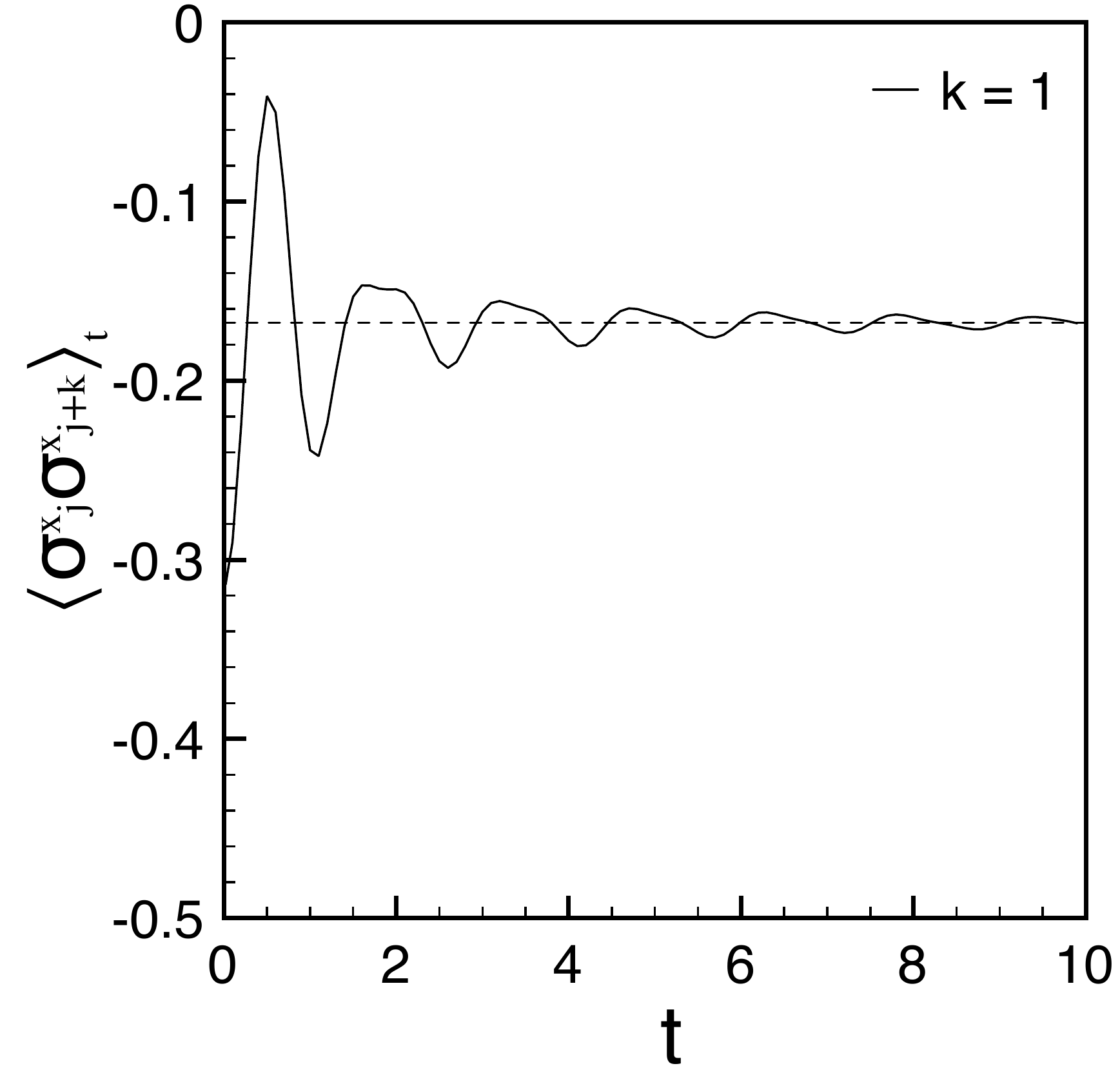}
\includegraphics[width=0.35\textwidth]{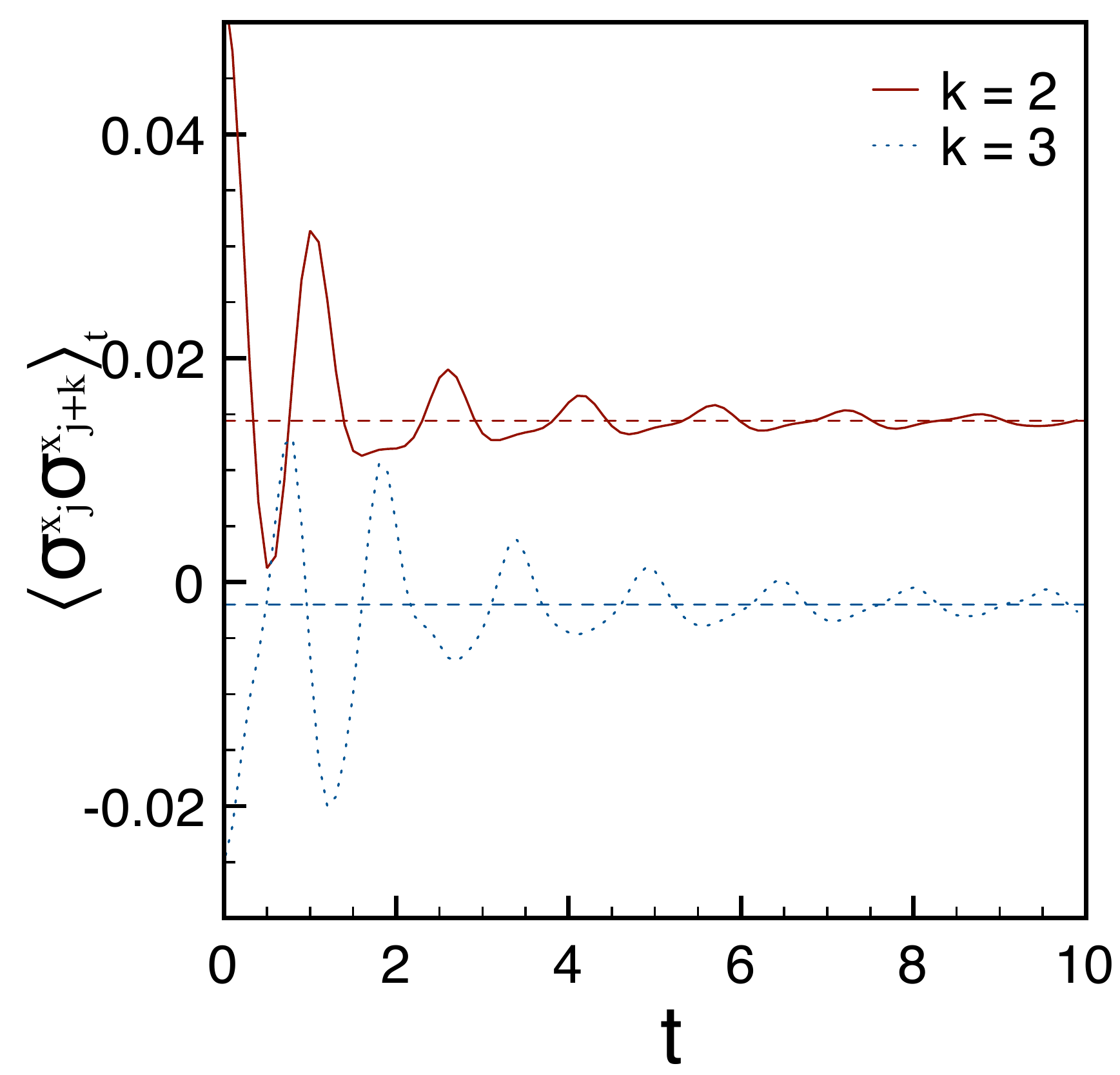}\\
\includegraphics[width=0.35\textwidth]{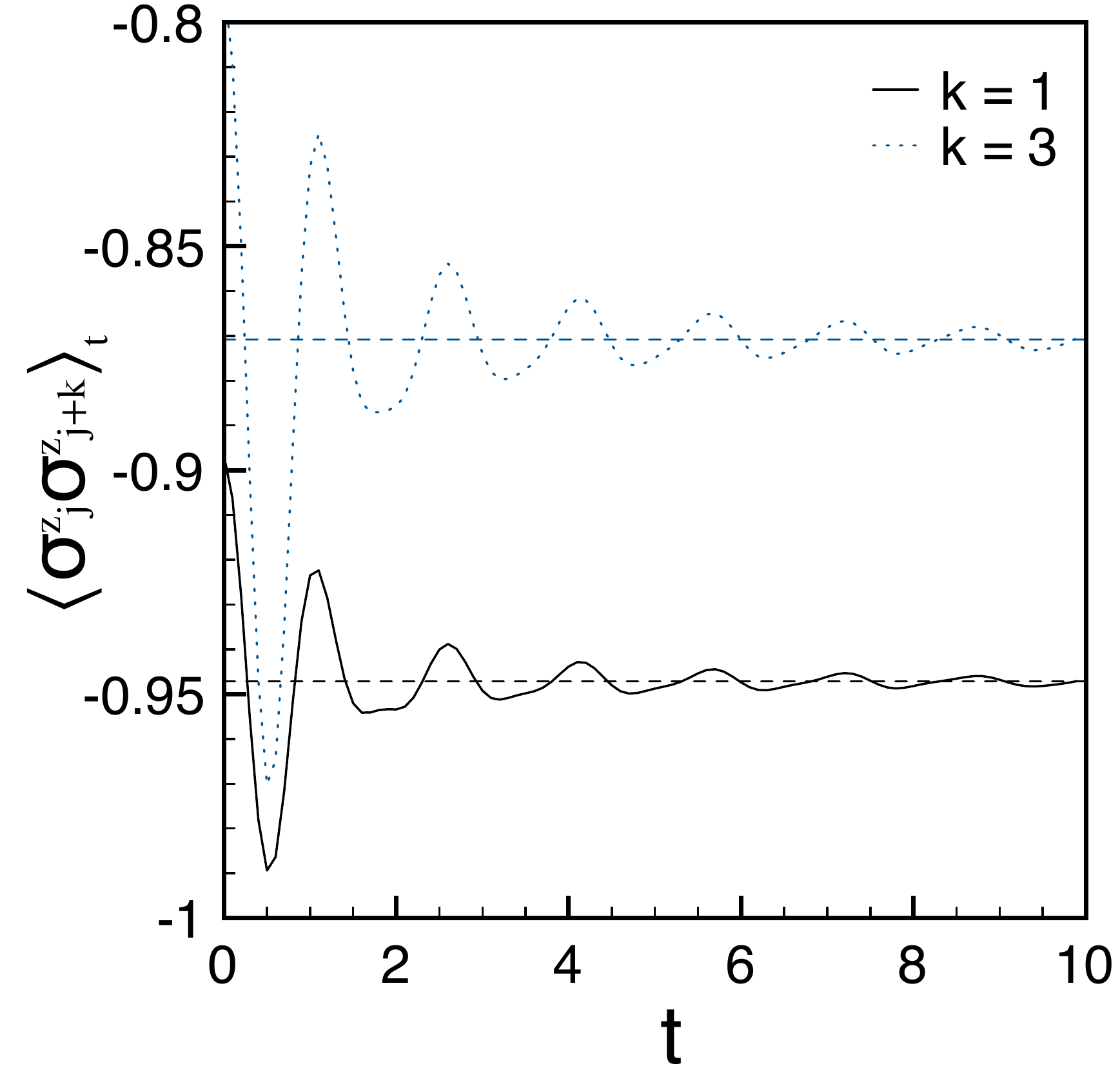}
\includegraphics[width=0.35\textwidth]{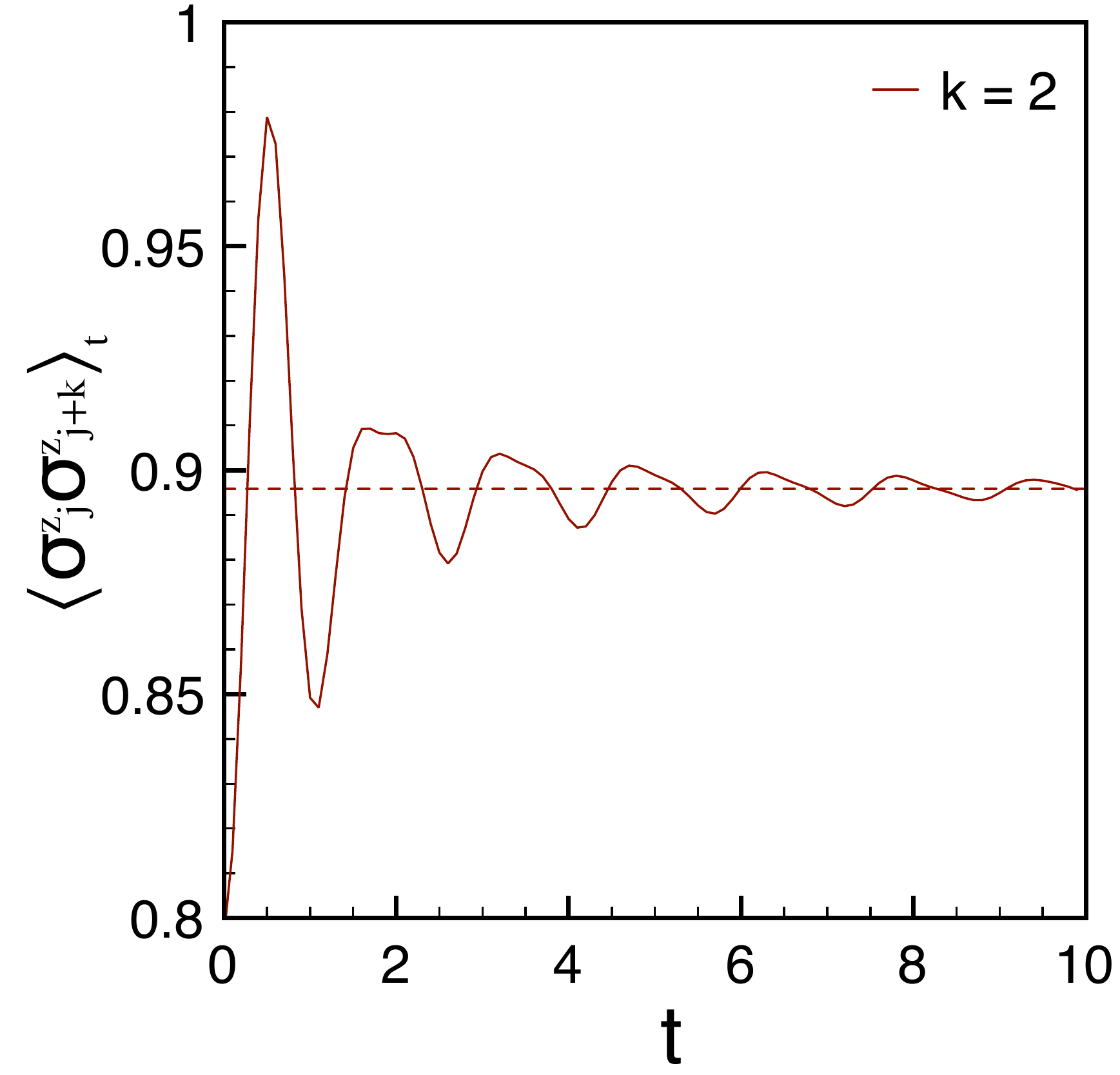}
\caption{\label{figD0toD6} Interaction quenches in the XXZ Hamiltonian
from $\Delta_{0}=3$ to $\Delta=6$. All data are from tDMRG
simulations for chains of length $L=64$.}  
\end{figure}



\begin{figure}[t]
\includegraphics[width=0.35\textwidth]{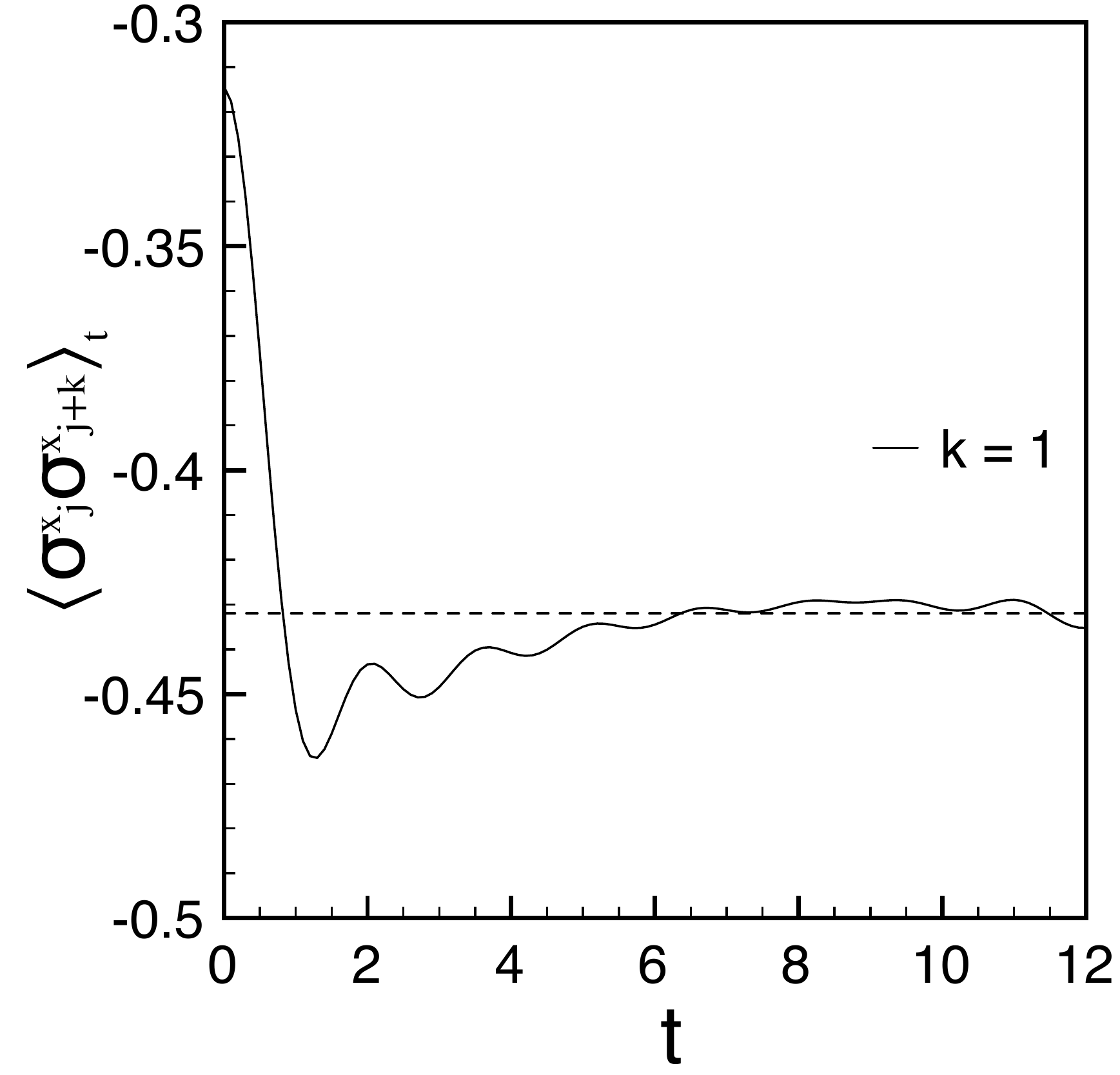}
\includegraphics[width=0.35\textwidth]{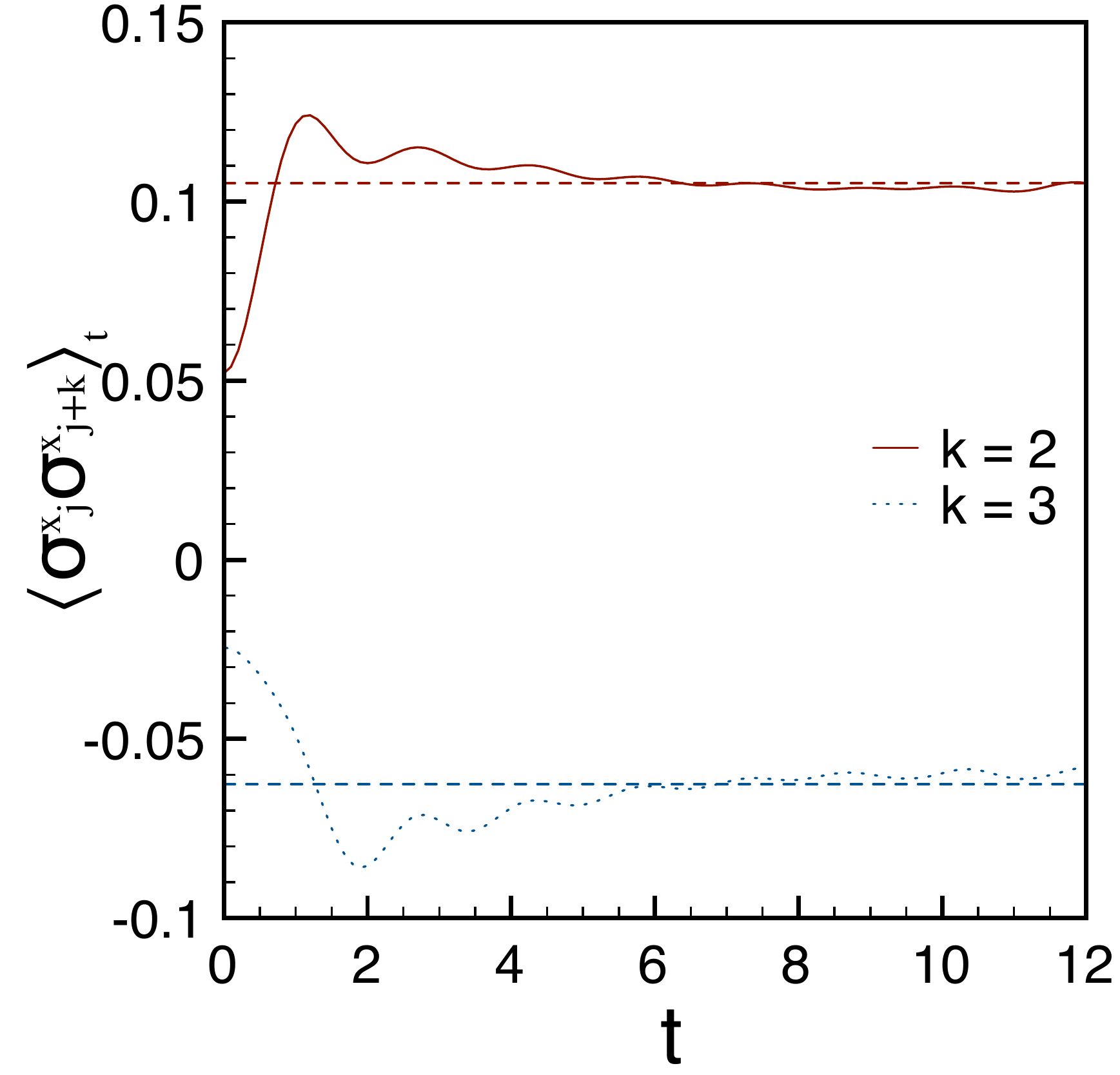}\\
\includegraphics[width=0.35\textwidth]{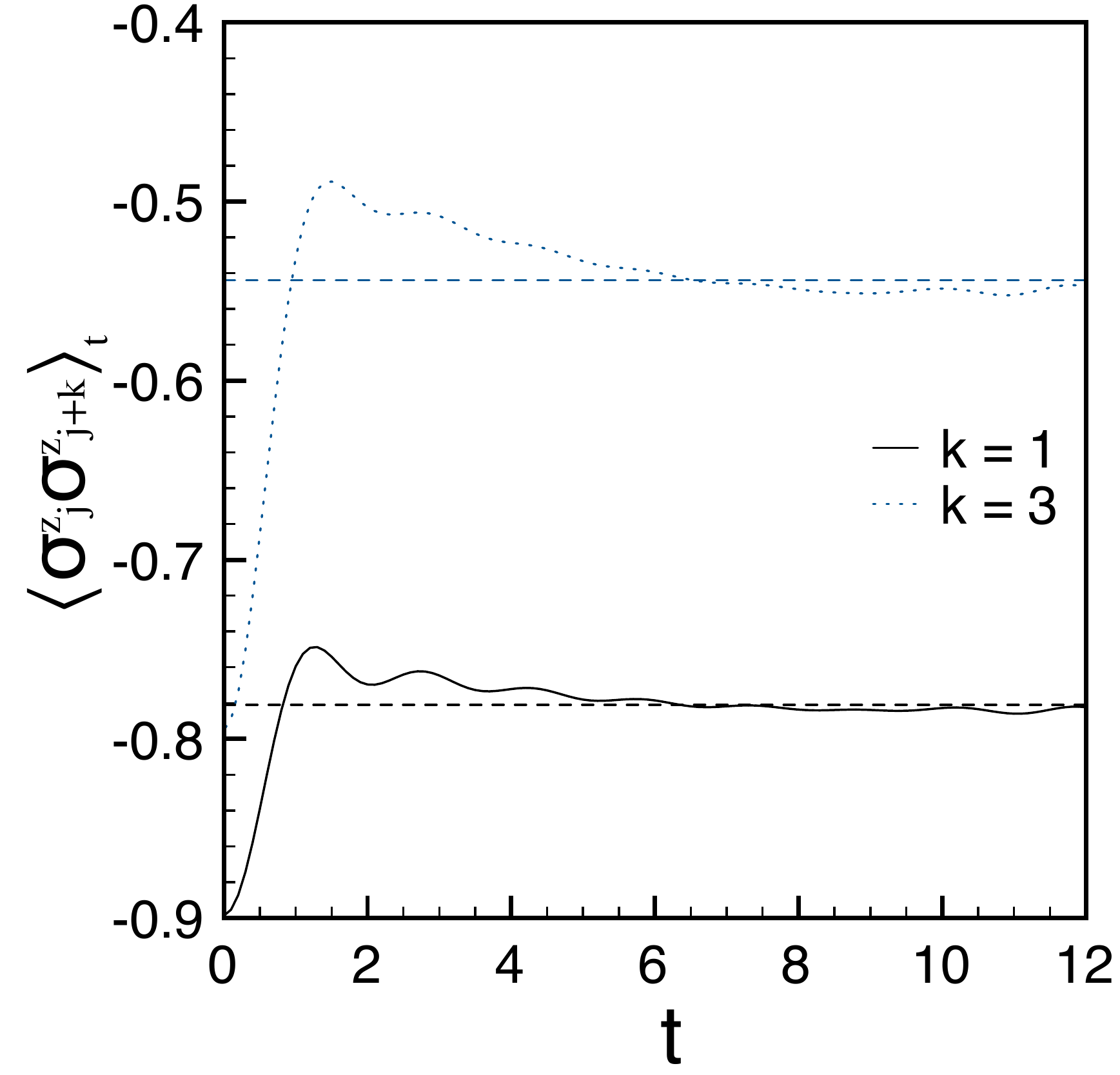}
\includegraphics[width=0.35\textwidth]{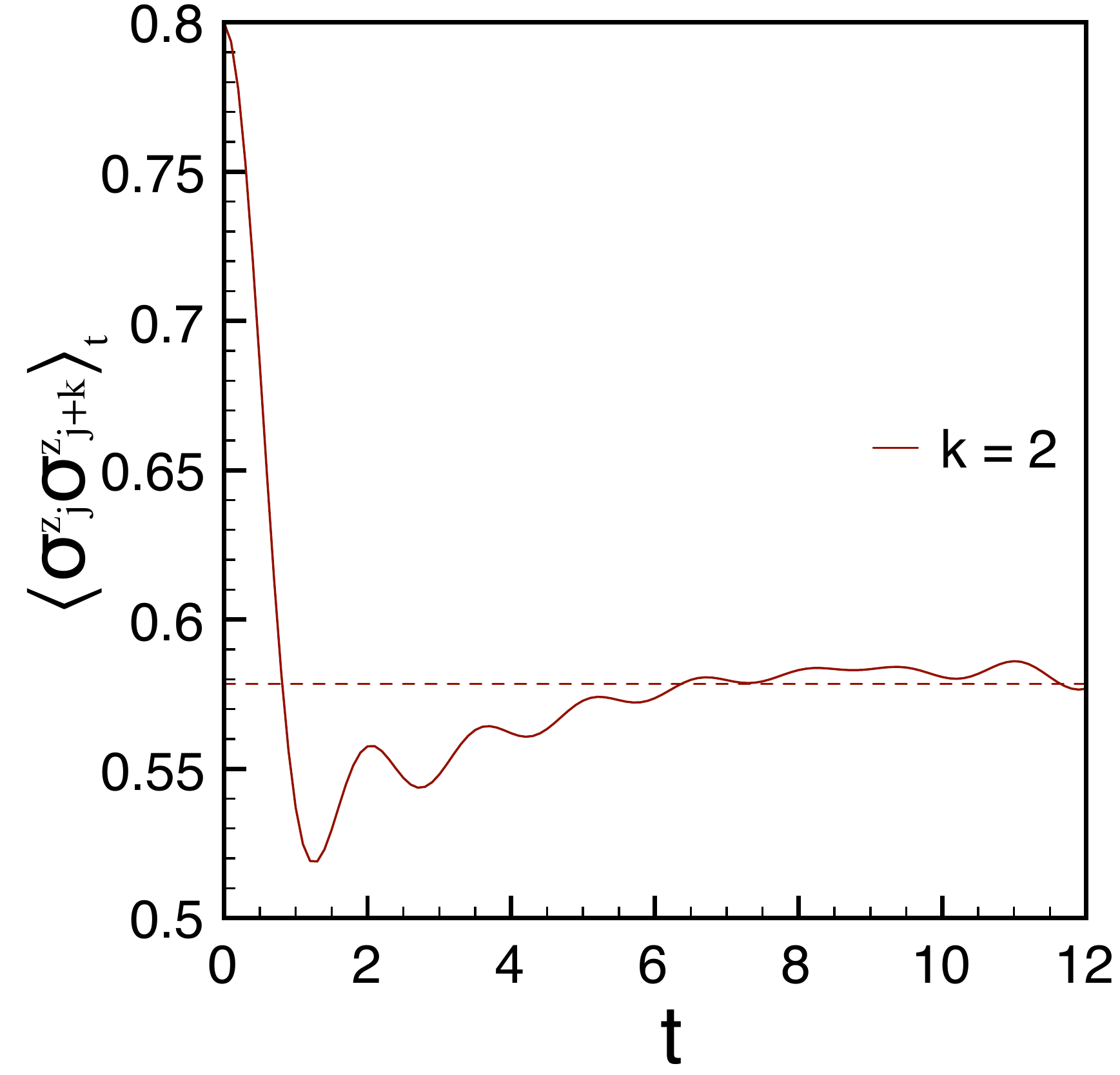}
\caption{\label{figD0toD2} Interaction quenches in the XXZ Hamiltonian
from $\Delta_{0}=3$ to $\Delta=2$. 
All data are obtained by tDMRG simulations for chains of length $L=64$.} 
\end{figure}
The final class of initial states we have considered are ground states
of the Heisenberg XXZ chain, i.e. \emph{interaction quenches}, where
we prepare the system in the ground state of the XXZ Hamiltonian at
parameter $\Delta_0$, and at time $t=0$ instantaneously quench it from 
$\Delta_0$ to $\Delta$. We have carried out tDMRG computations for a variety
of values of $\Delta_0$ and $\Delta$, and present some representative
results in Figs~\ref{figD0toD6} and \ref{figD0toD2}. The chain length
in these simulations is $L=64$. The initial state is selected by
running a static DMRG retaining $\chi_{0}=40$ states.  
After the quench, we perform the usual time-dependent routine
retaining at most $\chi=300$ states, which is enough to show
equilibration because the entanglement entropy grows very slowly  
(cf. Fig. \ref{figS}) and so these interaction quenches are less
computationally demanding than the quenches out of the initial
states we have considered above. It is clear from Figs~\ref{figD0toD6}
and \ref{figD0toD2} that all correlators relax in an oscillatory way
to the GGE predictions (dashed lines) at late times. Like in 
in all previous cases, the principal oscillation has a frequency
proportional to $\Delta$, but there are also less important
oscillations with higher frequencies. Therefore, for $\Delta=2$ these
oscillations slightly spoil the equilibration around the GGE values on
the time scale reported in the figure. 

\section{Conclusions}
We have considered quantum quenches from several initial states in the
spin-1/2 Heisenberg XXZ chain with Ising-like anisotropy $\Delta>1$.
In particular we considered
(a) Tilted N\'eel states;
(b) Majumdar-Ghosh dimer product states;
(c) Tilted ferromagnetic states;
(d) The ground state of the XXZ Hamiltonian for $\Delta_0>1$.
Following Ref.~[\onlinecite{FE_13b}] we constructed the corresponding
generalized Gibbs ensembles by means of the quantum transfer matrix
approach. We then determined the short-distance (up to distance three)
behaviour of spin-spin correlation functions in these ensembles.

We then considered the time evolution under the XXZ Hamiltonian when
starting in these initial states by means of numerical matrix product
techniques (i.e. tDMRG and iTEBD). In cases (a), (b) and (d) we
observed that on the accessible time-scales short-distance spin-spin
correlators appear to relax towards stationary values, which are in
good agreement with the GGE predictions. For tilted ferromagnetic
initial states the presence of an extraordinarily long relaxation time
precludes an analysis of the stationary behaviour by our numerical
methods. First and foremost, these results constitute a strong
test of the GGE predictions by independent methods. A second issue we
have focussed on is that of {\it symmetry restoration} after quantum quenches.
Most of the initial states we considered break symmetries of the XXZ
Hamiltonian and the generalized Gibbs ensemble. In order for the GGE
to be a valid description of the stationary state at late times, such
symmetries must be restored under time evolution. This is indeed what
we have observed in our numerical computations. To the best of our
knowledge this phenomenon was previously discussed only for the
transverse field Ising chain \cite{CEF,FE_13a}. Our results for the
XXZ chain show the general nature of this phenomenon.

They also raise many interesting open questions and problems:
\begin{itemize}
\item Our analysis has been restricted to the massive regime
$\Delta>1$. It will be very interesting to extend it to the critical
regime $|\Delta|<1$. For particular choices of initial states we
expect the quasi-local integrals of motion constructed recently\cite{prosen}
to come into play.

\item{} It would be interesting to consider initial states that break
the reflection symmetries of the Hamiltonian and lead to GGEs, in which
the parity-odd conservation play a role. In such cases we expect the
reflection symmetry not to be restored at late times.

\item We expect symmetry restoration to be a rather generic feature
for quenches in one dimensional systems. This is because spontaneous symmetry
breaking may occur only at zero temperature, and the finite energy
density present in the system after a quench plays a role very close to
a finite temperature. In higher dimensional models it ought to be
possible for spontaneous symmetry breaking to occur in GGEs describing
stationary states after quenches.
Some similar conclusions on symmetry restoration after a quench were also drawn in 
Ref. \cite{gc-11} by means of renormalisation group arguments and field theoretical methods 
(in imaginary time and analytically continuing the final results to real time). 

\item We have focussed on spin-spin correlators on short distances of
at most 3 sites. These can be generalized to longer distances
by combining the results of Refs~[\onlinecite{thermal}] for the thermal
case with our formalism of constructing the GGE for a given quench.

\item{} Finally, an analytic description of the full time evolution
after quantum quenches in interacting integrable models remains a largely
open problem. A possible approach to this problem is to use 
the overlaps between several initial states and arbitrary Bethe states
recently reported by Pozsgay \cite{p-13a} to determine the ``initial
data'' in the saddle-point approach of Ref.~[\onlinecite{ce-13}].

An alternative method is based on the Yudson representation
\cite{la-13,austen}, but at present this is restricted to the limit of zero
energy density (compared to the ground state of the post-quench
Hamiltonian). The generalization of this method to finite energy
densities remains an open problem. 
\end{itemize}

\section*{Acknowledgments}   
This work was supported by the EPSRC under grants EP/I032487/1 (FHLE
and MF) and EP/J014885/1 (FHLE and MF) and by the ERC under  Starting
Grant 279391 EDEQS (PC and MC).

\appendix

\section{Largest Eigenvalue of the Quantum Transfer Matrix}
\label{app:lambdamax}
In this section we discuss the validity of some assumptions concerning
the leading eigenvalue of our quantum transfer matrix, which underlie
the derivation of the system of equations \eqref{eq:systemh}. 
The latter was obtained in Ref.~[\onlinecite{FE_13b}]  by taking the
Trotter limit ($N\rightarrow\infty$) in the Bethe ansatz equations for
the leading eigenvalue of the quantum transfer matrix, which read
\be\label{BAE0}
\Bigl[\prod_{k=1}^y\frac{\sinh(w_j-u_{k;N}^{(y)})\sinh(w_j+\eta)}
{\sinh(w_j-u_{k;N}^{(y)}-\eta)\sinh(w_j)}\Bigr]^{N/y}
+\prod_{k=1}^N\frac{\sinh(w_j-w_k+\eta)}{\sinh(w_j-w_k-\eta)}=0\,\quad
j=1,\ldots,N. 
\ee
Here $u_{k;N}^{(y)}$ are inhomogeneities introduced into the transfer
matrix in order for it to give rise to the truncated GGE with only the
first $y$ conservation laws retained. The derivation of the integral
equations describing the largest eigenvalue of the quantum transfer
matrix was based on the following assumptions:
\begin{enumerate}[(a)]
\item \label{inter}The thermodynamic limit and the Trotter limit are interchangeable;\\
\item \label{gap}The leading eigenvalue of the quantum transfer matrix
  is non-degenerate and is separated from the subleading eigenvalues
  by a finite gap;
\item \label{Nroots}The largest eigenvalue is determined by a solution of the Bethe ansatz equations with $N$ roots. 
\end{enumerate}
In addition, the Trotter limit was taken by assuming that 
\begin{enumerate}[(d)]
\item the solutions of the Bethe ansatz equations for the largest eigenvalue of the quantum transfer matrix lie in the region $\mathrm{Re}[w_j]\in(-\eta/2,\eta/2)$.
\end{enumerate}
The system of equations \fr{eq:systemh} was finally obtained by taking
the limit $y\rightarrow\infty $. The description of the GGE by
considering a limiting procedure in which the number of retained
conservation laws is taken to infinity in the end of the calculation
was proposed in Ref.~[\onlinecite{FE_13a}] and has a sound physical
grounding. In the following we shall accept assumption~(\ref{inter})
as it is a central tenet of the quantum transfer matrix approach, but
scrutinize the remaining assumptions for a ``truncated
GGE''\cite{FE_13a} with three conservation laws. For simplicity
we restrict our analysis to parity invariant initial states, for which
the inhomogeneities can be chosen as 
\be
u_{j;N}^{(3)}=\frac{\sinh\eta}{2}\Bigl(-\lambda_1\frac{1}{N} +
(6\lambda_3)^{1/3}\frac{e^{2 \pi i j/3}}{N^{1/3}}\Bigr)\, . 
\ee
In spite of the unusual dependence on the Trotter number $N$, the
finite-$N$ corrections to the Lagrange multipliers scale as integer
powers of $1/N$. It is then reasonable to expect that the role
of the small parameter controlling large-$N$ expansions of physical
quantities will be played by $N^{-1}$ rather than $N^{-1/3}$.
This expectation is borne out by our direct calculations.

\begin{figure}[ht]
\includegraphics[width=0.45\textwidth]{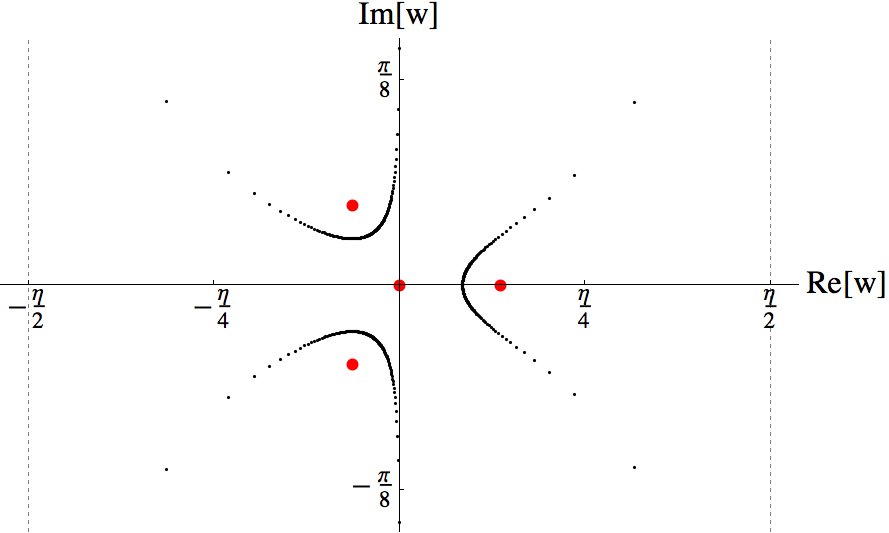}\qquad
\includegraphics[width=0.45\textwidth]{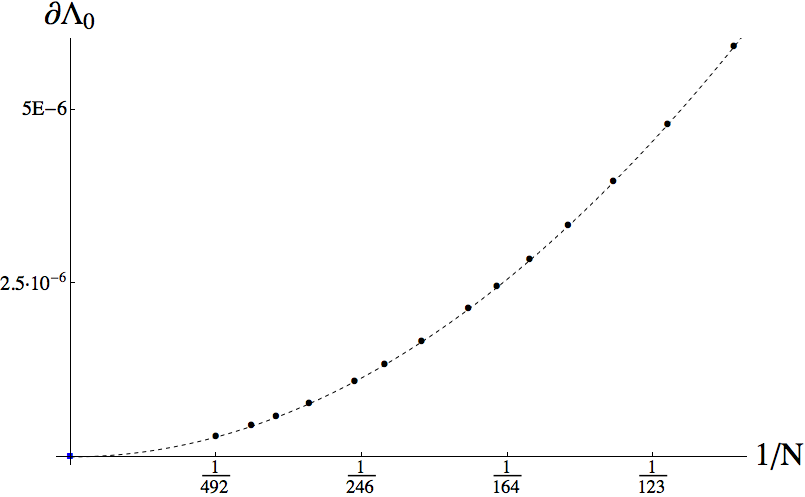}
\caption{Left: Bethe roots associated with the leading eigenvalue of
  the quantum transfer matrix for the quench from
  $|\rightarrow\rightarrow\dots\rangle$ with $\Delta=2$ and
  $N=492$. The dashed lines are part of the contour of integration
  used to derive the NLIEs in the Trotter limit. The red points are
  the inhomogeneities of the transfer matrix, which in the Trotter
  limit approach $0$. Right: The discrepancy $\delta
  \Lambda(N)\equiv\Lambda_0(N)-\Lambda_0$ between the the largest
  eigenvalue of the quantum transfer matrix for a finite number of
  roots and the eigenvalue as computed solving \eqref{eq:systemh}.} 
\label{fig:roots} 
\end{figure}

The first step in analyzing the spectrum of the transfer matrix is, as
always, to identify the root distribution of the leading eigenvalue by
comparing the eigenvalues obtained by solving the Bethe Ansatz
equations \fr{BAE0} to exact diagonalization data for small system sizes.
This analysis shows that for small systems both (\ref{gap}) and
(\ref{Nroots}) hold. The leading eigenvalue of the transfer matrix is
characterized by considering the logarithmic form of the Bethe Ansatz
equations, which for the quench in Fig.~\ref{fig:roots} and
$N=0\ {\rm mod}\ 6$ reads
\be\label{eq:bether}
\frac{N}{3}\sum_{k=1}^3\log\Bigl(\frac{s_k \sinh(w_j-u_{k;N}^{(3)})\sinh(w_j)}{\sinh(w_j-u_{k;N}^{(3)}-\eta)\sinh(w_j+\eta)}\Bigr)+N\log\Bigl(-\frac{\sinh^2(w_j+\eta)}{\sinh^2(w_j)}\Bigr)+\sum_{k=1}^N\log\Bigl(-\frac{\sinh(w_j-w_k+\eta)}{\sinh(w_j-w_k-\eta)}\Bigr)=2\pi i I_j\, ,
\ee
where $s_1=s_2=-1$ and $s_3=1$. The leading eigenvalue of the quantum
transfer matrix corresponds to the sequence of half integer numbers
\be
I_j=\frac{1-N}{2},\hdots,\frac{N-1}{2}\, .
\label{integers}
\ee 
As usual, we are now able to follow the solution \fr{integers} with
increasing $N$ by numerically solving the Bethe Ansatz equations
\fr{eq:bether}. We considered system sizes of a few hundred sites and
performed the following checks:
\begin{enumerate}
\item For large $N$, we checked whether $\mathrm{Re}[w_j]\in(-\eta/2,\eta/2)$;
\item We extrapolated the leading eigenvalue and the Fourier coefficients of $\rho(x)e^{\zeta(x)/2}$ and compared them with the corresponding quantities obtained from the solution of \eqref{eq:systemh} (re-adapted to the  truncated GGE).
\end{enumerate}
Fig.~\ref{fig:roots} shows the Bethe roots $w_j$ that solve \eqref{eq:bether} for a rather large value of $N$ for the quench from the ferromagnetic state in x-direction
$|\rightarrow\rightarrow\dots\rangle$  with $\Delta=2$. The roots lie
inside the integration contour used to derive the nonlinear integral
equations. In addition, the extrapolation of the leading eigenvalue of
the quantum transfer matrix in the Trotter limit  is in perfect
agreement with the value corresponding to the solution of
\eqref{eq:systemh}. An analogous discussion holds true for the Fourier
coefficients of $\rho e^{\zeta/2}$,  corroborating the assumptions we
made.

\section{List of explicit results for spin-spin correlators in the GGE}
\label{app:list}
In this appendix we list the GGE results for the spin-spin correlation functions
\fr{Faj} calculated by means of the quantum transfer matrix method as
described in the main part of the paper.

\vspace{0.5cm}
\begin{tabular}{|c | c c c | c c c|}
\hline
\multicolumn{7}{|c|}{30$^{\circ}$-N\'eel} \\
\hline
$\Delta$ & $F_{x1}$ & $F_{x2}$ & $F_{x3}$ & $F_{z1}$ & $F_{z2}$ & $F_{z3}$ \\
\hline
2 & -0.2951 & 0.06493 & 0.002095 & -0.5799 & 0.3076 & -0.1919 \\
4 & -0.1789 & 0.02855 & 0.009815 & -0.7231 & 0.5150 & -0.3864 \\
8 & -0.09891 & 0.01684 & 0.006760 & -0.7565 & 0.5705 & -0.4364 \\
\hline
\end{tabular} 

\vspace{0.5cm}
\begin{tabular}{|c | c c c | c c c|}
\hline
\multicolumn{7}{|c|}{20$^{\circ}$-N\'eel} \\
\hline
$\Delta$ & $F_{x1}$ & $F_{x2}$ & $F_{x3}$ & $F_{z1}$ & $F_{z2}$ & $F_{z3}$ \\
\hline
2 & -0.3194 & 0.06760 & -0.004129 & -0.6221 & 0.3553 & -0.2438 \\
4 & -0.2000 & 0.02499 & 0.004992 & -0.8122 & 0.6526 & -0.5552 \\
8 & -0.1068 & 0.008910 & 0.004371 & -0.8709 & 0.7571 & -0.6679 \\
\hline
\end{tabular} 

\vspace{0.5cm}
\begin{tabular}{|c | c c c | c c c|}
\hline
\multicolumn{7}{|c|}{10$^{\circ}$-N\'eel} \\
\hline
$\Delta$ & $F_{x1}$ & $F_{x2}$ & $F_{x3}$ & $F_{z1}$ & $F_{z2}$ & $F_{z3}$ \\
\hline
2 & -0.3367 & 0.07108 & -0.009721 & -0.6482 & 0.3872 & -0.2808 \\
4 & -0.2181 & 0.02592 & -0.0005958 & -0.8683 & 0.7483 & -0.6845 \\
8 & -0.1172 & 0.007326 & 0.001055 & -0.9443 & 0.8910 & -0.8537 \\
\hline
\end{tabular} 

\vspace{0.5cm}
\begin{tabular}{|c | c c c | c c c|}
\hline
\multicolumn{7}{|c|}{90$^{\circ}$-Ferromagnetic} \\
\hline
$\Delta$ & $F_{x1}$ & $F_{x2}$ & $F_{x3}$ & $F_{z1}$ & $F_{z2}$ & $F_{z3}$ \\
\hline
4 & 0.2391 & 0.1039 & 0.04256 & 0.1304 & 0.01049 & 0.08296 \\
\hline
\end{tabular} 

\vspace{0.5cm}
\begin{tabular}{|c | c c c | c c c|}
\hline
\multicolumn{7}{|c|}{60$^{\circ}$-Ferromagnetic} \\
\hline
$\Delta$ & $F_{x1}$ & $F_{x2}$ & $F_{x3}$ & $F_{z1}$ & $F_{z2}$ & $F_{z3}$ \\
\hline
4 & 0.2219 & 0.1322 & 0.07781 & 0.3266 & 0.2417 & 0.2735 \\
\hline
\end{tabular} 

\vspace{0.5cm}
\begin{tabular}{|c | c c c | c c c|}
\hline
\multicolumn{7}{|c|}{30$^{\circ}$-Ferromagnetic} \\
\hline
$\Delta$ & $F_{x1}$ & $F_{x2}$ & $F_{x3}$ & $F_{z1}$ & $F_{z2}$ & $F_{z3}$ \\
\hline
4 & 0.1064 & 0.09290 & 0.08058 & 0.7593 & 0.7452 & 0.7438 \\
\hline
\end{tabular}

\vspace{0.5cm}
\begin{tabular}{|c | c c c | c c c|}
\hline
\multicolumn{7}{|c|}{Majumdar Ghosh} \\
\hline
$\Delta$ & $F_{x1}$ & $F_{x2}$ & $F_{x3}$ & $F_{z1}$ & $F_{z2}$ & $F_{z3}$ \\
\hline
1.2 & -0.4799 & 0.1547 & -0.07218 & -0.5334 & 0.2172 & -0.1185 \\
1.4 & -0.4592 & 0.1377 & -0.05954 & -0.5583 & 0.2531 & -0.1428 \\
1.6 & -0.4399 & 0.1240 & -0.04998 & -0.5751 & 0.2793 & -0.1575 \\
2   &  -0.4081 & 0.1042 & -0.03752 & -0.5919 & 0.3080 & -0.1653\\ 
4 & -0.3317 & 0.06584 & -0.01802 & -0.5842 & 0.3045 & -0.1118 \\
8 & -0.2905 & 0.04834 & -0.01067 & -0.5524 & 0.2621 & -0.05291 \\
\hline
\end{tabular} 

\vspace{0.5cm}
\begin{tabular}{|c | c c c | c c c|}
\hline
\multicolumn{7}{|c|}{XXZ ground state: $\Delta_{0}=3$} \\
\hline
$\Delta$ & $F_{x1}$ & $F_{x2}$ & $F_{x3}$ & $F_{z1}$ & $F_{z2}$ & $F_{z3}$ \\
\hline
1.25 & -0.5102 & 0.1691 & -0.09802 & -0.5855 & 0.2668 & -0.1811 \\
1.5 & -0.4937 & 0.1488 & -0.09095 & -0.6597 & 0.3725 & -0.3010 \\
2 & -0.4320 & 0.1052 & -0.06263 & -0.7810 & 0.5784 & -0.5439 \\
2.5 & -0.3675 & 0.07300 & -0.03914 & -0.8560 & 0.7179 & -0.7054 \\
3 & -0.3148 & 0.05235 & -0.02452 & -0.8981 & 0.7990 & -0.7940 \\
4 & -0.2424 & 0.03031 & -0.01032 & -0.9344 & 0.8701 & -0.8622 \\
6 & -0.1677 & 0.01440 & -0.002004 & -0.9472 & 0.8958 & -0.8708 \\
\hline
\end{tabular}

\end{document}